\documentclass[11pt,a4paper]{article}
\pdfoutput=1
\usepackage{jheppub}
\usepackage{amsmath,amsfonts,amssymb,bbm, mathtools, mathrsfs}
\usepackage[scr=zapfc]{mathalfa}
\usepackage{hyperref, csquotes}
\usepackage{graphicx, color,subcaption,svg}
\usepackage{tikz,pgf,tikz-cd,float}
\usepackage{physics,tensor,upgreek}
\usepackage{enumitem}
\include{fdiagrams}

\newcommand{\one}{\mathbbm 1}
\newcommand{\R}{\mathbb R}
\newcommand{\C}{\mathbb C}
\newcommand{\e}{\textrm{e}}
\newcommand{\SUT}{\mathrm{SU}(2)}
\newcommand{\SUO}{\mathrm{SU}(1,1)}

\newcommand{\SL}{\text{SL$(2,\C)$}}

\newcommand{\spl}{\mathfrak{sl}\left(2,\C\right)}
\renewcommand{\TH}{\text{H}^3}
\newcommand{\OH}{\mathrm{H}^{1,2}}
\newcommand{\defeq}{\vcentcolon=}
\newcommand{\eqdef}{=\vcentcolon}

\newcommand{\qmarks}[1]{``#1''}

\newcommand{\GFT}{\mathrm{GFT}}
\newcommand{\GR}{\mathrm{GR}}
\newcommand{\CL}{\mathrm{CL}}
\renewcommand{\H}{\mathrm{H}}
\newcommand{\Pl}{\mathrm{Pl}}
\newcommand{\Hubble}{\mathscr{H}}

\newcommand{\mf}{\phi} 
\newcommand{\mm}{\pi_\phi} 
\newcommand{\pmm}{p_\phi} 
\newcommand{\rf}{\chi} 
\newcommand{\epsp}{\epsilon^+} 
\newcommand{\epsm}{\epsilon^-}
\newcommand{\pip}{\pi_0^+}
\newcommand{\pim}{\pi_0^-}
\newcommand{\tpip}{\tilde{\pi}_0^+}
\newcommand{\tpim}{\tilde{\pi}_0^-}
\newcommand{\slrcw}{\tilde{\sigma}} 
\newcommand{\tlrcw}{\tilde{\tau}} 

\makeatletter
\let\orgdescriptionlabel\descriptionlabel
\renewcommand*{\descriptionlabel}[1]{%
  \let\orglabel\label
  \let\label\@gobble
  \phantomsection
  \edef\@currentlabel{#1\unskip}
  \let\label\orglabel
  \orgdescriptionlabel{#1}%
}
\makeatother

\begin{document}
\title{Scalar Cosmological Perturbations from Quantum Entanglement within Lorentzian Quantum Gravity}

\author[a,b,c]{Alexander F. Jercher,}
\emailAdd{alexander.jercher@uni-jena.de}

\author[a,d]{Luca Marchetti,}
\emailAdd{luca.marchetti@unb.ca}

\author[a,b,e]{Andreas G. A. Pithis}
\emailAdd{andreas.pithis@physik.lmu.de}

\affiliation[a]{Arnold Sommerfeld Center for Theoretical Physics,\\ Ludwig-Maximilians-Universit\"at München \\ Theresienstrasse 37, 80333 M\"unchen, Germany, EU}
\affiliation[b]{Munich Center for Quantum Science and Technology (MCQST),\\ Schellingstr. 4, 80799 M\"unchen, Germany, EU}
\affiliation[c]{Theoretisch-Physikalisches Institut, Friedrich-Schiller-Universit\"{a}t Jena\\ Max-Wien-Platz 1, 07743 Jena, Germany, EU}
\affiliation[d]{Department of Mathematics and Statistics,\\ University of New Brunswick,\\ Fredericton, NB, Canada E3B 5A3}
\affiliation[e]{Center for Advanced Studies (CAS),\\ Ludwig-Maximilians-Universit\"at München \\ Seestrasse 13, 80802 München, Germany, EU}

\date{\today}

\begin{abstract}
{
We derive the dynamics of (isotropic) scalar perturbations from the mean-field hydrodynamics of full Lorentzian quantum gravity, as described by a two-sector (timelike and spacelike) Barrett-Crane group field theory (GFT) model. The rich causal structure of this model allows us to consistently implement in the quantum theory the causal properties of a physical Lorentzian reference frame composed of four minimally coupled, massless, and free scalar fields. Using this frame, we are able to effectively construct relational observables that are used to recover macroscopic cosmological quantities. In particular, small isotropic scalar inhomogeneities emerge as a result of (relational) nearest-neighbor two-body entanglement between degrees of freedom of the underlying quantum gravity theory. The dynamical equations we obtain for geometric and matter perturbations show remarkable agreement with those of classical general relativity for sub-Planckian modes. Quantum gravity effects produce important deviations from the classical general relativistic dynamics for trans-Planckian modes, which we show to be associated to sub-horizon scales in the physical reference frame we are employing.
}
\end{abstract}

\maketitle

\section{Introduction}\label{sec:Introduction}

The prevailing Lambda cold dark matter ($\Lambda$CDM) paradigm~\cite{Planck:2018vyg} gives a remarkably accurate description of the large-scale structure of our Universe in terms of a homogeneous and isotropic Friedmann-Lema\^{\i}tre-Robertson-Walker (FLRW) background spacetime, complemented by small, inhomogeneous perturbations thereof. This nearly homogeneous and isotropic geometry is sourced by a cosmic fluid permeating the Universe. Within the $\Lambda$CDM paradigm, this fluid is composed of standard and non-standard matter, in the form of a cosmological constant $\Lambda$ and cold dark matter.

However, despite its successes, this model falls short of providing a complete physical and conceptual clarification of critical open questions about our Universe touching for instance on the fate of the initial Big Bang singularity, the origin of cosmic structure, the enigmatic nature of dark energy and dark matter and the issue of the Hubble tension. Quantum gravity (QG) may help shed some light on these unanswered questions. On the other hand, cosmology, with its precision observations, presents itself as an incredibly promising testing ground for quantum gravity theories. For these reasons, extracting cosmological physics from QG is a key step to make substantial progress in both QG and theoretical cosmology.

Nevertheless, extracting cosmology from full QG is a formidable challenge, especially for approaches with \qmarks{pre-geometric} degrees of freedom that differ significantly from the continuum fields of standard cosmology. Thus, two main difficulties arise in this endeavor: (i) The transition from the microscopic realm of QG cosmological physics requires an appropriate definition of a semi-classical and continuum limit~\cite{Oriti:2018tym}. This possibly involves a coarse-graining procedure where macroscopic degrees of freedom are identified as effectively emerging from the underlying QG entities~\cite{Oriti:2018dsg}. (ii) Inherent to background independent approaches to QG is the absence of the conventional spacetime manifold structure. Consequently, notions of time evolution and spatial localization can only be understood in a relational sense~\cite{Rovelli:1990ph,Hoehn:2019fsy,Rovelli:2001bz,Dittrich:2005kc,Goeller:2022rsx}. 

Tensorial group field theories (TGFTs)~\cite{Freidel:2005qe,Oriti:2006se,Carrozza:2013oiy,Carrozza:2016vsq} constitute a promising framework that offers a versatile tool set to tackle both of the aforementioned challenges. Combinatorially, TGFTs can be seen as a higher-dimensional generalization of matrix models~\cite{DiFrancesco:1993cyw} and are closely related to tensor models~\cite{Gurau:2011xp,GurauBook,Gurau:2016cjo,Jercher:2022mky}. TGFT models which are enriched by quantum geometric degrees of freedom (encoded in group theoretic data) are called Group Field Theories (GFTs) ~\cite{Freidel:2005qe,Oriti:2011jm} and can be seen as quantum and statistical field theories of spacetime defined on a group manifold. Due to their quantum geometric interpretation, GFT models can be related to many other QG approaches, such as loop quantum gravity (LQG)~\cite{Ashtekar:2004eh,Oriti:2013aqa,Oriti:2014yla}, spin foam models~\cite{Perez:2003vx,Perez:2012wv}, simplicial gravity~\cite{Bonzom:2009hw,Baratin:2010wi,Baratin:2011tx,Baratin:2011hp,Finocchiaro:2018hks} or dynamical triangulations~\cite{Loll:1998aj,Ambjorn:2012jv,Jordan:2013sok,Loll:2019rdj}. 

It is generally expected that the continuum limit of TGFTs manifests itself in a phase of coarse-grained collective behavior~\cite{Oriti:2013jga,Gielen:2013kla,Oriti:2015rwa,Marchetti:2022nrf}. Although extremely challenging, substantial progress in exploring the phase diagram of TGFTs has been accomplished by adapting powerful tools from local statistical and quantum field theories, see for instance~\cite{Pithis:2018eaq,Marchetti:2020xvf,Benedetti:2015et,BenGeloun:2015ej,Benedetti:2016db,BenGeloun:2016kw,Carrozza:2016vsq,Pithis:2020sxm,Pithis:2020kio,Geloun:2023ray}. Primary examples of these techniques are the functional renormalization group (FRG) methodology~\cite{Berges:2000ew,Delamotte:2007pf,Kopietz:2010zz,Dupuis:2020fhh} and Landau-Ginzburg mean-field theory~\cite{Zinn-Justin:2002ecy,Zinn-Justin:2007uvz,Kopietz:2010zz}. Of particular relevance for this article, the Landau-Ginzburg method has been applied recently to quantum geometric TGFT models with Lorentzian signature, finding a remarkable robustness of the mean-field approximation~\cite{Marchetti:2022igl,Marchetti:2022nrf}.

Inspired by the physics of Bose-Einstein condensates~\cite{pitaevskii2016bose}, the GFT condensate cosmology program~\cite{Gielen:2013kla,Oriti:2016acw,Gielen:2016dss,Pithis:2019tvp,Marchetti:2020umh} models this mean-field description of GFTs in terms of coherent peaked states (CPSs) subject to the classical GFT equation of motion~\cite{Gielen:2013naa,Gielen:2014ila,Gielen:2014uga,Oriti:2016ueo,Oriti:2016qtz,Marchetti:2020umh,Marchetti:2020qsq}. In this way, this program has successfully reproduced the Friedmann dynamics for spatially homogeneous and isotropic flat geometries at late relational times, while at early relational times, the initial Big Bang singularity of classical cosmology is replaced by a Big Bounce~\cite{Oriti:2016ueo,Oriti:2016qtz,Marchetti:2020umh,Marchetti:2020qsq}. Of particular relevance to our present work, all of these results, initially obtained using an Engle-Pereira-Rovelli-Livine-like (EPRL-like) GFT model~\cite{Baratin:2011hp,BenGeloun:2010qkf}, have been independently reproduced through an extended formulation of the Barrett-Crane (BC) model~\cite{Jercher:2021bie}. This observation hints at a potential universal behavior of different microscopic GFT models after coarse-graining in the continuum limit~\cite{Marchetti:2022nrf}. 

Scalar cosmological perturbations in the GFT condensate cosmology framework have been studied in~\cite{Gielen:2022iuu,Gielen:2017eco,Gielen:2018xph,Gerhardt:2018byq}, and most notably in~\cite{Marchetti:2021gcv} where the dynamics of general relativity (GR) in the super-horizon limit of large perturbation wavelength are reproduced. For non-negligible wave vectors $k$, qualitative deviations from classical results arise in the spatial derivative term of the perturbation equations. Building on the interpretations for this mismatch proposed in~\cite{Marchetti:2021gcv}, here, we suggest that it ultimately originates from: (i) an insufficient coupling between the physical reference frame and the causal structure of the underlying geometry; and (ii) a lack of quantum gravitational correlations generating the macroscopic inhomogeneities.

In this paper, {we address these issues by making use} of the richer causal structures available in the extended Barrett-Crane GFT model. {We derive the dynamics of scalar cosmological perturbations which are remarkably closer to those of GR and thereby significantly improve the results of \cite{Marchetti:2021gcv}.} To {accomplish} this, we first establish a connection between the causal character of the quantum geometry and that of the clock and rods, thus making the Lorentzian interpretation of the physical frame manifest. Building on this interplay, we introduce perturbed coherent peaked states that capture the collective behavior of spacelike and timelike tetrahedra. At the background level, these states reproduce the homogeneous cosmological dynamics obtained in previous studies. Perturbations on the other hand are encoded in the two-body quantum entanglement within and between the spacelike and timelike sector. This procedure differs significantly from the purely spacelike perturbations of~\cite{Marchetti:2021gcv} and can be seen as out-of-condensate perturbations. Finally, we extract and study the dynamics of cosmological perturbations from the mean-field quantum dynamics associated to such perturbed CPS, and we compare our results to predictions of classical GR. We refer the reader who is interested only in the cosmological dynamics and not its derivations directly to Sec.~\ref{sec:Solutions of GFT and GR perturbations}. 

The article is organized as follows: We begin by setting up the complete BC model with spacelike and timelike tetrahedra in Sec.~\ref{sec:fockspacestructure}, introducing in particular the two-sector Fock space. In Sec.~\ref{sec:Coupling reference and matter fields}, we couple four reference and one matter field to the model, proposing a restriction to couple clocks and rods to the causality of the underlying geometry. Within the causally extended setting, CPS and their relational dynamics are introduced in Sec.~\ref{sec:Coherent peaked states and perturbations}, first at the background and then at the level of first-order perturbations. In Secs.~\ref{sec:Geometric observables}--\ref{sec:Dynamics of matter observables}, we study the expectation values of relevant operators with respect to the condensate state and derive dynamical equations thereof in an effective  relational fashion. Solutions of these equations are analyzed in Sec.~\ref{sec:Solutions of GFT and GR perturbations} and compared to solutions of the classical perturbation equations from GR. 

\section{Complete Barrett-Crane model: spacelike and timelike tetrahedra}\label{sec:The complete BC model}

{After a brief discussion on the status of causality in GFTs, we introduce the Barrett-Crane TGFT model with spacelike and timelike tetrahedra in Sec.~\ref{sec:fockspacestructure}. Thereafter, the coupling of matter and reference fields to this model is discussed in Sec.~\ref{sec:Coupling reference and matter fields}.} In particular, we present the possibility to relate the causal nature of clocks and rods to the causal structure of the underlying quantum geometry.

\subsection{Spacelike and timelike tetrahedra}\label{sec:fockspacestructure}

As introduced in Sec.~\ref{sec:Introduction}, GFTs are quantum and statistical field theories, defined on a group manifold $G$, endowed with a quantum geometric interpretation. One-particle excitations of a $d$-dimensional simplicial GFT are interpreted as $(d-1)$-dimensional simplices which, under non-local interactions, form $d$-dimensional simplices that collectively build up spacetime. A choice of group $G$, dimension $d$ and GFT-action $S_{\mathrm{GFT}}$ defines a specific GFT model, the partition function of which can be perturbatively expanded in terms of Feynman amplitudes dual to simplicial pseudo-manifolds~\cite{Gurau:2011xp,Carrozza:2013oiy,Carrozza:2016vsq}. Depending on the representation employed, these amplitudes can be associated to either spin foam or simplicial gravity amplitudes.

{Basic prerequisite for this work is a quantum geometric GFT model with an accessible causal structure, the importance of which is highlighted in approaches like causal dynamical triangulations~\cite{Ambjorn:2012jv,Jordan:2013sok,Loll:2019rdj} or causal set theory~\cite{Surya:2019ndm}. The issue of properly encoding microcausality within GFTs, spin foam models and LQG has been scarcely studied (see e.g. Refs.~\cite{Barrett:1999qw,Perez:2000ep,Livine:2002rh,Alexandrov:2005ar,Speziale:2013ifa,Neiman:2012fu}) and only recently aroused interest once again. This has been triggered by studies on the asymptotics~\cite{Simao:2021qno,Kaminski:2017eew,Liu:2018gfc,Han:2021rjo,Han:2021bln,Han:2021kll} of the Conrady-Hnybida (CH) extension~\cite{Conrady:2010kc,Conrady:2010vx} of the EPRL spin foam model~\cite{Rovelli:2011eq,Perez:2012wv} which includes spacelike and timelike tetrahedra to encode Lorentzian quantum geometries. In addition, in the context of effective spin foams~\cite{Asante:2020qpa,Asante:2020iwm,Asante:2021zzh}, the path integral for Lorentzian quantum gravity has been studied, shedding light on causality violating configurations~\cite{Asante:2021phx,Dittrich:2023rcr,Jercher:2023csk}. Finally, forming the basis of the present work, a completion of the Lorentzian Barrett-Crane GFT and spin foam model~\cite{Barrett:1999qw,Perez:2000ep,Perez:2000ec} with $d=4$ and $G = \SL$ has been developed in~\cite{Jercher:2022mky} which also includes timelike and lightlike tetrahedra. 
}

\paragraph{The Lorentzian Barrett-Crane GFT model.} 

Restricting to spacelike tetrahedra and to homogeneous and isotropic condensates, it has been shown in~\cite{Jercher:2021bie} that at late times flat Friedmann dynamics emerge for the volume of the universe. In this work, we go beyond this setting and include, as a minimal extension, also timelike tetrahedra while excluding lightlike tetrahedra from the outset (Assumption \ref{ass:ds1}). Indeed, as we will argue below, {timelike} tetrahedra are necessary to properly couple the reference fields according to the signature of the quantum geometric building blocks (see Sec.~\ref{sec:Coupling reference and matter fields} for more details). As the results of Sec.~\ref{sec:Perturbation equations from quantum gravity} show, this restricted set of causal configurations is already sufficient to yield GR-like cosmological perturbations. 

Within the extension to spacelike and timelike tetrahedra, the group fields are functions $\varphi(g_v,X_{\alpha})$ with $\alpha\in\{+,-\}$ assigning a spacelike, respectively a timelike signature. The four group elements $g_v = (g_1,g_2,g_3,g_4)$ are elements of $\SL$ and $X_\alpha$ is a normal vector with the according signature. More precisely, $X_\alpha$ is an element of the homogeneous space $\SL/\mathrm{U}^{(\alpha)}$, where $\mathrm{U}^{(+)} = \SUT$ and $\mathrm{U}^{(-)} = \SUO$ are the stabilizer subgroups of the respective normal vectors
\begin{equation}
X_+ = (1,0,0,0),\qquad X_- = (0,0,0,1).
\end{equation}
The fields $\varphi_\pm$ exhibit two defining symmetries
\begin{align}
\varphi(g_v,X_\alpha) &= \varphi(g_v h^{-1},h\cdot X_\alpha),\quad\forall h \in\SL,\label{eq:closure}\\[7pt]
\varphi(g_v,X_{\alpha}) &= \varphi(g_vu_v,X_{\alpha}),\quad \forall u_1,...,u_4\in\mathrm{U}_{X_{\alpha}},\label{eq:simplicity}
\end{align}
referred to as closure and simplicity constraints, respectively, or both together as geometricity constraints. $\mathrm{U}_{X_{\alpha}}$ denotes the stabilizer subgroup of $\SL$ with respect to the normal vector $X_\alpha$, which is isomorphic to $\mathrm{U}^{(\alpha)}$. Based on the ideas of~\cite{Baratin:2011tx,Baratin:2011hp}, extending the domain $\SL^4$ by the normal vector allows to impose the constraints in a covariant and commuting fashion. Consequently, $X_\alpha$ is only considered as an auxiliary non-dynamical variable that does not carry intrinsic geometric information.

Following the introduction of the fields $\varphi$, the model is then defined by its action $S[\varphi,\bar{\varphi}]$, which decomposes into a kinetic part
\begin{equation}\label{eq:general GFT kinetic action}
K[\varphi,\bar{\varphi}] =
\sum_{\alpha}\int\limits_{\SL^8}\dd{g_v}\dd{g_w}\int\limits_{\SL/\mathrm{U}^{(\alpha)}}\dd{X_{\alpha}}\bar{\varphi}(g_v,X_{\alpha})\mathcal{K}_\alpha(g_v,g_w)\varphi(g_w,X_{\alpha}),
\end{equation}
with kernels $\mathcal{K}_\alpha$ and an interaction part $V[\varphi,\bar{\varphi}]$, the latter of which is explicitly given in~\cite{Jercher:2021bie,Jercher:2022mky}. A priori, the interaction $V$ does incorporate all possible simplicial interactions composed of altogether five tetrahedra of spacelike and timelike signature.

For explicit computations as well as to connect to the spin foam formalism~\cite{Perez:2003vx,Perez:2012wv}, the spin-representation of the group field is a crucial tool that is going to be utilized heavily in this work. Following the constructions of~\cite{Jercher:2022mky}, the expansion of $\varphi(g_v,X_\alpha)$ in terms of unitary irreducible $\SL$-representation labels $(\rho,\nu)\in\R\times\mathbb{N}/2$ of the principal series is given by
\begin{align}
\varphi(g_v,X_+) &= \int\dd{\rho_v}\sum_{j_v m_v l_v n_v}\varphi^{\rho_v}_{+,j_v m_v}\prod_{i=1}^4 \rho_i^2 D^{(\rho_i,0)}_{j_i m_i l_i m_i}(g_i g_X)\bar{\mathcal{I}}^{\rho_i,+}_{l_i m_i},\label{eq:spacelike group field spin rep}\\[7pt]
\varphi(g_v,X_-) &= \int\dd{\rho_v}\sum_{\nu_v}\sum_{j_v m_v l_v n_v}\varphi^{\rho_v\nu_v}_{-,j_v m_v}\prod_{i=1}^4 \left(\rho_i^2\delta_{\nu_i,0}+\nu_i^2\delta(\rho_i)\chi_{\nu_i}\right)D^{(\rho_i,\nu_i)}_{j_i m_i l_i n_i}(g_ig_X)\bar{\mathcal{I}}^{\rho_i\nu_i,-}_{l_i n_i}.\label{eq:timelike group field spin rep}
\end{align}
The continuous label $\rho\in\R$ is associated to spacelike faces, irrespective of the containing tetrahedron being spacelike or timelike. $\nu\in 2\mathbb{N}^+$ is associated to timelike faces, which are necessarily contained in timelike tetrahedra. These conditions follow from the the simplicity constraint of the Barrett-Crane model~\cite{Jercher:2021bie,Jercher:2022mky}, derived in the framework of integral geometry~\cite{VilenkinBook}. $\mathcal{I}^\pm$ are invariant symbols that ensure the constraints of Eqs.~\eqref{eq:closure} and~\eqref{eq:simplicity} which, upon integration over the normal vector, yield generalized Barrett-Crane intertwiners $B^{\rho_v\nu_v,\alpha}_{l_v n_v}$ defined in~\cite{Jercher:2022mky}.
\paragraph{Extended Fock space.}
The extension of the GFT model to include also timelike tetrahedra necessitates the definition of an extended Fock space structure. That is because the individual field operators $\hat{\varphi}(g_v,X_+)$ and $\hat{\varphi}(g_v,X_-)$ are defined on different domains and act on different Fock spaces. These two sectors, denoted by $\mathcal{F}_+$ and $\mathcal{F}_-$, respectively, are defined as
\begin{equation}
\mathcal{F}_\pm \defeq \bigoplus_{N = 0}^\infty \mathrm{sym}\left(\mathcal{H}_\pm^{(1)}\otimes...\otimes \mathcal{H}_\pm^{(N)}\right),
\end{equation}
where the one-particle Hilbert spaces for spacelike and timelike tetrahedra are given by
\begin{equation}
\mathcal{H}_+ \defeq L^2\left(\SL^4\times\TH/\sim_+\right),
\end{equation}
(see also~\cite{Jercher:2021bie}) and
\begin{equation}
\mathcal{H}_-\defeq L^2\left(\SL^4\times \OH/\sim_-\right).
\end{equation}
Here, $\sim_{\pm}$ denotes the imposition of geometricity constraints with respect to a timelike and spacelike normal, respectively. These normal vectors are elements of the two-sheeted and one-sheeted hyperboloids, $X_+\in\TH$ and $X_- \in\OH$.

The total Fock space $\mathcal{F}$ of the theory is constructed as the tensor product of $\mathcal{F}_+$ and $\mathcal{F}_-$, i.e.
\begin{equation}
\mathcal{F}\defeq \mathcal{F}_+\otimes\mathcal{F}_- = \bigoplus_{N_{\text{tot}}}^{\infty}\bigoplus_{N+M = N_{\text{tot}}}\mathrm{sym}\left(\mathcal{H}_+^{\otimes N}\right)\otimes\mathrm{sym}\left(\mathcal{H}_-^{\otimes M}\right).
\end{equation}
{As usual in quantum field theory, the linear structure of the individual and total Fock spaces is strictly only possible when interactions, as introduced for the complete BC model in ~\cite{Jercher:2022mky}, are neglected, as assumed hereafter. For a discussion of this matter for the single-sector Fock space, we refer to~\cite{Gielen:2013naa}.} Creation and annihilation operators of spacelike and timelike tetrahedra, abbreviated as $\hat{\varphi}_{\pm}^\dagger$ and $\hat{\varphi}_\pm$, respectively, are defined in terms of the creation and annihilation operators of the respective sectors. We frequently suppress the trivial action on the opposite sector, e.g. $\hat{\varphi}_+ \equiv \hat{\varphi}_+\otimes \one_-$. Following the usual commutation rules, the operators $\hat{\varphi}_{\pm}$ satisfy the algebra
\begin{equation}\label{eq:commutation relations 1}
\comm{\hat{\varphi}_{\pm}}{\hat{\varphi}_{\pm}^{\dagger}} = \one_{\pm},\qquad \comm{\hat{\varphi}_{\pm}}{\hat{\varphi}_{\pm}} = \comm{\hat{\varphi}_{\pm}^{\dagger}}{\hat{\varphi}_{\pm}^{\dagger}} = 0,
\end{equation}
where $\one_\pm$ is the identity on $\mathcal{F}_{\pm}$ respecting closure and simplicity constraints. Notice that, by construction, operators of different sectors mutually commute
\begin{equation}\label{eq:commutation relations 2}
\comm{\varphi_{\pm}}{\varphi_{\mp}^{\dagger}} = \comm{\hat{\varphi}_{\pm}}{\hat{\varphi}_{\mp}} = \comm{\hat{\varphi}_{\pm}^{\dagger}}{\hat{\varphi}_{\mp}^{\dagger}} = 0.
\end{equation}
The vacuum state $\ket{\emptyset}$ of the total Fock space is naturally defined as the state which is annihilated by both, $\hat{\varphi}_+$ and $\hat{\varphi}_-$. It therefore corresponds to the tensor product of the respective vacuum states, i.e. $\ket{\emptyset} = \ket{\emptyset}_+\otimes\ket{\emptyset}_-$.
\paragraph{Operators.}
Building up on the Fock space structure we introduced, operators are in general defined as convolutions of kernels with creation and annihilation operators (see~\cite{Oriti:2016qtz,Oriti:2013aqa,Marchetti:2020umh} for further details).\footnote{If the group is non-compact and the field symmetries yield empty group integrations, a regularization procedure is required, for which we refer to~\cite{Jercher:2021bie,Gielen:2013naa} (Assumption~\ref{ass:ks5}).} For the purposes of this work, we are particularly interested in one- and two-body operators on both sectors, i.e. the spacelike and timelike ones. The most important one-body operators are the $\alpha$-number operator
\begin{equation}\label{eq:number operator}
\hat{N}_\alpha = \int\dd{g_v}\dd{X_\alpha}\hat{\varphi}^{\dagger}(g_v,X_\alpha)\hat{\varphi}(g_v,X_\alpha),
\end{equation}
and the spatial $3$-volume operator
\begin{equation}\label{eq:volume operator}
\hat{V} = \int\dd{g_v}\dd{X_+}\hat{\varphi}^\dagger(g_v,X_+)V(g_v)\hat{\varphi}(g_v,X_+).
\end{equation}
In spin-representation, the kernel $V$ of the volume operator is given in analogy to the eigenvalues of the LQG volume operator~\cite{Rovelli:1994ge,Barbieri:1997ks,Brunnemann:2004xi,Ding:2009jq}. In case of isotropy of the representation labels $(\rho_i \equiv \rho)$ the kernel scales as $V\sim\rho^{3/2}$~\cite{Oriti:2016qtz,Marchetti:2020umh,Jercher:2021bie}.

A two-body operator $\hat{O}_{\alpha\beta}$ that describes a non-trivial correlation between the sectors $\alpha$ and $\beta$ is generally given by
\begin{equation}\label{eq:two-body operator}
\hat{O}_{\alpha\beta} = \int\dd{g_v}\dd{X_\alpha}\dd{g_w}\dd{X_\beta}O(g_v,X_\alpha,g_w,X_\beta)\hat{\varphi}^{\dagger}(g_v,X_\alpha)\times\hat{\varphi}^{\dagger}(g_w,X_\beta),
\end{equation}
where $\times$ symbolizes operator multiplication, ``$\cdot$'', if $\alpha=\beta$ or a tensor product, ``$\otimes$'', if $\alpha\neq \beta$. Notice that, $\hat{O}_{\alpha\beta}$ does not factorize in general, thus creating an entangled state when acting on a product state in $\mathcal{F}_+\otimes\mathcal{F}_-$. In Sec.~\ref{sec:Perturbed coherent peaked states}, we introduce three such operators, $\hat{\delta\Phi}, \hat{\delta\Psi}$ and $\hat{\delta\Xi}$, which encode two-body quantum entanglement, constituting the source of cosmological perturbations that we later derive.

\subsection{Coupling scalar fields: Matter and physical Lorentzian reference frame}\label{sec:Coupling reference and matter fields}

{As mentioned in Sec.~\ref{sec:Introduction}, in background independent theories, physical observables are naturally understood as relational, localizing dynamical degrees of freedom with respect to other dynamical degrees of freedom. However, the implementation of a relational description in full general relativity is quite complicated, even more so for quantum gravity theories. This is especially true for approaches characterized by new fundamental pre-geometric degrees of freedom, since relationality -- as we usually understand it -- is tightly related to the emergence of continuum notions~\cite{Marchetti:2020umh}.}
In the context of classical homogeneous cosmology, the simplest implementation of the relational strategy involves {a minimally coupled massless free (MCMF)} scalar field, which serves as a relational clock~\cite{Domagala:2010bm,Giesel:2012rb}. If inhomogeneities are taken into account, three additional degrees of freedom, serving as relational rods, are required. This can be achieved by including three more {MCMF} \qmarks{rod} scalar fields \cite{Giesel:2012rb}. Another explicit example for such a physical reference frame is given by Brown-Kucha\v{r} dust~\cite{Brown:1994py,Kuchar:1995xn,Bicak:1997bx}, which has for instance been employed to define a fully relational cosmological perturbation theory~\cite{Giesel:2007wi,Giesel:2007wk}.

{In this work, following~\cite{Marchetti:2021gcv}, we implement the relational strategy by using a physical Lorentzian reference frame} {composed of four MCMF scalar fields, serving as the dynamical clock and rods of our system.} Furthermore, we introduce an additional {MCMF} \qmarks{matter} scalar field which is assumed to dominate the field content of the emergent cosmology (Assumption~\ref{ass:ds5}). Following the strategy of~\cite{Oriti:2016qtz,Li:2017uao,Gielen:2018fqv}, the scalar fields are coupled to the GFT in such a way that the Feynman amplitudes correspond to simplicial gravity path integrals with a minimally coupled scalar field placed on dual vertices of the triangulation and propagating along dual edges (see~\ref{ass:ds2}). As shown in detail in Refs.~\cite{Oriti:2016qtz,Li:2017uao}, this is realized by extending the domain of the group field by the scalar field values,
\begin{equation}
\varphi(g_v,X_\alpha)\longrightarrow \varphi(g_v,X_\alpha,\rf^\mu,\mf),
\end{equation}
where $\rf^\mu = (\rf^0,\vb*{\rf})$ are the four reference fields with $\rf^0$ being the clock and $\vb*{\rf}$ being the three rods, and where $\mf$ denotes the additional \enquote{matter} scalar field. {We use $\mu,\nu,...\in\{0,1,2,3\}$ as indices for the reference fields as well as for harmonic coordinates, discussed in Appendix~\ref{sec:Classical perturbation theory}. Lower Latin letters $i,j,...\in\{1,2,3\}$ denote hereby spatial components.  For other than harmonic coordinates, spacetime indices are given by lower Latin letters $a,b,... \in\{0,1,2,3\}$.} 

The two kinetic kernels $\mathcal{K}_\alpha$, entering Eq.~\eqref{eq:general GFT kinetic action}, are extended to
\begin{equation}
\mathcal{K}_\pm(g_v,g_w) \longrightarrow \mathcal{K}_\pm(g_v,g_w,(\rf^\mu-\rf^{\prime \mu})^2,(\mf-\mf')^2),
\end{equation}
respecting the translation and reflection invariance of the classical actions for the fields $\rf^\mu$ and $\mf$. Importantly, it is the kinetic kernels which encode the propagation of the scalar fields along the simplicial complex. Since the scalar field is understood to be constant on a single simplex, the five group fields entering the vertex action $V[\varphi,\bar{\varphi}]$ carry the same scalar field value. In this sense, the interactions are local with respect to the scalar field degrees of freedom~\cite{Li:2017uao,Jercher:2021bie}. 

Notice that the Fock space structure introduced in Sec.~\ref{sec:fockspacestructure} naturally extends to the case where such scalar fields are present, such that the non-zero commutation relations of Eq.~\eqref{eq:commutation relations 1} are given by
\begin{equation}
\comm{\hat{\varphi}_\pm(\rf^\mu,\mf)}{\hat{\varphi}^{\dagger}_\pm(\rf^{\prime\mu},\mf')}=\one_\pm\delta^{(4)}(\rf^\mu-{\rf'}^\mu)\delta(\mf-\mf').
\end{equation}
Also, operators now include an integration over the full domain including the scalar field values. For instance, the $\alpha$-number operator is now defined as
\begin{equation}
\hat{N}_\alpha = \int\dd{g_v}\dd{X_\alpha}\dd[4]{\rf}\dd{\mf}\;\hat{\varphi}^\dagger(g_v,X_\alpha,\rf^\mu,\mf)\hat{\varphi}(g_v,X_\alpha,\rf^\mu,\mf).
\end{equation}
For further details, we refer to~\cite{Marchetti:2020umh,Jercher:2021bie}.

\paragraph{Clock and rods.}
In GR, perturbation equations clearly distinguish between derivatives with respect to (clock) time and (rods) space. As we can see explicitly from e.g.\ equation \eqref{eq:classical relative perturbed volume equation}, it is not just a matter of signature: when the physical frame is made of four minimally coupled, massless and free scalar fields, their harmonic behavior imposes a different relative weight of $a^4$ between (relational) space and time derivatives. Since symmetries on field space of the classical action are naturally reflected at the level of the kernels of the GFT action \cite{Oriti:2016qtz}, one would naively expect that, by imposing Lorentz symmetry at the level of the frame field action, one would recover effective equations which at least show the right signature of temporal and spatial derivative terms. However, as shown in~\cite{Marchetti:2021gcv}, this seems not to be the case: the signature of effective equations turns out to be independent on the symmetries of the frame action, and to be fixed essentially only by the parameters of the CPSs. On top of that, in the effective perturbation equations derived in~\cite{Marchetti:2021gcv}, temporal and spatial derivatives enter with the same weight. As emphasized already in~\cite{Marchetti:2021gcv}, this property of the effective equations is the source of a crucial mismatch with GR when perturbations momenta are $k> 0$. 

Both of the above results seem to suggest a difficulty in distinguishing between rods and clock at the level of the underlying GFT. This difficulty, however, may be solved by carefully coupling the frame according 
to the causal structure of the underlying geometry. To do so, in this work we take advantage of the extended structure of the present model, which, crucially, includes both spacelike and timelike tetrahedra. 

In the next two following paragraphs, we restrict our attention to the reference fields, since we do not intend to impose the same causality conditions on the matter field $\mf$. 

\paragraph{Classical and discrete perspective.}

In the continuum, the action of the four reference fields is given by
\begin{equation}
S[\chi] = \frac{1}{2}\int\dd[4]{x}\sqrt{-g}\sum_{\mu = 0}^3 g^{ab}\partial_a \chi^\mu\partial_b \chi^\mu
\end{equation}
and we enforce our assumption on the signature of clock and rods in terms of the conditions
\begin{subequations}\label{eq:dchi}
\begin{align}
& g(\partial\chi^0,\partial\chi^0) < 0,\label{eq:dchi0}\\[7pt]
& g(\partial\chi^i,\partial\chi^i) > 0,\label{eq:dchii}
\end{align}
\end{subequations}
meaning that the clock has a timelike gradient and rods have a spacelike gradient. {Notice that here we are not imposing a Lorentz symmetry (on field space) of the action. Indeed, as already argued in \cite{Marchetti:2021gcv}, only this choice for the action guarantees the appropriate sign for the energy density of the four fields.} We note two points here: First, the conditions on the gradients are not implied by the Klein-Gordon equation, but constitute indeed an additional physical requirement. Second, despite the point-particle intuition, a massless scalar field does not necessarily have a lightlike gradient. One of the simplest counter examples is that of a massless scalar field in a homogeneous background that can be used as a clock, therefore having a timelike gradient.\footnote{The authors thank S. Gielen and D. Oriti for a clarifying exchange on this matter.} 

Introducing a discretization of the continuum scalar field theory on a $2$-complex $\Gamma$, where the fields are placed on dual vertices $v\in\Gamma$, the action can be written as~
\begin{equation}\label{eq:discrete scalar field action}
S_{\Gamma}[\chi] = \frac{1}{2}\sum_{(vv')\in\Gamma}V_{(vv')}^{*(4)}\sum_{\mu=0}^3\left(\frac{\rf^\mu_v-\rf^\mu_{v'}}{l_{(vv')}}\right)^2,
\end{equation}
with $V_{(vv')}^{*(4)}$ being the Voronoi $4$-volume dual to the dual edge $(vv')$ and with $l_{(vv')}$ being the dual edge length~\cite{Hamber2009}. The sum over $(vv')$ denotes the sum over all dual edges which already suggests that, after quantization, the scalar field dynamics are going to be encoded in the edge amplitudes, i.e. in the kinetic kernels of the GFT. {Notice, that the discretization of a continuum field theory is always accompanied by ambiguities in the construction of discrete derivatives and dual geometric quantities~\cite{Hamber2009}. A necessary condition for a discretization to be viable is that it exhibits the correct continuum limit~\cite{Thurigen:2015uc}}.

\begin{figure}
    \centering
    \includegraphics[width=0.7\textwidth]{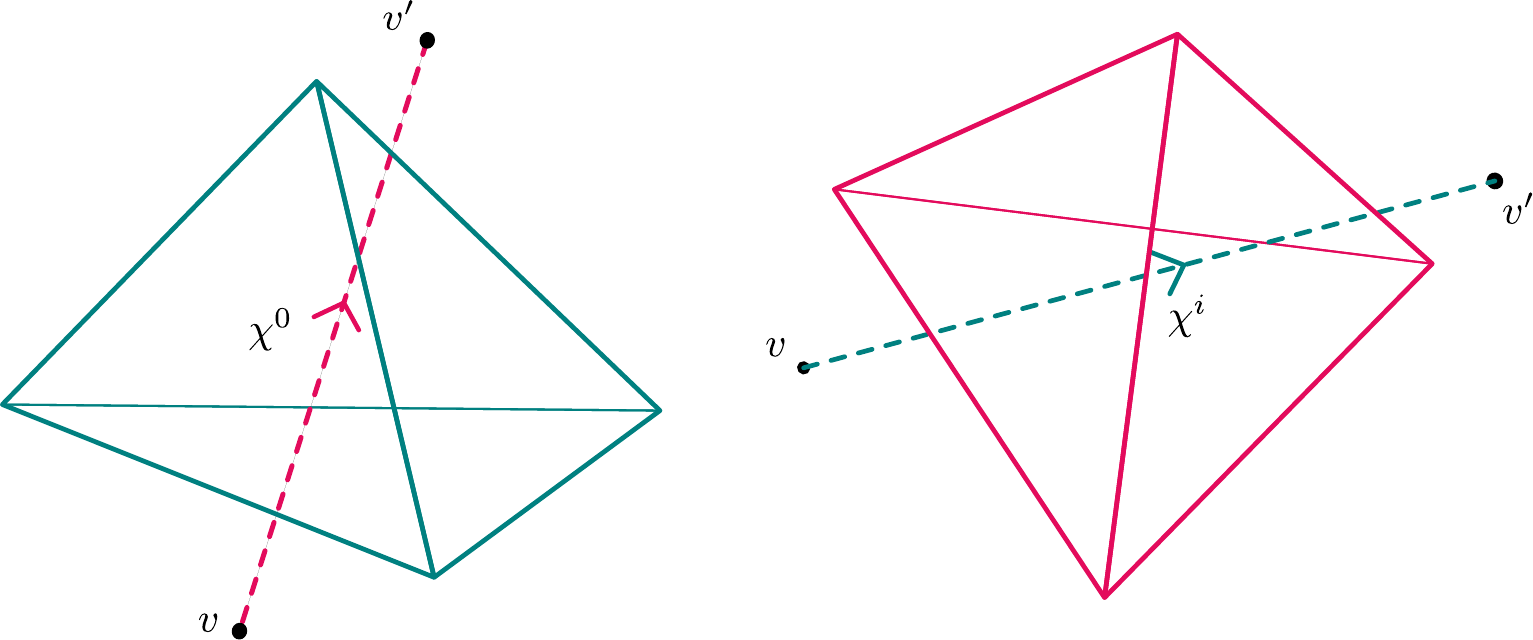}
    \caption{Left: A spacelike tetrahedron (teal) and its corresponding timelike dual edge (pink), connecting the dual vertices $v$ and $v'$. Following the restriction in Eq.~\eqref{eq:discrete clock propagatiom}, the clock $\rf^0$ only propagates along timelike dual edges. Right: A timelike tetrahedron (pink) and its corresponding spacelike dual edge (teal). In this case, Eq.~\eqref{eq:discrete rod propagation} imposes that rods $\rf^i$ only propagate along spacelike dual edges.}
    \label{fig:spacetime_tetrahedra}
\end{figure}

Given the Lorentzian structure of the original continuum manifold, $\Gamma$ is the dual complex of a Lorentzian discretization. This implies that edges $(vv')$ which are dual to spacelike and timelike tetrahedra are timelike and spacelike, respectively, for which a visual intuition is given in Fig.~\ref{fig:spacetime_tetrahedra}. {Notice that (i) the propagation of the scalar field is only sensitive to the signature of the dual edges and (ii) lightlike dual edges are excluded as a consequence of lightlike tetrahedra being excluded from the outset.} Based on this, the discrete scalar field action in Eq.~\eqref{eq:discrete scalar field action} can then be split into spacelike and timelike dual edges
\begin{equation}
S_\Gamma[\rf] = S_\Gamma^+[\rf]+S_\Gamma^-[\rf],
\end{equation}
defined by
\begin{equation}
S_{\Gamma}[\chi] = \frac{1}{2}\sum_{\mu = 0}^3\left[\sum_{(vv')\text{ t.l.}}w_{vv'}(\chi_v^{\mu}-\chi_{v'}^{\mu})^2+\sum_{(vv')\text{ s.l.}}w_{vv'}(\chi_v^{\mu}-\chi_{v'}^{\mu})^2\right],
\end{equation}
where $w_{vv'}$ denotes the geometric coefficients $V^{*(4)}_{(vv')}/l_{(vv')}^2$. Clearly, both types of reference fields propagate a priori on both types of dual edges. In analogy to the continuum Equations~\eqref{eq:dchi}, we propose to align the causal character of the reference frame with that of geometry by introducing the conditions
\begin{subequations}\label{eq:discrete frame propagation}
\begin{align}
\chi^0_v-\chi^0_{v'} &= 0,\quad \text{for }(vv')\text{ spacelike}\label{eq:discrete clock propagatiom},\\[7pt]
\chi^i_v-\chi^i_{v'} &= 0,\quad \text{for }(vv')\text{ timelike}\label{eq:discrete rod propagation}.
\end{align}
\end{subequations}
Formulated geometrically, the clock propagates along timelike dual edges and rods propagate along spacelike dual edges, as depicted in Fig.~\ref{fig:spacetime_tetrahedra}.\footnote{Notice that these conditions enforce the gradients of clocks and rods to have only temporal, respectively spatial, entries, which is a stronger condition than requiring the signature to be timelike or spacelike.} As a result, the discrete scalar field action splits into a clock and a rod part, associated to the signature of the respective dual edges. In the following, we discuss the realization of the conditions~\eqref{eq:discrete frame propagation} at the level of the GFT coupling.

\paragraph{Restriction of kinetic kernels.} Proceeding in parallel to~\cite{Li:2017uao}, the scalar field coupling is obtained by considering the simplicial gravity path integral on a given complex $\Gamma$ for the coupled gravity-matter system. As geometric quantities, dual edge lengths and tetrahedron volumes can be re-written in terms of bivector variables $B\in\spl$. Then, the GFT model which generates these amplitudes is derived, showing that the details of propagation are encoded in the kinetic kernel, while the GFT interaction is local with respect to the scalar fields. In particular, the details of the discretized geometric quantities $L^*_{vv'}$ and $V_{vv'}^{(3)}$ are encoded in the kinetic kernels, which we keep implicitly defined for the rest of this work.

As a result of the assignment of clocks to timelike dual edges and rods to spacelike dual edges above, the kinetic kernels $\mathcal{K}_\pm$ have a restricted dependence, given by
\begin{subequations}
\label{eqn:restrictionkinetic}
\begin{align}
\mathcal{K}_+(g_v,g_w, (\rf-\rf')^2) &= \mathcal{K}_+(g_v,g_w, (\rf^0-\rf^{0\prime})^2),\label{eq:spacelike kernel restriction}\\[7pt]
\mathcal{K}_-(g_v,g_w, (\rf-\rf')^2) &= \mathcal{K}_-(g_v,g_w, \abs{\vb*{\rf}-\vb*{\rf}'}^2)\label{eq:timelike kernel restriction}.
\end{align}
\end{subequations}
{This structure of the kinetic kernels corresponds to a strong imposition of the classical discrete conditions in Eq.~\eqref{eq:discrete frame propagation}. Clearly, a weaker imposition (e.g.\ via a Gaussian) including quantum fluctuations around the classical behavior is also possible.} For the rest of this work, we assume equations \eqref{eqn:restrictionkinetic} to hold (Assumption~\ref{ass:ds2}). Notice again, that the matter field $\mf$ is a priori not affected by this restriction. 

\section{Coherent peaked states and perturbations}\label{sec:Coherent peaked states and perturbations}

A crucial ingredient for the extraction of cosmological physics is the identification of states that can be associated with continuum, classical physics. Since GFTs are many-body quantum field theories of atoms of spacetime, by analogy with condensed matter systems, one would naturally expect these states to exhibit some form of collective behavior. The simplest form of such collective behavior is captured by coherent (or condensate) states. Importantly, strong evidence has recently been provided for the existence of such a condensate phase in quantum geometric TGFTs~\cite{Marchetti:2022igl,Marchetti:2022nrf}. One-body condensates of spacelike tetrahedra (whose condensate wavefunction encodes the macroscopic physics of the system) have indeed been used to derive an effective cosmological dynamics that exhibits a resolution of the big bang into a big bounce~\cite{Oriti:2016qtz,Marchetti:2020umh,Jercher:2021bie} and offers intriguing phenomenological implications such as dynamical isotropization~\cite{Pithis:2016cxg}, emergent inflation~\cite{deCesare:2016rsf,DeSousa:2023tja} or a late-time de Sitter phase~\cite{Oriti:2021rvm}. Motivated by these encouraging results and given the need to introduce timelike tetrahedra to improve the frame coupling, we propose to describe the background component of our collective states as the tensor product of a spacelike and a timelike one-body condensate
\begin{equation}
\ket{\sigma;x^0}\otimes\ket{\tau;x^0,\vb*{x}} = \mathcal{N}_\sigma\mathcal{N}_\tau\e^{\hat{\sigma}\otimes\one+\one\otimes\hat{\tau}}\ket{\emptyset},
\end{equation}
where the spacelike and timelike condensate states (whose details are discussed in Secs.~\ref{sec:Spacelike CPS} and~\ref{sec:Timelike CPS}) are denoted as $\ket{\sigma;x^0}$ and $\ket{\tau;x^0,\vb*{x}}$ respectively. At the right-hand-side of the above equation, we have rewritten the two condensate states as exponentials of the one-body operators $\hat{\sigma}$ and $\hat{\tau}$, respectively acting on the vacuum of $\mathcal{F}_+\otimes\mathcal{F}_-$. Finally, $\mathcal{N}_\sigma$ and $\mathcal{N}_\tau$ are normalization factors. {We emphasize that the timelike condensate state $\ket{\tau;x^0,\vb*{x}}$ is part of the background since the spatial peaking on $\vb*{x}$ only enters the peaking function and not the reduced condensate wavefunction $\tlrcw$, as discussed in detail down below.}

Given the above choice of background states, the most natural way to describe cosmological inhomogeneities would seem to be to include small inhomogeneous perturbations of the one-body condensate wavefunctions (since this is where the macroscopic physics of the system is encoded). However, as it will become clear in Sec.~\ref{sec:Background equations of motion}, the mean-field dynamics of the two one-body condensate wavefunctions turn out to be completely decoupled (at least in the regime of negligible interactions that we will consider below). Therefore, this choice would produce results that are effectively equivalent to those found in~\cite{Marchetti:2021gcv}, and thus eventually lead to the puzzling indistinguishably between clocks and rods discussed in the previous section and to the consequent mismatch with GR.

For this reason, and following the intriguing idea that non-trivial geometries are associated with entanglement of quantum gravity degrees of freedom, in this paper we propose an alternative description of cosmological inhomogeneities in terms of correlations of the underlying GFT quanta. Since, however, cosmological inhomogeneities are a macroscopic phenomenon from the QG perspective, the perturbations describing them should be realized in a collective manner at the level of the GFT. Therefore, we consider states of the form
\begin{equation}\label{eqn:perturbedstates}
\ket{\Delta;x^0,\vb*{x}} = \mathcal{N}_\Delta\exp(\hat{\sigma}\otimes\one+\one\otimes\hat{\tau}+\hat{\delta\Phi}\otimes\one+\hat{\delta\Psi}+\one\otimes\hat{\delta\Xi})\ket{\emptyset},
\end{equation}
where the perturbations are encoded in the operators $\hat{\delta\Phi}$, $\hat{\delta\Psi}$ and $\hat{\delta\Xi}$ (see also \ref{ass:ks1}). In general, these can be a combination of $n$-body operators, each of which encoding $n$-body correlations within and in between the spacelike and timelike sectors. However, in the following, we will restrict to the simplest non-trivial case, i.e.\ $2$-body operators. Also, as the perturbations are assumed to be small, the final form of states we will employ are a linearized version of Eq.~\eqref{eqn:perturbedstates}. Notice that in the picture we propose, we do not consider perturbations at the level of quantum geometric operators but rather at the level of the states, chosen appropriately to describe GR-like perturbations.  

Finally, in Sec.~\ref{sec:Effective relational dynamics}, we derive and discuss the effective relational dynamics of the linearized perturbed states. 

\subsection{Coherent peaked states for spacelike and timelike tetrahedra}\label{sec:CPS for spacelike and timelike tetrahedra}

As discussed above, our working assumption is that the spacelike and timelike sector of the background structure separate. Since the total Fock space $\mathcal{F}$, introduced in Sec.~\ref{sec:fockspacestructure}, is given in terms of a tensor product of $\mathcal{F}_+$ and $\mathcal{F}_-$, the states of the background are therefore product states. 

\subsubsection{Spacelike CPS}\label{sec:Spacelike CPS}

On the spacelike Fock space $\mathcal{F}_+$, we introduce the coherent peaked state
\begin{equation}
\begin{aligned}
&\ket{\sigma_{\epsp,\pip};x^0} \\[7pt]
=&\; \mathcal{N}_\sigma\exp\left(\int\dd{g_v}\dd{X_+}\dd[4]{\rf}\dd{\mf}\;\sigma_{\epsp,\pip;x^0}(g_v,X_+,\rf^0,\mf)\hat{\varphi}^\dagger(g_v,X_+,\rf^\mu,\mf)\right)\ket{\emptyset},
\end{aligned}
\end{equation}
which is assumed to be normalized via the factor $\mathcal{N}_\sigma$~\cite{Marchetti:2020umh,Jercher:2021bie}. Key ingredient is the condensate wavefunction $\sigma_{\epsp,\pip;x^0}$, with $x^0$ being the reference field value on which the state is peaked and with $\epsp$ and $\pip$ characterizing the peaking properties (see~\cite{Marchetti:2020qsq,Marchetti:2020umh} for further details on the formalism of coherent peaked states). It can be understood as a mean-field, since it is the expectation value of the spacelike group field operator
\begin{equation}
\bra{\sigma_{\epsp,\pip};x^0}\hat{\varphi}(g_v,X_+,\rf^\mu,\mf)\ket{\sigma_{\epsp,\pip};x^0} = \sigma_{\epsp,\pip;x^0}(g_v,X_+,\rf^0,\mf).
\end{equation}
The condensate wavefunction factorizes as
\begin{equation}\label{eq:spacelike condensate wavefunction}
\sigma_{\epsp,\pip;x^0}(g_v,X_+,\rf^0,\mf) = \eta_{\epsp}(\rf^0-x^0;\pip)\tilde{\sigma}(g_v,X_+,\rf^0,\mf),
\end{equation}
wherein
\begin{equation}
\eta_{\epsp}(\rf^0-x^0;\pip) = \mathcal{N}_{\epsp}\exp\left(-\frac{(\rf^0-x^0)^2}{2\epsp}\right)\e^{i\pip(\rf^0-x^0)}
\end{equation}
encodes the Gaussian peaking on the reference field value $x^0$ together with a phase factor that ensures finiteness of the reference field momenta~\cite{Marchetti:2020umh,Jercher:2021bie} (see also \ref{ass:ks3}). $\mathcal{N}_\epsp$ is a normalization factor of the Gaussian function. The remaining geometric information is carried by the reduced condensate wavefunction $\tilde{\sigma}(g_v,X_+,\rf^0,\mf)$.

\paragraph{Choice of spacelike peaking.} We have chosen a particular peaking for the spacelike condensate wavefunction as well as a specific dependence on the components of the reference fields $\rf^\mu$. Recall that the group field $\varphi(g_v,X_+,\rf^\mu,\mf)$ does depend on all components of the reference fields, even if the kinetic kernel is restricted according to Eq.~\eqref{eq:spacelike kernel restriction}. However, since we want the background condensates to be associated to  perfectly homogeneous geometries, the condensate wavefunction is peaked only on a chosen clock-value and the reduced condensate wavefunction is a function of the clock only (Assumption~\ref{ass:ks3}). 

\paragraph{Symmetries of \pmb{$\tilde{\sigma}$}.} Importantly, the reduced condensate wavefunction $\tilde{\sigma}$ satisfies additional symmetries besides that of the group field, given in Eqs.~\eqref{eq:closure} and~\eqref{eq:simplicity}. First, notice that the embedding of a single tetrahedron in Minkowski space, as dictated by the group field, is not a gauge-invariant information. In fact, we argue that the embedding information should be realized in a relational fashion, which is ensured by the clock-peaking. As argued in~\cite{Jercher:2021bie}, so-called adjoint covariance
\begin{equation}\label{eq:adjoint covariance}
\slrcw(g_v,X_+,\rf^0,\mf) = \slrcw(hg_vh^{-1},h\cdot X_+,\rf^0,\mf),\qquad\forall h\in\SL,
\end{equation}
can be understood as averaging over the embedding of a single tetrahedron, therefore eliminating this gauge-variant information. The resulting domain of the reduced condensate wavefunction carries the correct number of degrees of freedom, corresponding to the homogeneous spatial metric at a relational instance of time, i.e.  it is diffeomorphic to minisuperspace~\cite{Gielen:2014ila,Jercher:2021bie,Marchetti:2020umh}. We refer to this property as \textit{relational homogeneity}~\cite{Marchetti:2020umh}.\footnote{Previously, equivalence of minisuperspace and the domain of the condensate wavefunction has been established using only the geometric variables~\cite{Gielen:2014ila,Oriti:2016qtz,Jercher:2021bie}. In this work, however, we advocate for a relational notion of such an equivalence.} The remaining two conditions are most clearly seen in spin-representation. A priori, the four spacelike faces of a tetrahedron, labelled by $\rho_i$, take different values. Imposing however that the tetrahedra are equilateral, which is often referred to as an isotropy condition~\cite{Oriti:2016qtz,Jercher:2021bie}, we fix for the remainder all $\rho_i$ to be equal (Assumption~\ref{ass:ks2}).\footnote{Notice that while this notion of isotropy seems natural from a geometric point of view, a different restriction onto the reduced condensate wavefunction has been explored~\cite{Pithis:2016cxg}. However, at the level of the background dynamics this produces physically equivalent results which is line with naive universality arguments.} Furthermore, the $\SL$-intertwiner labels arising from Eq.~\eqref{eq:adjoint covariance} are fixed~\cite{Jercher:2021bie}. To simplify matters even further, we assume that the condensate is dominated by a single spin label $\rho$, as justified by~\cite{Gielen:2016uft,Pithis:2016cxg,Jercher:2021bie} (Assumption~\ref{ass:dc2}). As a result of all of these additional conditions, the reduced condensate wavefunction has a spin-representation given by
\begin{equation}
\tilde{\sigma}_\rho(\rf^0,\mf) \equiv \tilde{\sigma}(\rf^0,\mf),
\end{equation} 
where we suppress the fixed label $\rho$ in the notation for the remainder. Following this introduction of coherent states for the spacelike background, we elaborate in the following on the timelike background.

\subsubsection{Timelike CPS}\label{sec:Timelike CPS}

Following the arguments of the introduction of this section, we assume the timelike background to be described by a condensate which, as it turns out in Sec.~\ref{sec:Perturbation equations from quantum gravity}, proves sufficient to capture GR-like perturbations. Following this idea, we denote the condensate state on $\mathcal{F}_-$ as 
\begin{equation}
\begin{aligned}
&\ket{\tau_{\epsm,\pim,\delta,\pi_x};x^0,\vb*{x}} \\[7pt]
=&\;
\mathcal{N}_\tau\exp\left(\int\dd{g_v}\dd{X_-}\dd[4]{\rf}\dd{\mf}\;\tau_{\epsm,\pim,\delta,\pi_x;x^0,\vb*{x}}(g_v,X_-,\rf^\mu,\mf)\hat{\varphi}^\dagger(g_v,X_-,\rf^\mu,\mf)\right)\ket{\emptyset},
\end{aligned}
\end{equation}
which is now an eigenstate of the timelike group field operator, again normalized by the factor $\mathcal{N}_\tau$. Similar to the spacelike case, the timelike condensate wavefunction $\tau_{\epsm,\pim,\delta,\pi_x;x^0,\vb*{x}}$ factorizes according to
\begin{equation}\label{eq:timelike condensate wavefunction}
\tau_{\epsm,\pim,\delta,\pi_x;x^0,\vb*{x}}(g_v,X_-,\rf^\mu,\mf) = \eta_{\epsm}(\rf^0-x^0,\pim)\eta_{\delta}(\abs{\vb*{\rf}-\vb*{x}},\pi_x)\tilde{\tau}(g_v,X_-,\rf^0,\mf),
\end{equation}
where $\tilde{\tau}(g_v,X_-,\rf^0,\mf)$ is the timelike reduced condensate wavefunction. Besides a clock-peaking, the timelike condensate is also peaked on the rod variables $\vb*{\rf}$ via
\begin{equation}
\eta_{\delta}(\abs{\vb*{\rf}-\vb*{x}},\pi_x) = \mathcal{N}_\delta\exp\left(-\frac{\abs{\vb*{\rf}-\vb*{x}}^2}{2\delta}\right)\e^{i\pi_x\abs{\vb*{\rf} - \vb*{x}}}.
\end{equation}
We have chosen an isotropic peaking of the rod variables with the same parameters $\delta$ and $\pi_x$ for every spatial direction, following the strategy of~\cite{Marchetti:2021gcv} (see also \ref{ass:ks2} and~\ref{ass:ks3}). 

\paragraph{Choice of timelike peaking.} Since the timelike condensate is associated to the background structure, the reduced condensate wavefunction $\tilde{\tau}(g_v,\rf^0,\mf,X_-)$ only depends on the relational clock $\rf^0$. A peaking on rod variables is added for the timelike condensate to associate spatial derivatives to the timelike sector, as we discuss in more detail in Sec.~\ref{sec:Perturbed equations of motion} (see also Assumption~\ref{ass:ks3}). 

\paragraph{Symmetries of \pmb{$\tilde{\tau}$}.} In addition to the symmetries of $\hat{\varphi}(g_v,X_-,\rf^\mu,\mf)$, we introduce additional conditions to the timelike reduced condensate wavefunction $\tilde{\tau}$, similar to $\slrcw$ above. Most importantly, these restrictions ensure that $\tlrcw$ carries the correct degrees of freedom. First, $\tilde{\tau}$ also satisfies adjoint covariance
\begin{equation}
\tlrcw(g_v,X_-,\rf^0,\mf) = \tlrcw(hg_vh^{-1},h\cdot X_-,\rf^0,\mf),\qquad\forall h\in\SL,
\end{equation}
with the resulting $\SL$-intertwiner label in spin-representation being fixed. As a result, the domain of $\tlrcw$ corresponds to the metric degrees of freedom on a $(2+1)$-dimensional slice at a given instance of relational time. Therefore, the number of degrees of freedom of the spacelike and timelike condensates is the same, which is important for the later analysis. Since timelike tetrahedra admit an arbitrary mixture of spacelike and timelike faces, the reduced condensate wavefunction $\tilde{\tau}$ carries a priori all possible combinations of $(\rho,0)$ and $(0,\nu)$ labels, as Eq.~\eqref{eq:timelike group field spin rep} indicates. In the following, we are going to restrict to the case where the condensate wavefunction only carries spacelike faces and fix the corresponding label to the same label $\rho$ as for $\tilde{\sigma}$ (Assumptions~\ref{ass:ks2},~\ref{ass:ks4} and~\ref{ass:dc2}). {Besides a simplification of the dynamics, there are two further reasons to restrict to spacelike faces only. First, current developments in the Landau-Ginzburg mean-field analysis of the complete Barrett-Crane model~\cite{Marchetti:2022igl,Marchetti:2022nrf,Jercher:2024abc} suggest that a condensate phase for timelike tetrahedra exists if the faces are all spacelike. Second, as detailed in~\cite{Jercher:2022mky}, correlations between spacelike and timelike tetrahedra can only be mediated via spacelike faces. Hence, correlations between the spacelike and timelike sector, introduced below, are only possible if the faces carry the same signature.} 

As a result, the timelike reduced condensate wavefunction in spin-representation is of the form
\begin{equation}
\tilde{\tau}_\rho(\rf^0,\mf) \equiv \tilde{\tau}(\rf^0,\mf),
\end{equation}
where we again suppress the fixed label $\rho$ in the remainder.  

In summary, the background structure on the total Fock space is defined by the state 
\begin{equation}
\ket{\sigma_{\epsp,\pip};x^0}\otimes \ket{\tau_{\epsm,\pim,\delta,\pi_x};x^0,\vb*{x}},
\end{equation}
on which the group field operators act accordingly. The effective relational dynamics of this background state is computed in Sec.~\ref{sec:Background equations of motion}.

\subsection{Perturbed coherent peaked states}\label{sec:Perturbed coherent peaked states}

Following the introduction of this section, inhomogeneities are encoded in the perturbed coherent peaked state of Eq.~\eqref{eqn:perturbedstates}, with the three $2$-body operators $\hat{\delta\Phi}\otimes\one_-,\hat{\delta\Psi}$ and $\one_+\otimes\hat{\delta\Xi}$ sourcing a quantum entanglement within and between the spacelike and timelike sectors. {These three operators are defined in analogy to $\hat{O}_{++},\hat{O}_{+-}$ and $\hat{O}_{--}$ in Eq.~\eqref{eq:two-body operator}, respectively. The bi-local functions $\mathcal{O}_{
\alpha\beta}$ are referred to as kernels.} 

A priori, the kernels $\delta\Phi,\delta\Psi$ and $\delta\Xi$ that define the three operators above are bi-local functions on the respective domains. On both copies of the domain, we impose the same restrictions as for the spacelike and timelike reduced condensate wavefunctions, respectively (see Assumptions~\ref{ass:ks3},~\ref{ass:ks4} and~\ref{ass:dc2}).  As a result, the spin-representation of the three kernels is explicitly given by
\begin{equation}
\delta\Phi(\rf^\mu,\mf,\rf^{\prime\mu},\mf'),\qquad \delta\Psi(\rf^\mu,\mf,\rf^{\prime\mu},\mf')\qquad\text{and} \qquad \delta\Xi(\rf^\mu,\mf,\rf^{\prime\mu},\mf'),
\end{equation}
where we suppressed the dependence on the fixed $\SL$-representation label $\rho$ in the notation.\footnote{Notice, that the functions $\delta\Phi,\delta\Psi$ and $\delta\Xi$ carry a dependence on the clock and on the rods. Together with the restriction in Eq.~\eqref{eq:locality condition} below, where the two copies of reference fields are identified via a $\delta$-distribution, the domain of these kernels corresponds to the metric degrees of freedom at a relational spacetime point. Thus, the two-body correlations describe perturbations of the relational notion of homogeneity by a direct rod-dependence.} Crucially, the three two-body operators do not factorize into one-body operators if the kernels do not factorize accordingly. As a result, acting with $\hat{\delta\Phi},\hat{\delta\Psi}$ and $\hat{\delta\Xi}$ on the tensor product of spacelike and timelike condensate creates an entangled state in the respective sectors.

Finally, since we are interested in small perturbations, we employ the linearized form of $\ket{\Delta;x^0,\vb*{x}}$, given by
\begin{equation}\label{eq:linearized perturbed states}
\ket{\Delta;x^0,\vb*{x}} \approx \mathcal{N}_\Delta\left(\one + \hat{\delta\Phi}+\hat{\delta\Psi}+\hat{\delta\Xi}\right)\ket{\sigma;x^0}\otimes\ket{\tau,x^0,\vb*{x}}.
\end{equation}
This is the state that we are going to employ for computing the cosmological dynamics, including perturbations. In the following two sections, we show how to obtain effective relational dynamics as the expectation value of the GFT equations of motion and how to connect macroscopic quantities such as the $3$-volume to the expectation value of quantum geometric operators. 

\subsection{Effective relational dynamics of perturbed CPS}\label{sec:Effective relational dynamics}

Average relational dynamics of GFT condensates are, at a mean-field level, obtained by taking the expectation value of the GFT equations of motion with respect to the macroscopic state $\ket{\Delta;x^0,\vb*{x}}$, see also \ref{ass:ds3}. Due to the presence of two fields, corresponding to spacelike and timelike tetrahedra, there are two effective relational equations of motion, which are given by
\begin{equation}\label{eq:lowest SDE}
\expval**{\fdv{S[\hat{\varphi},\hat{\varphi}^{\dagger}]}{\hat{\varphi}(g_v,X_\alpha,x^\mu,\mf)}}{\Delta;x^0,\vb*{x}}= 0,
\end{equation}
for each signature $\alpha\in\{+,-\}$. Notice, that these equations correspond to the first of an infinite tower of Schwinger-Dyson equations~\cite{Oriti:2016qtz,Gielen:2013naa}. {Ultimately, the dynamics of the perturbed CPS $\ket{\Delta;x^0\vb*{x}}$ will govern the dynamics of cosmological observables obtained as the expectation value of GFT operators with respect to $\ket{\Delta;x^0\vb*{x}}$. However, as it requires particular care, we dedicate to this step Secs.~\ref{sec:Geometric observables}--\ref{sec:Dynamics of matter observables} while focusing here only on the dynamics of the condensate.}

For the remainder of this work, we assume negligible interactions (Assumption~\ref{ass:ds4}), which was shown in~\cite{Oriti:2016qtz,Jercher:2021bie} to be a valid approximation at late but not very late times, see also~\cite{Gielen:2013naa} for a discussion. Notice, that the argument provided therein also applies to the timelike sector. One of the crucial consequences of this assumption is that higher orders of the Schwinger-Dyson equations reduce to powers of the lowest order equations~\eqref{eq:lowest SDE}, as we show in Appendix~\ref{sec:Going beyond mean-field in the absence of interactions}. Hence, solutions of Eq.~\eqref{eq:lowest SDE} solve also all higher orders. We comment on this matter in more detail in Sec.~\ref{sec:Discussion and Conclusion}. Notice that despite negligible interactions, the spacelike and timelike sectors get coupled via the spacelike-timelike quantum correlation $\delta\Psi$, as we show in detail in Sec.~\ref{sec:Perturbed equations of motion}. 

Due to the presence of perturbations, the two equations of motion can be separated into a zeroth-order background part and a first-order perturbation part, which we discuss separately in Secs.~\ref{sec:Background equations of motion} and~\ref{sec:Perturbed equations of motion}, respectively.

\subsubsection{Background equations of motion}\label{sec:Background equations of motion}

At background level, the two equations of motion in spin-representation are given by
\begin{align}
0 &= \int\dd{\rf^0}\dd{\mf'}\mathcal{K}_+\left(\rf^0,(\mf-\mf')^2\right)\sigma(\rf^0+x^0,\mf'),\label{eq:spacelike bkg eom}\\[7pt]
0 &= \int\dd[4]{\rf}\dd{\mf'}\mathcal{K}_-\left(\vb*{\rf},(\mf-\mf')^2\right)\tau(\rf^0+x^0,\vb*{\rf}+\vb*{x},\mf'),\label{eq:timelike bkg eom}
\end{align}
where we note that {due to spatial homogeneity, the background equation of motion on the spacelike sector would actually contain an empty integration over the rods $\rf^i$ which we henceforth regularize by introducing a fiducial cell of finite volume.} For a further analysis, we perform a Fourier transform of the matter field variables $\mf\rightarrow\mm$. Following~\cite{Marchetti:2021gcv}, we assume a peaking of both condensate wavefunctions on a fixed scalar field momentum $\pmm$, realized by a Gaussian peaking. Since the scalar field is minimally coupled, its canonical conjugate momentum is constant at the classical continuum level. We translate this idea to the present context 
by peaking on a fixed value $\pmm$ (Assumption~\ref{ass:kc1}). Exploiting furthermore the peaking properties of $\sigma$ and $\tau$, defined in Eqs.~\eqref{eq:spacelike condensate wavefunction} and~\eqref{eq:timelike condensate wavefunction}, respectively, we obtain the dynamical equations for the reduced condensate wavefunctions $\slrcw$ and $\tlrcw$
\begin{align}
\partial_0^2\slrcw(x^0,\pmm)-2i\tpip\partial_0\slrcw(x^0,\pmm)-E_+^2(\mm)\slrcw(x^0,\pmm) &= 0,\\[7pt]
\partial_0^2\tlrcw(x^0,\pmm)-2i\tpim\partial_0\tlrcw(x^0,\pmm)-E_-^2\tlrcw(x^0,\pmm) &= 0,
\end{align}
where the quantities $E_\pm$ and $\tilde{\pi}_0^\pm$ are defined in Appendix~\ref{sec:Derivation of background equations}, to which we refer for further details. 

Following the procedure of~\cite{Oriti:2016qtz,Marchetti:2020umh,Marchetti:2021gcv}, the reduced condensate wavefunctions can be decomposed into a radial and angular part, denoted as $r_\alpha(x^0,\pmm)$ and $\theta_\alpha(x^0,\pmm)$, respectively. Splitting the resulting equations into real and imaginary part, one obtains
\begin{align}
 r''_\alpha(x^0,\pmm)-\frac{Q_\alpha^2(\pmm)}{r_\alpha^3(x^0,\pmm)}-\mu_\alpha^2(\pmm)r_\alpha(x^0,\pmm) &= 0,\\[7pt]
 \theta_\alpha'(x^0,\pmm) -\tilde{\pi}_0^\alpha-\frac{Q_\alpha(\pmm)}{r_\alpha^2(x^0,\pmm)} &= 0,
\end{align}
where a prime denotes differentiation with respect to $x^0$, $Q_\alpha$ are integration constants and the $\mu_\alpha$ are defined as $\mu_\alpha^2(\pmm) \defeq E_\alpha^2(\pmm)-(\tilde{\pi}_0^\alpha)^2$. As demonstrated in Appendix~\ref{sec:Derivation of background equations}, $E_-$ and thus $\mu_-$ are actually independent of the peaked matter momentum $\pmm$.

\paragraph{Classical limit.} As elaborated previously~\cite{Oriti:2016qtz,Marchetti:2020umh,Marchetti:2020qsq}, the semi-classical limit of the condensate is obtained at late relational time scales where the moduli of the condensate wavefunctions $r_\alpha$ are dominant with respect to $Q_\alpha$ and $\mu_\alpha$ but where interactions are still negligible (see also~\ref{ass:dc1}). Furthermore, it has been shown in~\cite{Pithis:2016cxg}, that in this limit, expectation values of for instance the volume operator are sharply peaked, providing a highly non-trivial consistency check for the semi-classical interpretation. In this limit, the background equations of motion simplify significantly, yielding solutions
\begin{align}
\slrcw(x^0,\pmm) &= \slrcw_0\e^{(\mu_+ + i\tpip)x^0},\label{eqn:bkgsigma}\\[7pt]
\tlrcw(x^0,\pmm) &= \tlrcw_0\e^{(\mu_- + i\tpim)x^0},\label{eqn:bkgtau}
\end{align}
where $\slrcw_0$ and $\tlrcw_0$ are determined by initial conditions. {These two equations will be heavily employed to derive the dynamics of cosmological quantities at the background level such as the volume dynamics in Eq.~\eqref{eq:bkg volume dynamics}, the dynamics of spacelike and timelike particle number in Eqs.~\eqref{eq:bkg N+ dynamics} and~\eqref{eq:bkg N- dynamics} or the dynamics of the background scalar field in Eq.~\eqref{eq:mf background}. Furthermore, the dynamics of the cosmological observables at first order in perturbations crucially depend on the background condensate solutions above.}

This concludes the effective background equations of motion. In the next section, we compute the effective equations of motion at first order in perturbations.

\subsubsection{Perturbed equations of motion}\label{sec:Perturbed equations of motion}

At first order in perturbations, there are two equations, one for the spacelike and one for the timelike sector. Since there are three distinct two-body correlations $\delta\Phi, \delta\Psi$ and $\delta\Xi$, there is a dynamical freedom for one of the variables if one assumes negligible interactions and works within a first-order perturbative framework. In the following, we utilize this freedom to relate the functions $\delta\Phi$ and $\delta\Psi$ via an arbitrary function $\mathrm{f}(\rf^\mu)$ that will be ultimately fixed by matching the perturbed volume to the classical quantity of GR. We provide a physical interpretation of this assumption in Sec.~\ref{sec:Number of tetrahedra} in terms of the perturbations of the timelike number operator, $\delta N_-$.

\paragraph{Spacelike perturbed dynamics.} Dynamics of the spacelike sector at first order of perturbations are governed by
\begin{equation}\label{eq:pert eom sl}
\begin{aligned}
0 &= \int\dd[4]{\rf}\dd{\mf'}\mathcal{K}_{+}((\rf^0)^2,(\mf-\mf')^2)\int\dd[4]{\rf'}\dd{\mf''}\Bigg{[}\delta\Psi(\rf^\mu+x^\mu,\mf',\rf^{\mu\prime},\mf'')\bar{\tau}(\rf^{\mu\prime},\mf'')\\[7pt]
&+\delta\Phi(\rf^\mu+x^\mu,\mf',\rf^{\mu\prime},\mf'')\bar{\sigma}(\rf^{0\prime},\mf'') \Bigg{]}.
\end{aligned}
\end{equation}
As a first simplification, we choose the bi-local kernel $\delta\Psi$ to depend only on one copy of relational frame data (Assumption~\ref{ass:kc2}), i.e.
\begin{align}\label{eq:locality condition}
\delta\Psi(\rf^\mu,\mm,\rf^{\mu\prime},\mm') &= \delta\Psi(\rf^\mu,\mm)\delta^{(4)}(\rf^\mu-\rf^{\prime\mu})\delta(\mm-\mm').
\end{align}
From a simplicial gravity perspective, locality with respect to the reference fields $\rf^\mu$ corresponds to correlations only within the same $4$-simplex,  which can be compared to nearest-neighbor interactions in statistical spin systems. For the momenta of the matter field $\mm$, the condition is interpreted as momentum conservation across tetrahedra of the same $4$-simplex. Next, we exploit the dynamical freedom for one of the perturbation functions by imposing the relation (Assumption~\ref{ass:dc3})
\begin{equation}\label{eq:relation of dPsi and dPhi}
\delta\Phi(\rf^\mu,\mm) = \mathrm{f}(\rf^\mu)\delta\Psi(\rf^\mu,\mm),  
\end{equation}
with the complex valued function $\mathrm{f}$ defined as
\begin{equation}\label{eq:def of f}
\mathrm{f}(\rf^0,\vb*{\rf}) = f(\rf^0)\e^{i\theta_f(\rf^0)}\abs{\eta_\delta(\abs{\vb*{\rf}-\vb*{x}})}\e^{2i\pip\rf^0}.
\end{equation}
Here, $f$ and $\theta_f$ are real functions that only depend on the reference clock $\rf^0$. In addition, we consider the following relations between the peaking parameters $\epsilon^\pm$ and $\pi_0^\pm$ of the different sectors
\begin{equation}\label{eq:peaking parameter relations}
\epsp = \epsm,\qquad  \pip = -\pim,
\end{equation}
also entering Assumption~\ref{ass:kc3}. Besides simplifying the spacelike equations of motion at first order in perturbations, we show in Sec.~\ref{sec:Geometric observables} that the expression of the perturbed $3$-volume $\delta V$ takes a manageable form under Eq.~\eqref{eq:peaking parameter relations}.

Within this set of choices, the peaking properties of the spacelike and timelike condensate wavefunctions yield the following equation of motion for $\delta\Psi$
\begin{equation}\label{eq:dPsi equation}
\begin{aligned}
0 &= \partial_0^2\left(\delta\Psi(J_0\bar{\tlrcw}+f\e^{i\theta_f}\bar{\slrcw})\right)-2i\tpip\partial_0\left(\delta\Psi(J_0\bar{\tlrcw}+f\e^{i\theta_f}\bar{\slrcw})\right)\\[7pt]
&- E_+^2\delta\Psi(J_0\bar{\tlrcw}+f\e^{i\theta_f}\bar{\slrcw}) +\alpha\bar{\tlrcw}\nabla_{\vb*{x}}^2\delta\Psi,
\end{aligned}
\end{equation}
for which a detailed derivation is provided in Appendix~\ref{sec:Derivation of perturbation equations}. All of the functions above depend on the peaked value of the matter momentum, $\pmm$, but we suppress that dependence for notational clarity. As written out explicitly before, the two reduced condensate wavefunctions $\slrcw$ and $\tlrcw$ depend on the relational time $x^0$ since they are part of the background, while in contrast, $\delta\Psi(x^\mu,\pmm)$ depends on all four reference field values. The remaining coefficients $E_+(\pmm), \tpip, J_0$ and $\alpha$ are defined in Appendix~\ref{sec:Derivation of perturbation equations} and are entirely determined by peaking parameters and the peaked matter momentum $\pmm$. 

Solving the background equations of motion and inserting them in the first-order perturbation equation, one obtains an equation for $\delta\Psi$ which is of the general form
\begin{equation}
0 = \delta\Psi''+t_1(x^0,\pmm)\delta\Psi'+t_0(x^0,\pmm)\delta\Psi+s_2(x^0,\pmm)\nabla_{\vb*{x}}^2\delta\Psi,
\end{equation}
with complex 
functions $t_i(x^0,\pmm)$ and $s_2(x^0,\pmm)$. The conditions on these coefficients to yield GR-like perturbation equations are discussed in Sec.~\ref{sec:Perturbation equations from quantum gravity}.

\paragraph{Timelike perturbed dynamics.} As we will explicitly see in the next section, the dynamics of observables other than the spatial $3$-volume, such as the matter field, its momentum or the total number operator, are governed by the equations of motion of both sectors, spacelike and timelike. For this reason, in the following, we study the perturbed equations of motion on the timelike sector which, in spin-representation, are given by
\begin{equation}\label{eq:pert eom tl}
\begin{aligned}
0 &= \int\dd[4]{\rf}\dd{\mf'}\mathcal{K}_{-}(\abs{\vb*{\rf}}^2,(\mf-\mf')^2)\int\dd[4]{\rf'}\dd{\mf''}\Bigg{[}\delta\Psi(\rf^{0\prime},\vb*{\rf}',\mf',\rf^\mu+x^\mu,\mf'')\bar{\sigma}(\rf^{0\prime},\mf'')\\[7pt]
&+\delta\Xi(\rf^\mu+x^\mu,\mf',\rf^{0\prime},\vb*{\rf}',\mf'')\bar{\tau}(\rf^{\mu\prime},\mf'') \Bigg{]}.
\end{aligned}
\end{equation}
Using the peaking properties of $\sigma$ and $\tau$, the locality condition in Eq.~\eqref{eq:locality condition}, the relation of peaking parameters in Eq.~\eqref{eq:peaking parameter relations}, as well as the classical background equations of motion in Eqs.~\eqref{eqn:bkgsigma} and~\eqref{eqn:bkgtau}, one obtains
\begin{equation}\label{eq:dXi equation}
\begin{aligned}
0 &=  \bar{\slrcw}\int\dd[3]{\rf}\mathcal{K}_-(\abs{\vb*{\rf}},\pmm^2)\left[\partial_0^2\delta\Psi+2\mu_+\partial_0\delta\Psi-\frac{\mathcal{K}_+^{(2)}}{\mathcal{K}_+^{(0)}}\delta\Psi\right]\\[7pt]
&+ 
\bar{\tlrcw}\mathcal{K}_-^{(0)}J_{0,\vb*{0}}\Bigg{[}\partial_0^2\delta\Xi+2\mu_-\partial_0\delta\Xi-\beta\delta\Xi+\gamma\nabla_{\vb*{x}}^2\delta\Xi\Bigg{]},
\end{aligned}
\end{equation}
for which a derivation is given in Appendix~\ref{sec:Derivation of perturbation equations}, including a definition of the parameters $\beta$ and $\gamma$. Since the space-dependence of the first term is integrated out, solutions $\delta\Xi$ need to be space-independent, i.e. $\delta\Xi(x^\mu,\pmm) \equiv \delta\Xi(x^0,\pmm)$. Hence, the space-derivative acting on $\delta\Xi$ vanishes and the equation reduces to a second-order inhomogeneous ordinary differential equation. In particular, the pure time dependence of $\delta\Xi$ will have important consequences for the behavior of timelike particle number perturbations $\delta N_-$, discussed in detail in Sec.~\ref{sec:Number of tetrahedra}. 

\section{Dynamics of observables}\label{sec:Perturbation equations from quantum gravity}

In the spirit of obtaining cosmology as a hydrodynamic limit of QG, classical cosmological quantities are associated with averages on the above condensate states of appropriate one-body observables defined within the GFT Fock space. Importantly, this can only hold under the assumption that quantum fluctuations of such observables on the states of interest are small. As emphasized in  \cite{Marchetti:2020qsq,Gielen:2019kae}, this classicality requirement is automatically satisfied at late relational times. For this reason, in the following we will focus only on this regime. 

Observables of interest for cosmological applications can be roughly divided into two categories: geometric observables (such as volume, area, curvature, etc.) and matter observables (such as the scalar field operators and their momenta). However, as one might expect, not all operators available in the quantum theory fall into these two categories. A particularly important example of \qmarks{observables} that have no classical counterpart are number operators, i.e.\ operators that count the number of GFT (timelike and/or spacelike) quanta. In fact, the classical limit turns out to be associated with a large number of quanta in the condensate and is thus directly controlled by the above quantities.

More concretely, one can compute the expectation value of a second-quantized operator $\hat{\mathcal{O}}$ on the perturbed condensate states $\ket{\Delta;x^0,\vb*{x}}$ by using the algebra of creation and annihilation operators. The result can be split generically as
\begin{equation}
\mathcal{O}(x^0,\vb*{x})\equiv \expval{\hat{\mathcal{O}}}{\Delta;x^0,\vb*{x}} = \bar{\mathcal{O}}(x^0)+\delta\mathcal{O}(x^0,\vb*{x}),
\end{equation}
where $\bar{\mathcal{O}}$ and $\delta\mathcal{O}$ are background and perturbed contributions, respectively. Note that, due to the peaking properties of the states $\ket{\Delta;x^0,\vb*{x}}$, the above expectation value is localized in relational space and time. In this sense, the quantities obtained are effective relational observables. As such, their dynamics should be compared, at least in an appropriate limit, with the dynamics of the corresponding classical cosmological relational observables. Since these are gauge-invariant extensions of gauge-fixed quantities, one could alternatively compare the dynamics of the above expectation values with the dynamics of the corresponding classical cosmological observables in harmonic gauge (since the physical frame used to localize quantities is in fact harmonic)~\cite{Gielen:2013naa,Oriti:2016qtz,Marchetti:2020umh,Marchetti:2021gcv}, as derived in Appendix~\ref{sec:Classical perturbation theory}. 

{
In Sec.~\ref{sec:Geometric observables}$-$\ref{sec:Dynamics of matter observables}, respectively, we will compute expectation values and dynamics of geometric, number, and matter operators. Then, in Sec.~\ref{sec:mukhanovsasaki} we will use the results of the previous sections to derive the dynamics of an appropriate \qmarks{curvature-like variable}. Finally, in Sec.~\ref{sec:Solutions of GFT and GR perturbations}, we will compare the effective GFT perturbation dynamics with the GR ones in harmonic gauge.}

\subsection{Volume operator and related geometric observables}\label{sec:Geometric observables}

Classically, the geometry of a flat, slightly inhomogeneous universe is characterized by the line element
\begin{equation}\label{eq:perturbed line elementmaintext}
\dd{s}^2 = -a^6(1+2A)\dd{t}^2+a^4\partial_iB\dd{t}\dd{x^i}+a^2\left((1-2\psi)\delta_{ij}+2\partial_i\partial_j E\right)\dd{x}^i\dd{x}^j,
\end{equation}
where we have considered only scalar perturbations captured by the small functions $A$, $B$, $\psi$ and $E$. Note that, for the reasons explained above, Eq.~\eqref{eq:perturbed line elementmaintext} is written in harmonic gauge. At the background level, this means working in harmonic ($\bar{N}^2=a^6$), rather than proper ($\bar{N}^2=1$) or conformal ($\bar{N}^2=a^2$) time, while at the level of perturbations this forces the functions $A$, $B$, $\psi$ and $E$ to satisfy the constraints \eqref{eq:pert HGC} (see Appendix~\ref{sec:Classical perturbation theory} for more details).

{Note that isotropic information on the spatial geometry is fully captured by the local $3$-volume, $ \sqrt{-g_{(3)}}=a^3(1-3\psi+\nabla^2E)$. In this section, we will mainly focus on computing the expectation value of the corresponding $3$-volume GFT operator. Note that while this is clearly sufficient to characterize the full geometry of a homogeneous and isotropic universe (in particular, the Hubble parameter is $\Hubble=\bar{V}'/(3\bar{V})$), it is a strong restriction at the perturbation level, as it only captures information related to a specific combination of $\psi$ and $E$. To extract a full-fledged cosmological perturbation theory from GFTs, it is therefore imperative to construct more sophisticated geometric observables. This is an avenue of research that has been only tentatively explored \cite{Gielen:2021vdd, Calcinari:2022iss}. We will return to this topic in Sec.~\ref{sec:Discussion and Conclusion}.
}

Using the choices on the perturbation functions $\delta\Phi$ and $\delta\Psi$ in Eqs.~\eqref{eq:locality condition} and~\eqref{eq:relation of dPsi and dPhi}, respectively, the expectation value of $\hat{V}$ {on the perturbed condensate state $\ket{\Delta;x^0,\vb*{x}}$ is given by}
\begin{equation}
\begin{aligned}
 \expval{\hat{V}}{\Delta;x^0,\vb*{x}} &= v\int\dd[4]{\rf}\dd{\mm}\bar{\sigma}(\rf^0,\mm)\sigma(\rf^0,\mm)\\[7pt]
&\quad+ 2v\mathfrak{Re}\left\{\int\dd[4]{\rf}\dd{\mm}\delta\Psi(\rf^\mu,\mm)\bar{\sigma}(\rf^0,\mm)\bar{\tau}(\rf^{\mu},\mm)\right\}\\[7pt]
&\quad+2v\mathfrak{Re}\left\{\int\dd[4]{\rf}\dd{\mm}\mathrm{f}(\rf^\mu)\delta\Psi(\rf^\mu,\mm)\bar{\sigma}(\rf^0,\mm)\bar{\sigma}(\rf^{0},\mm)\right\},
\end{aligned}
\end{equation}
where $v$ is similar to a volume eigenvalue~\cite{Oriti:2016qtz,Jercher:2021bie}, scaling as $\rho^{3/2}$. Applying Eq.~\eqref{eq:peaking parameter relations} and exploiting the peaking properties of $\sigma$ and $\tau$, the volume expectation value becomes
\begin{equation}\label{eq:volume expval}
\begin{aligned}
& \expval{\hat{V}}{\Delta;x^0,\vb*{x}} =  v\abs{\slrcw(x^0,\pmm)}^2\int\dd[3]{\chi}\\[7pt]
&\quad+2v\mathfrak{Re}\left\{J_0\delta\Psi(x^\mu,\pmm)\bar{\slrcw}(x^0,\pmm)\bar{\tlrcw}(x^0,\pmm)+\frac{J_2}{2}\bar{\slrcw}(x^0,\pmm)\bar{\tlrcw}(x^0,\pmm)\nabla_{\vb*{x}}^2\delta\Psi(x^\mu,\pmm)\right\}\\[7pt]
&\quad+2v\mathfrak{Re}\left\{\delta\Psi(x^\mu,\pmm)f(x^0)\e^{i\theta_f(x^0)}\bar{\slrcw}(x^0,\pmm)\bar{\slrcw}(x^0,\pmm)\right\}.
\end{aligned}
\end{equation}
The first term, containing an empty rod-integration (see Assumption~\ref{ass:ks5}), defines the background volume 
\begin{equation}\label{eq:background volume}
\bar{V} = v\abs{\slrcw(x^0,\pmm)}^2,
\end{equation}
while the remaining two contributions make up the perturbations of the volume
\begin{equation}
\begin{aligned}\label{eq:pert volume expval}
&\delta V(x^\mu,\pmm) = 2v\mathfrak{Re}\left\{\delta\Psi(x^\mu,\pmm)f(x^0)\e^{i\theta_f(x^0)}\bar{\slrcw}(x^0,\pmm)\bar{\slrcw}(x^0,\pmm)\right\}\\[7pt]
&\quad+2v\mathfrak{Re}\left\{J_0\delta\Psi(x^\mu,\pmm)\bar{\slrcw}(x^0,\pmm)\bar{\tlrcw}(x^0,\pmm)+\frac{J_2}{2}\bar{\slrcw}(x^0,\pmm)\bar{\tlrcw}(x^0,\pmm)\nabla_{\vb*{x}}^2\delta\Psi(x^\mu,\pmm)\right\}.
\end{aligned}
\end{equation}
From the dynamics derived in Sec.~\ref{sec:Effective relational dynamics}, we can straightforwardly obtain the dynamics of the background and the perturbed averaged volume. This will be done in the next two paragraphs.

\paragraph{Background volume.} At the level of the background, the expectation value of the spatial volume operator in a classical limit (see Assumption~\ref{ass:dc1}) is given by 
\begin{equation}\label{eq:background volume classical limit}
\bar{V}(x^0,\pmm) = \slrcw_0^2\e^{2\mu_+x^0}.
\end{equation}
Performing derivatives with respect to the clock field value $x^0$, $\bar{V}$ satisfies
\begin{equation}\label{eq:bkg volume dynamics}
\frac{\bar{V}'}{3\bar{V}} = \frac{2}{3}\mu_+(\pmm),\qquad \left(\frac{\bar{V}'}{3\bar{V}}\right)' = 0.
\end{equation}
{In terms of the Hubble parameter $\Hubble = \bar{V}'/(3\bar{V})$ in harmonic gauge,}  these equations read
\begin{equation}
\Hubble^2 = \frac{4}{9}\mu_+^2(\pmm),\qquad \Hubble'  =0.
\end{equation}
{In GR, the Hubble parameter in harmonic gauge for a universe filled with MCMF scalar fields is also constant. Thus, the GFT background dynamics matches the classical GR ones if (see Sec.~\ref{sec:Classical perturbation theory - geometry})} 
\begin{equation}\label{eq:mu_+ matching}
\mu_+^2(\pmm) = \frac{3}{8 M_{\Pl}^2}\bar{\pi}_\mf^2,
\end{equation}
where $\bar{\pi}_\mf$ is the constant momentum of the matter field $\mf$ at background level. 

The factor of the Planck mass {in Eq.~\eqref{eq:mu_+ matching}} present at the level of the classical Einstein equations (see Eq.~\eqref{eq:classical 3H^2 eq}) and thus needs to be accounted for in the matching procedure. Working with fields of energy dimension $1$, $[\mf] = 1$, and thus with conjugate momenta of energy dimension $2$, $[\mm] = 2$, the matching is therefore consistent with the fact that $[\mu_+]$ is required to be $1$.

\paragraph{Volume perturbations.} To study the dynamics of $\delta V$ (defined in Eq.~\eqref{eq:pert volume expval}) it is convenient to perform a split of the complex-valued function $\delta\Psi$ into its modulus and phase, $\delta\Psi = R(x^\mu,\pmm)\e^{i\Theta}$. We pose the condition that this phase is in fact constant, see also Assumption~\ref{ass:dc4}.\footnote{Assuming instead a time-dependent phase and splitting the equation into real and imaginary part, one finds $\Theta' = c/R^2$ with some time-dependent factor $c$. Since $R$ is however space-dependent and we require $\Theta$ to be only time-dependent, the function $c$ must vanish and we conclude that $\Theta$ is in fact constant.} As a result, the overall phases of the first and second term inside the real parts of $\delta V$ are respectively given by
\begin{align}
\theta_1 &= \Theta + \theta_f(x^0) -2\tpip x^0,\\[7pt]
\theta_2 &= \Theta.
\end{align}
Exploiting once more the dynamical freedom on $\delta\Phi$, and thus on the function $\mathrm{f}(\rf^0,\vb*{\rf})$ entering Eq.~\eqref{eq:relation of dPsi and dPhi}, we set $\theta_f = \frac{\pi}{2}+2\tpip x^0$. In momentum space of the rod variable, which we consider for the remainder of Section~\ref{sec:Perturbation equations from quantum gravity}, the resulting form of $\delta V$ is given by
\begin{equation}\label{eq:dV=AR}
\frac{\delta V(x^0,k)}{2v\slrcw_0\tlrcw_0} = \left[\cos(\Theta)\e^{(\mu_++\mu_-)x^0}\left(J_0-\frac{J_2}{2}k^2\right)+\sin(\Theta)\e^{2\mu_+ x^0}\frac{\slrcw_0}{\tlrcw_0}f\right]R. 
\end{equation}
Put in this form, the perturbed volume $\delta V$ is directly related to the modulus $R$ by a time- and momentum-dependent factor $A$, 
\begin{equation}
\frac{\delta V(x^0,k)}{2v\slrcw_0\tlrcw_0} \eqdef A(x^0,k)R.
\end{equation}
Therefore, the dynamics of $\delta V$ are essentially governed by the dynamics of $R$, which we discuss next.

Introducing the function
\begin{equation}
 g_f\defeq \left(\slrcw_0 f\e^{\mu_+x^0}+J_0\tlrcw_0\e^{\mu_- x^0}\right),
\end{equation}
the dynamics of $\delta\Psi = R\e^{i\Theta}$ for a constant phase $\Theta$ are given by
\begin{equation}\label{eq:R equation}
  0 = R'' +2\frac{g_f'}{g_f}R'+\left(\frac{g_f''}{g_f}-\mu_+^2\right)R -\frac{\alpha\tlrcw_0\e^{\mu_-x^0}}{g_f}k^2 R\,,  
\end{equation}
which straightforwardly follows from Eq.~\eqref{eq:dPsi equation} and the derivations of Sec.~\ref{sec:Perturbed equations of motion}. Combining Eqs.~\eqref{eq:dV=AR} and~\eqref{eq:R equation}, the dynamical equation for the perturbed volume $\delta V$ is given by
\begin{equation}\label{eq:dV with A}
    \delta V'' +\left[2\frac{g_f'}{g_f}-2\frac{A'}{A}\right]\delta V'+\left[\frac{g_f''}{g_f}-\mu_+^2+2\left(\frac{A'}{A}\right)^2-\frac{A''}{A}-2\frac{g_f'}{g_f}\frac{A'}{A}\right]\delta V-\frac{\alpha\tlrcw_0\e^{\mu_-x^0}}{g_f} k^2\delta V = 0\,.
\end{equation}
The above equation, and thus any solution of it, clearly depends on the function $g_f$ encoding the aforementioned mean-field dynamical freedom. Remarkably, however, this freedom can be fixed entirely by requiring the above equation to take the same functional form (at least in the late time, classical regime) of the corresponding GR one, given in Eq.~\eqref{eq:classical perturbed volume equation}.

To see this explicitly, we start from the spatial derivative term, whose pre-factor $a^4$, as mentioned in Sec.~\ref{sec:Introduction}, could not be recovered by considering a perturbed condensate of only spacelike tetrahedra~\cite{Marchetti:2021gcv}. Exactly because of the additional timelike degrees of freedom, and thus of the above dynamical freedom, here we can easily recover the appropriate pre-factor, by simply requiring the function $g_f$ to satisfy
\begin{equation}
-\frac{\alpha\tlrcw_0\e^{\mu_- x^0}}{g_f} = a^4 = \slrcw_0^{8/3}\e^{8\mu_+ x^0/3},
\end{equation}
where $a$ is the scale factor. The above condition corresponds to the following choice of $f$:
\begin{equation}
f = -\frac{\tlrcw_0}{\slrcw_0}\e^{(\mu_- -\mu_+)x^0}\left(J_0+\alpha a^{-4}\right),
\end{equation}
fixing the aforementioned dynamical freedom completely (see Assumption~\ref{ass:dc3}).\footnote{Note that the initial conditions for scale factor are chosen such that the present day value at time $x^0_*$ is normalized, i.e. $a(x^0_*) = 1$. Therefore, $a < 1$ for all times $x < x^0_*$ and therefore, the volume factor $a^{-4}$ in the equation above is not negligible.} 

As a result of this fixing, the function $g_f$ satisfies the following derivative properties
\begin{equation}\label{eq:g_f derivatives}
\frac{g_f'}{g_f} = \mu_- -\frac{8}{3}\mu_+,\qquad \frac{g_f''}{g_f} = \left(\mu_- -\frac{8}{3}\mu_+\right)^2.
\end{equation}
Inserting the expression of $f$ into the function $A(x^0,k)$, one obtains
\begin{equation}
\frac{A'}{A} = \mu_++\mu_- +\frac{8}{3}\mu_+\frac{\alpha\frac{\tlrcw_0}{\slrcw_0}\sin(\Theta)a^{-4}}{\cos(\Theta)\left(J_0-\frac{J_2}{2}k^2\right)-\frac{\tlrcw_0}{\slrcw_0}\sin(\Theta)\left(J_0+\alpha a^{-4}\right)}.
\end{equation}
As we see from the above equation, in general $A$ is a complicated function of the momenta $k$. As a consequence, the same holds for  the factors in front of $\delta V'$ and $\delta V$ in Eq.~\eqref{eq:dV with A}. This is in sharp contrast to what happens in GR, see again Eq.~\eqref{eq:classical perturbed volume equation}. However, this undesired $k$-dependence can be easily removed by choosing $\Theta = n\frac{\pi}{2}$ with odd integer $n$ and assuming that $J_0$ is negligible with respect to $\alpha a^{-4}$ (Assumptions~\ref{ass:kc3},~\ref{ass:dc4} and~\ref{ass:dc5}). {Notice that this is equivalent to $(\delta\pi_x/\epsilon\pi_0^+)^2 a^{-4}\gg 1$ which is ensured by the condition $\pi_x\gg\pi_0^+$.} Under these assumptions, the derivatives of $A$ take the form
\begin{equation}\label{eq:A derivatives}
\frac{A'}{A} = -\frac{5}{3}\mu_++\mu_-,\qquad \frac{A''}{A} = \left(-\frac{5}{3}\mu_++\mu_-\right)^2.
\end{equation}
Combining Eqs.~\eqref{eq:g_f derivatives} and~\eqref{eq:A derivatives}, the perturbed volume equation attains the form
\begin{equation}\label{eq:GFT perturbed volume equation}
\delta V'' - 3\Hubble\delta V' +a^4 k^2\delta V = 0,
\end{equation}
where we identified $\Hubble = \frac{2}{3}\mu_+$ from the background equations. Expressed instead in terms of the ratio $\delta V/\bar{V}$, the relative perturbed volume equation is given by
\begin{equation}\label{eq:GFT perturbed relative volume equation}
\left(\frac{\delta V}{\bar{V}}\right)''+3\Hubble\left(\frac{\delta V}{\bar{V}}\right)' +a^4 k^2\left(\frac{\delta V}{\bar{V}}\right) = 0.
\end{equation}
Remarkably, the two coefficients in front of the zeroth and first derivative term in Eq.~\eqref{eq:GFT perturbed volume equation} are both completely fixed by the background parameter $\mu_+$.\footnote{The values of these two coefficients is a direct consequence of matching the spatial derivative term. 
If the exponent of $a$ is chosen to be $\lambda\in\R$ instead of $4$, the first derivative coefficient is given by $-2\mu_+(2\lambda+1)$. Since the $a^4$-factor is crucial for obtaining the appropriate behavior of perturbations, we fix $\lambda = 4$.} In fact, the parameter $\mu_-$, characterizing the behavior of the timelike condensate, does not enter the perturbed volume equation at all. 
Even though the inclusion of a timelike condensate allows to nicely match the functional form of Eq.~\eqref{eq:GFT perturbed relative volume equation} (in particular solving all the issues reported in \cite{Marchetti:2021gcv}) with that of the GR Eq.~\eqref{eqn:grvolumeperturbations}, the GFT volume perturbation equation does show some new intriguing features, which we investigate in detail in Sec.~\ref{sec:Solutions of GFT and GR perturbations}.

\paragraph{Scale factor observables.} Before closing this section, we would like to emphasize that the cosmological equations obtained at the level of background and perturbations do not depend on the fact that we choose the $3$-volume to encode the (scalar, isotropic) geometric information. In fact, due to homogeneity and isotropy at the level of the background condensate, one could heuristically consider some geometric observable $\mathcal{O}$ whose expectation value is associated to $\abs{\slrcw}^2$ at the background level and which would classically be interpreted as an appropriate power of the scale factor: $\left\langle\mathcal{O}\right\rangle\sim a^d$. Examples other than the volume would be length for $d=1$ or area for $d=2$, {although we remark that here we do not attempt to provide a rigorous definition for these operators, but rather to offer heuristic insights based on their collective nature (i.e.\ their properties as second quantized GFT operators).} Re-iterating the same derivation as above but now with general $d$, the relation of $\mu_+$ and $\Hubble$ is given by
\begin{equation}
\Hubble = \frac{\mathcal{O}'}{\mathcal{O}d} = \frac{2\mu_+}{d}.
\end{equation}
At the perturbed level, the GFT equations would show exactly the same behavior as for the volume, namely
\begin{equation}
\delta\mathcal{O}_{\GFT}''-d\Hubble\delta\mathcal{O}_\GFT' + a^4 k^2\delta\mathcal{O}_{\GFT} = 0.
\end{equation}
Perturbations of the classical quantity, $\delta \mathcal{O}_\GR,$ would instead be described by
\begin{equation}
\delta\mathcal{O}_{\GR}''-2d\Hubble\delta\mathcal{O}_\GR'+d^2\Hubble^2\delta\mathcal{O}_\GR + a^4 k^2\delta\mathcal{O}_\GR = 0\,,
\end{equation}
which shows that the differences in the equations of $\delta V$ from GFT and GR do not arise from considering the \qmarks{wrong} geometric observable.

\subsection{Number of quanta}\label{sec:Number of tetrahedra} 

The number operator, introduced in Eq.~\eqref{eq:number operator}, is clearly the simplest second-quantized operator available in the Fock space. However, as we mentioned above, it is extremely important to characterize the classical and continuum limit of the QG system. On the two-sector Fock space, one can define individual number operators $\hat{N}_\alpha$, counting the number of spacelike and timelike tetrahedra, respectively, or the total number operator $\hat{N} = \hat{N}_+ +\hat{N}_-$. Forming the expectation value of $\hat{N}_\alpha$ with respect to $\ket{\Delta;x^0,\vb*{x}}$, the contributions of the background and perturbations are respectively given by
\begin{align}
\bar{N}_+ &= \abs{\slrcw(x^0,\pmm)}^2,\\[7pt]
\bar{N}_- &= \abs{\tlrcw(x^0,\pmm)}^2,
\end{align}
and
\begin{align}
\delta N_+ &= 2\mathfrak{Re}\left\{\int\dd[4]{\rf}\dd{\mm}\delta\Psi(\rf^\mu,\mm)\left[\bar{\sigma}(\rf^0,\mm)\bar{\tau}(\rf^\mu,\mm)+\mathrm{f}(\rf^\mu)\bar{\sigma}^2(\rf^0,\mm)\right]\right\},\\[7pt]
\delta N_- &= 2\mathfrak{Re}\left\{\int\dd[4]{\rf}\dd{\mm}\left[\delta\Psi(\rf^\mu,\mm)\bar{\sigma}(\rf^0,\mm)\bar{\tau}(\rf^\mu,\mm)+\delta\Xi(\rf^\mu,\mm)\bar{\tau}^2(\rf^\mu,\mm)\right]\right\},\label{eq:dN_-}
\end{align}
{wherein a regularization of an empty rod integration entering $\bar{N}_+$ is understood.} By considering a single-spin condensate, the number operator on the spacelike sector is directly related to the volume operator by the factor of $v$, $\hat{V} = v\hat{N}_+$. Therefore, the expectation values of $\hat{N}_+$ and $\hat{V}$ are related by $v$ at every order of perturbations.

The expectation value of the timelike number operator will be particularly important in the following for two main reasons. First, the matching conditions of the volume perturbations that we derived in Sec.~\ref{sec:Geometric observables} can be interpreted as a condition on $\delta N_-$, detailed in the last paragraph below. Second, the number operators enter the expressions for the matter observables which we analyze in Sec.~\ref{sec:Dynamics of matter observables}. 

\paragraph{Background: spacelike sector.} The number of spacelike tetrahedra $\bar{N}_+$ at background level satisfies
\begin{equation}\label{eq:bkg N+ dynamics}
\frac{\bar{N}_+'}{\bar{N}_+} = 2\mu_+,\qquad \left(\frac{\bar{N}_+'}{\bar{N}_+}\right)' = 0,
\end{equation}
where $\mu_+$ is related to the scalar field momentum $\bar{\pi}_\mf$ as determined by Eq.~\eqref{eq:mu_+ matching}. {Following from the single spin assumption, spacelike particle number and volume are proportional to each other. Thus, the exponential form of $\bar{N}_+$ reflects the form of the scale factor (or the $3$-volume) which is exponential in the relational clock $\rf^0$.} 

\paragraph{Background: timelike sector.} At the background level, the number of timelike tetrahedra satisfies the equations
\begin{equation}\label{eq:bkg N- dynamics}
\frac{\bar{N}_-'}{\bar{N}_-} = 2\mu_-,\qquad \left(\frac{\bar{N}_-'}{\bar{N}_-}\right)' = 0.
\end{equation}
Since the spatial background geometry is fully determined by the spacelike condensate, there are a priori no matching conditions for the parameter $\mu_-$ with respect to an observable of classical GR. This is also due to a lack of GFT-observables that characterize the geometry of timelike slices. {Such observables could also help in deciding whether the timelike condensate state is actually sufficient to characterize a timelike slice in a spatially, but not temporally homogeneous setting.} Further research might reveal additional constraints on {$\mu_-$}, as we discuss in more detail in Sec.~\ref{sec:Discussion and Conclusion}.

{We also remark that the exponential behavior of $\bar{N}_+$ and $\bar{N}_-$ guarantees expectation values of background observables to be sharply peaked, necessary for a classical interpretation. For further references on such a \emph{classicalization}, see~\cite{Pithis:2016cxg,Marchetti:2020qsq,Gielen:2019kae}.}

\paragraph{Perturbations: spacelike sector.} At first order of perturbations, $\delta N_+$ is related to $\delta V$ by a constant factor of $v$. This implies in particular that
\begin{equation}
\frac{\delta N_+}{\bar{N}_+} = \frac{\delta V}{\bar{V}}.
\end{equation}
Given the dynamics of $\delta V/\bar{V}$ in Eq.~\eqref{eq:GFT perturbed volume equation} after matching with GR, the ratio $\delta N_+/\bar{N}_+$ satisfies
\begin{equation}
\left(\frac{\delta N_+}{\bar{N}_+}\right)''+3\Hubble\left(\frac{\delta N_+}{\bar{N}_+}\right)'+ a^4 k^2\left(\frac{\delta N_+}{\bar{N}_+}\right) = 0.
\end{equation}

\paragraph{Perturbations: timelike sector and interpretation of matching conditions.} In  Secs.~\ref{sec:Perturbed equations of motion} and~\ref{sec:Geometric observables}  we introduced some important conditions on the perturbed condensate wavefunction. This allowed us to simplify the intricate equations of motions of $\delta\Psi$ and to match the GR functional form of the perturbed volume equations. Remarkably, there is a direct physical interpretation of these conditions in terms of the perturbed timelike particle number, which we detail in the following.

Considering $\delta N_-$ in Eq.~\eqref{eq:dN_-}, we perform again a split of all the complex-valued quantities into a modulus and a phase. For the first term, $\delta\Psi\bar{\sigma}\bar{\tau}$, the choice of peaking parameters in Eq.~\eqref{eq:peaking parameter relations} leads to a phase which only consists of $\Theta$. Therefore, by setting $\Theta = n\frac{\pi}{2}$ with $n$ an odd integer, this contribution to $\delta N_-$ vanishes with the remaining term being
\begin{equation}
\delta N_- = 2\mathfrak{Re}\left\{\int\dd{\rf^0}\delta\Xi(\rf^0,\mm)\bar{\tlrcw}^2(\rf^0,\mm)\eta^2_{\epsp}(\rf^0-x^0;\pip)\right\}.
\end{equation}
Since $\delta\Xi$ is only time-dependent, as we have shown in Sec.~\ref{sec:Perturbed equations of motion}, it follows that $\delta N_-(x^\mu,\pmm)\equiv\delta N_-(x^0,\pmm)$ only depends on the relational time. Thus, from a relational perspective, the perturbation of the timelike tetrahedra number can be absorbed into the background and does not contribute to the space-dependent first-order perturbations. 

Although this triviality of perturbations in the number of timelike tetrahedra seems to suggest some sort of effective \qmarks{irrelevance} of purely timelike correlations, we note that the above result is due to matching conditions imposed only on a spacelike operator (i.e., the volume). It is conceivable that by considering classical matching conditions on both spacelike and timelike observables, purely timelike correlations would play a more important role.

\subsection{Dynamics of matter observables}\label{sec:Dynamics of matter observables}

In this section, we will derive the dynamics of the \qmarks{matter} (i.e.\ the only non-frame) scalar field $\phi$. Its classical relational dynamics is captured by
the expectation values of suitably defined matter and momentum operators
\begin{align}
\hat{\Upphi}_\alpha &= \frac{1}{i}\int\dd{g_v}\dd{X_\alpha}\dd{\rf^\mu}\dd{\mm}\hat{\varphi}^\dagger(g_v,X_\alpha,\rf^\mu,\mm)\pdv{}{\mm}\hat{\varphi}(g_v,X_\alpha,\rf^\mu,\mm),\label{eq:mf operator}\\[7pt]
\hat{\Pi}^\alpha_\mf &= \int\dd{g_v}\dd{X_\alpha}\dd{\rf^\mu}\dd{\mm}\hat{\varphi}^\dagger(g_v,X_\alpha,\rf^\mu,\mm)\:\mm\:\hat{\varphi}(g_v,X_\alpha,\rf^\mu,\mm)\label{eq:mm operator}.
\end{align}
{Notice that the scalar field operator is denoted by $\hat{\Upphi}$ which is not to be confused with the spacelike-spacelike perturbation $\hat{\delta\Phi}$ and its function $\Phi$.} We also remind that, in contrast to the reference fields $\rf^\mu$, we do not assume a priori that the scalar field propagates only along dual edges of a certain causal character. 

In perfect analogy with Secs.~\ref{sec:Geometric observables} and~\ref{sec:Number of tetrahedra}, we separate the expectation value of the above operators on the condensate states $\ket{\Delta;x^0,\vb*{x}}$ in background and perturbations.
Expectation values of $\hat{\Upphi}_\alpha$ at the background and perturbed level evaluate to
\begin{align}
    \bar{\Upphi}_+ &= \frac{1}{i}\bar{\slrcw}(x^0,\mm)\eval{\pdv{}{\mm}\slrcw(x^0,\mm)}_{\mm=\pmm},\\[7pt]
    \bar{\Upphi}_- &= \frac{1}{i}\bar{\tlrcw}(x^0,\mm)\eval{\pdv{}{\mm}\tlrcw(x^0,\mm)}_{\mm=\pmm},
\end{align}
and 
\begin{align}
\delta\Upphi_+ &= \frac{1}{i}\int\dd[4]{\rf}\dd{\mm}\left[\bar{\sigma}\partial_{\mm}(\delta\Phi\bar{\sigma})+\bar{\delta\Phi}\sigma\partial_{\mm}\sigma+\bar{\sigma}\partial_{\mm}(\delta\Psi\bar{\tau})+\bar{\delta\Psi}\tau\partial_{\mm}\sigma\right],\label{eq:dphi+ pre}\\[7pt]
\delta\Upphi_- &= \frac{1}{i}\int\dd[4]{\rf}\dd{\mm}\left[\bar{\tau}\partial_{\mm}(\delta\Psi\bar{\sigma})+\bar{\delta\Psi}\sigma\partial_{\mm}\tau+\bar{\tau}\partial_{\mm}(\delta\Xi\bar{\tau})+\bar{\delta\Xi}\tau\partial_{\mm}\tau\right],\label{eq:dphi- pre}
\end{align}
respectively. Since we work in momentum space while peaking on momentum, the operators $\hat{\Pi}_\mf^\alpha$ and  $\hat{N}_\alpha$ are closely defined and thus, the corresponding expectation values are simply given by
\begin{equation}
\bar{\Pi}_\mf^\alpha = \bar{N}_\alpha(x^0,\pmm)\pmm,
\end{equation}
and
\begin{equation}
\delta\Pi_\phi^\alpha = \pmm\delta N_\alpha.
\end{equation}
In the following two paragraphs, we analyze the dynamics of these expectation values and suggest a matching to the quantities $\mf$ and $\mm$ of general relativity.

\paragraph{Background part.} To compute $\bar{\Upphi}_\alpha$, we recall the decomposition of the condensate wavefunctions into radial and angular part, $r_{\alpha}(x^0,\mm)$ and $\theta_\alpha(x^0,\mm)$, respectively. Keeping only dominant contributions in $r_\alpha$, one obtains
\begin{equation}
\bar{\Upphi}_\alpha = \bar{N}_\alpha\eval{\partial_{\mm}\theta_\alpha}_{\mm = \pmm}.
\end{equation}
Solutions of the background phases $\theta_\alpha$ are given by
\begin{equation}
\theta_\alpha = \tilde{\pi}^\alpha x^0 - \frac{Q_\alpha}{\mu_\alpha r_\alpha^2} + C_\alpha,
\end{equation}
where $Q_\alpha$ and $C_\alpha$ are integration constants. Then, the zeroth order expectation value of $\hat{\Upphi}_\alpha$ is given by
\begin{equation}
\bar{\Upphi}_\alpha = \eval{-\partial_{\mm}\left(\frac{Q_\alpha}{\mu_\alpha}\right)+2\frac{Q_\alpha}{\mu_\alpha r_\alpha^2}(\partial_{\mm}{\mu_\alpha})x^0+\bar{N}_\alpha\partial_{\mm}{C_\alpha}}_{\mm=\pmm}.
\end{equation}
As a consequence of the peaking properties of $\sigma$ and $\tau$, the timelike condensate parameter $\mu_-$ is independent of $\mm$, i.e. $\partial_{\mm}\mu_- = 0$. If we choose in addition $C_\alpha$ to be independent of $\mm$, $\bar{\Upphi}_\alpha$ is an intensive quantity for both $\alpha$, as one would expect for a scalar field:
\begin{equation}\label{eqn:phibar+}
\bar{\Upphi}_+ = \eval{-\partial_{\mm}\left(\frac{Q_+}{\mu_+}\right)+2\frac{Q_+}{\mu_+}(\partial_{\mm}{\mu_+})x^0}_{\mm=\pmm},\qquad \bar{\Upphi}_- = \eval{-\frac{1}{\mu_-}\partial_{\mm}Q_-}_{\mm=\pmm}.
\end{equation}

In order to connect these expectation values to the scalar field variable $\mf$ of GR, one needs to define a way to combine the expectation values $\Upphi_\alpha$. To that end, we notice that the scalar field is intensive and canonically conjugate to the extensive quantity $\hat{\Pi}_\phi$. In analogy to the chemical potential in statistical physics, one possible way to combine $\Upphi_+$ and $\Upphi_-$ is to consider the weighted sum
\begin{equation}\label{eq:mf as weighted sum}
\mf = \Upphi_+\frac{N_+}{N}+\Upphi_-\frac{N_-}{N},
\end{equation}
where all the quantities appearing are the full expectation values, containing zeroth- and first-order terms. $N$ denotes the expectation value of the total number of GFT particles, i.e. $N = N_+ + N_-$. Expanding all the quantities to linear order, we identify the background scalar field as
\begin{equation}
\bar{\phi} = \bar{\Upphi}_+\frac{\bar{N}_+}{\bar{N}} + \bar{\Upphi}_-\frac{\bar{N}_-}{\bar{N}}.
\end{equation}
{Assuming that $\bar{N}_+\gg\bar{N}_-$ at late times, corresponding to $\mu_+ > \mu_-$ (see also Assumption~\ref{ass:dc6}) and reflecting that the background is predominantly characterized by the spatial geometry}, the matter field can be approximated as
\begin{equation}
\bar{\mf} \approx \bar{\Upphi}_+.
\end{equation}
{Using Eq.~\eqref{eqn:phibar+}, we see that the scalar field is linear in relational time, as expected classically.
Thus, we can easily match the classical GR background equations for $\bar{\mf}$:  imposing $Q_+ = \mm^2$, yields}
\begin{equation}\label{eq:mf background}
\bar{\mf}' = \pmm,\qquad \bar{\mf}'' = 0,
\end{equation}
{as required}. Besides the relation $\mu_+ > \mu_-$, the background matching does not impose any further conditions on $Q_-$ and the precise form of $\mu_-$. 

For $\Pi_\mf^\alpha$, we notice that this quantity grows with the system size, given by the respective number of tetrahedra $\bar{N}_\alpha$. At lowest order, we therefore identify the classical quantity $\bar{\pi}_\mf$ as%
\begin{equation}
\bar{\pi}_\mf = \frac{\bar{\Pi}_\mf^+ + \bar{\Pi}_\mf^-}{\bar{N}} = \frac{\bar{N}_+ + \bar{N}_-}{\bar{N}_+ + \bar{N}_-} \pmm= \pmm,
\end{equation}
which corresponds to the peaked matter momentum $\pmm$. With this identification, the GFT parameter $\mu_+$ can be expressed by the peaked matter momentum as
\begin{equation}
M_{\Pl}^2\:\mu_+^2(\pmm) = \frac{8}{3}\bar{\pi}_\mf^2 = \frac{8}{3}\pmm^2,
\end{equation}
where again a factor of Planck mass has been added to ensure the correct energy dimensions.

\paragraph{First-order perturbations.} Given the expectation values $\delta\Phi_\alpha$ in Eqs.~\eqref{eq:dphi+ pre} and~\eqref{eq:dphi- pre}, we perform a partial integration in $\mm$ and only keep dominating terms, yielding
\begin{align}
\delta\Upphi_+ &= 2\mathfrak{Re}\left\{\int\dd[4]{\rf}\dd{\mm}\left[\delta\Phi\bar{\sigma}^2\partial_{\mm}\theta_++\delta\Psi\bar{\tau}^2\partial_{\mm}\theta_+\right]\right\},\\[7pt]
\delta\Upphi_- &= 2\mathfrak{Re}\left\{\int\dd[4]{\rf}\dd{\mm}\left[\delta\Psi\bar{\sigma}^2\partial_{\mm}\theta_-+\delta\Xi\bar{\tau}^2\partial_{\mm}\theta_-\right]\right\}.
\end{align}
Using the relation of $\delta\Phi$ and $\delta\Psi$ in Eq.~\eqref{eq:relation of dPsi and dPhi}, as well as the assumptions on the peaking parameters of $\sigma$ and $\tau$ in Eq.~\eqref{eq:peaking parameter relations}, the first-order expectation value $\delta\Upphi_+$ evaluates to
\begin{equation}\label{eq:deltaUpphi_+}
\delta\Upphi_+ = \eval{\delta N_+(x^\mu,\mm)\partial_{\mm}\theta_+}_{\mm=\pmm} = \frac{\delta N_+}{\bar{N}_+}\bar{\phi}.
\end{equation}
In contrast to $\delta\Upphi_+$, the evaluation of $\delta\Upphi_-$ is more intricate since the peaking properties of $\bar{\tau}^2$ yield a time derivative expansion when integrating over the reference field. However, as we show next, the perturbed scalar field $\delta\mf$ does not explicitly depend on $\delta\Upphi_-$ under the assumption that $\mu_+>\mu_-$. Following the definition of $\mf$ in Eq.~\eqref{eq:mf as weighted sum}, at linear order in perturbations, one obtains
\begin{equation}
\delta\phi \approx \bar{\phi}\left(\frac{\delta N_+-\delta N_-}{\bar{N}_+}\right)+\bar{\Upphi}_-\frac{\delta N_-}{\bar{N}_+}.
\end{equation}
Since the timelike number perturbation $\delta N_-$ is only time-dependent, and therefore part of the background, the factors of $\delta N_-/\bar{N}_+$ are negligible and one is left with
\begin{equation}
\delta\mf = \left(\frac{\delta V}{\bar{V}}\right)\bar{\mf}.
\end{equation}
Applying Eqs.~\eqref{eq:GFT perturbed relative volume equation} and~\eqref{eq:mf background} for $\delta V/\bar{V}$ and $\bar{\mf}$, respectively, the dynamical equation for $\delta\mf$ from GFT is given by
\begin{equation}\label{eq:GFT perturbed mf equation}
\delta\mf''+a^4 k^2\delta\mf = \left(-3\Hubble\bar{\phi}+2\bar{\phi}'\right)\left(\frac{\delta V}{\bar{V}}\right)'.
\end{equation}
Notice that the right-hand side of this partial differential equation constitutes a source term that is absent in the classical equation of $\mf_\GR$, given in Eq.~\eqref{eq:classical mf perturbation equation}, formulated in harmonic gauge. We discuss the different features of GFT and GR solutions in Sec.~\ref{sec:Solutions of GFT and GR perturbations}.

Let us consider now the first-order matter momentum variable $\delta\Pi_\mf^\alpha$ which, as for the background variable, scales with the system size. In order to connect this quantity to the intrinsic quantity $\delta\mm$ of GR, dividing $\delta\Pi_\mf^\alpha$ by the particle number is required. In principle, there are two different ways to do so, both of which we present in the following. 

First, one can define $\delta\mm$ as the first-order term of
\begin{equation}
\delta\mm\overset{(1)}{=}\frac{\Pi_\mf^++\Pi_\mf^-}{N_++N_-} = 0,
\end{equation}
where all the quantities entering this expression contain both, zeroth- and first-order perturbations. However, in this case $\delta\mm = 0$. Operatively, this could be interpreted as a perturbation of the background momentum $\bar{\pi}_\mf$. Since this is a constant of motion, any such perturbation would vanish by construction.

Alternatively, one could perturb only the momenta and keep the particle numbers at zeroth order. In this case, $\delta\mm$ is given by
\begin{equation}
\delta\mm \overset{(2)}{=} \frac{\delta\Pi_{\mf}^+ + \delta\Pi_{\mf}^-}{\bar{N}_++\bar{N}_-} \approx \pmm\frac{\delta N_+}{\bar{N}_+} = \pmm\frac{\delta V}{\bar{V}}.
\end{equation}

None of the options above offer a matching to the classical perturbed momentum variable $\delta\pi_\mf^0$, defined in Eq.~\eqref{eq:classical perturbed mm 0} as the $0$-component of the conjugate momentum of $\mf$ at linear order. The main difficulty in matching these two quantities is that the classical equation~\eqref{eq:classical perturbed mm 0} depends on the perturbation of the lapse function, $A$. To recover this quantity from the fundamental QG theory, one would need additional (relational) geometric operators other than the volume. We will return to this issue when in Sec.~\ref{sec:Discussion and Conclusion}. 

\subsection{A Mukhanov-Sasaki-like equation}\label{sec:mukhanovsasaki}

In classical cosmology, physical information is encoded in perturbatively gauge-invariant quantities (see \cite{Riotto:2002yw,Baumann:2018muz,Brandenberger:2003vk,maggiorebook,Dodelson:2003ft,gorbunov} for a review), such as the Bardeen \cite{bardeen1,bardeen2} and the curvature perturbation variables $\zeta$ \cite{bardeen2,brandenberger1} and $\mathcal{R}$ \cite{lyth}, defined in Eq~\eqref{eq:classical R}. The latter, usually called comoving curvature perturbation, is especially important in inflationary physics, being proportional to the so-called Mukhanov-Sasaki variable \cite{Mukhanov:1988jd,mukhanov2,sasaki}. 

However, as discussed in Appendix~\ref{sec:Classical perturbation theory}, $\mathcal{R}$ cannot be constructed out of volume and matter observables only. {That is because, as emphasized already at the beginning of Sec.~\ref{sec:Geometric observables}, the volume $\delta V/\bar{V}$ is composed of both perturbation functions $\psi$ and $E$. In order to single out the function $E$ and identifying it with expectation values of GFT operators, one would have to relax isotropy (see Assumption \ref{ass:ks2}) and introduce anisotropic observables, such as the areas of orthogonal two-surfaces. Until such operators are defined, the importance of which we highlight in Sec.~\ref{sec:Discussion}, we introduce a \qmarks{curvature-like} variable $\tilde{\mathcal{R}}$
\begin{equation}
\tilde{\mathcal{R}}\defeq -\frac{\delta V}{3\bar{V}}+\Hubble\frac{\delta\mf}{\bar{\mf}'},
\end{equation}
in analogy to $\mathcal{R}$.} As remarked above, classically, $\mathcal{R}$ is gauge-invariant under infinitesimal transformations $x^\mu\mapsto x^\mu+\xi^\mu$. In contrast, $\tilde{\mathcal{R}}$, as defined above (see also Eq.~\eqref{eq:classical Rtilde}), is classically gauge-invariant only in the super-horizon limit. In the context of GFT however, the quantity $\tilde{\mathcal{R}}_\GFT$ defined above is obtained by combining effectively relational observables (obtained via averages on CPSs), and thus it is (effectively) gauge-invariant by construction. Still, as for the volume and matter observables, its dynamics can be directly compared with that of GR in harmonic gauge.

Applying the GFT dynamics for $\delta V/\bar{V}$ and $\delta\mf$ to $\tilde{\mathcal{R}}$, given in Eqs.~\eqref{eq:GFT perturbed relative volume equation} and~\eqref{eq:GFT perturbed mf equation}, respectively, yields
\begin{equation}\label{eq:GFT perturbed R}
\tilde{\mathcal{R}}'' + a^4 k^2\tilde{\mathcal{R}} = \left[3\Hubble-\frac{1}{4 M_{\Pl}^2}\left(\bar{\mf}^2\right)'\right]\left(\frac{\delta V}{\bar{V}}\right)'.
\end{equation}
Similar to the perturbed matter equation from GFT, the differential equation for $\tilde{\mathcal{R}}$ contains a source term on the right-hand side that is not present in the classical case, Eq.~\eqref{eq:classical R perturbation equation}. We explicitly compare the solutions of $\tilde{\mathcal{R}}_\GR$ and $\tilde{\mathcal{R}}_\GFT$ in Sec.~\ref{sec:Solutions of GFT and GR perturbations}, showing that the discrepancies are negligible under certain assumptions on the initial conditions.  

We close this section by presenting a different expression for the Mukhanov-Sasaki-like equation which is closer to standard cosmology formulations. As mentioned above, since Eq.~\eqref{eq:GFT perturbed R} is expressed in a fully relational fashion using the physical scalar reference frame, it is most naturally compared to the GR-equation formulated in harmonic coordinates, given in Eq.~\eqref{eq:classical R perturbation equation}. However, in standard cosmology, it is common practice to use conformal-longitudinal coordinates instead. Since we work in a manifestly coordinate-independent setting, we need to introduce a parametrization of the reference fields to make the connection to this representation in GFT. Adapted to the scalar reference frame, we introduce harmonic coordinates and then change to the conformal-longitudinal system (see Appendix~\ref{sec:Change of gauge} for more details). As a relational quantity, $\tilde{\mathcal{R}}_\GFT$ is  manifestly gauge-invariant and thus behaves as a scalar under the parametrization change.
Moreover, as it is a first-order quantity, only the background change from harmonic time to conformal time $\tau$ matters, yielding
\begin{equation}
\dv[2]{\tilde{\mathcal{R}}}{\tau}+2\mathcal{H}\dv{\tilde{\mathcal{R}}}{\tau}+k^2\tilde{\mathcal{R}} = \left[3\mathcal{H}-\frac{1}{4 M_{\Pl}^2}\dv{}{\tau}\left(\bar{\mf}^2\right)\right]\dv{}{\tau}\left(\frac{\delta V}{\Bar{V}}\right),
\end{equation}
where $\mathcal{H}$ is the Hubble parameter with respect to conformal time. 

\subsection{Solutions of GFT and GR perturbations}\label{sec:Solutions of GFT and GR perturbations}

In this section, we provide a direct comparison of emergent perturbations from GFT and classical perturbations from GR. In particular, we study solutions of the GFT and GR equations to see how the differences in the respective differential equations are reflected in their solutions. This allows to determine conditions on initial values under which the GFT curvature-like variable $\tilde{\mathcal{R}}_\GFT$ shows good agreement with $\tilde{\mathcal{R}}_\GR$ even for intermediate and sub-horizon modes $a^2k\gtrsim\Hubble$.

\begin{figure}
    \centering
    \begin{subfigure}{0.5\textwidth}
    \includegraphics[width=\linewidth]{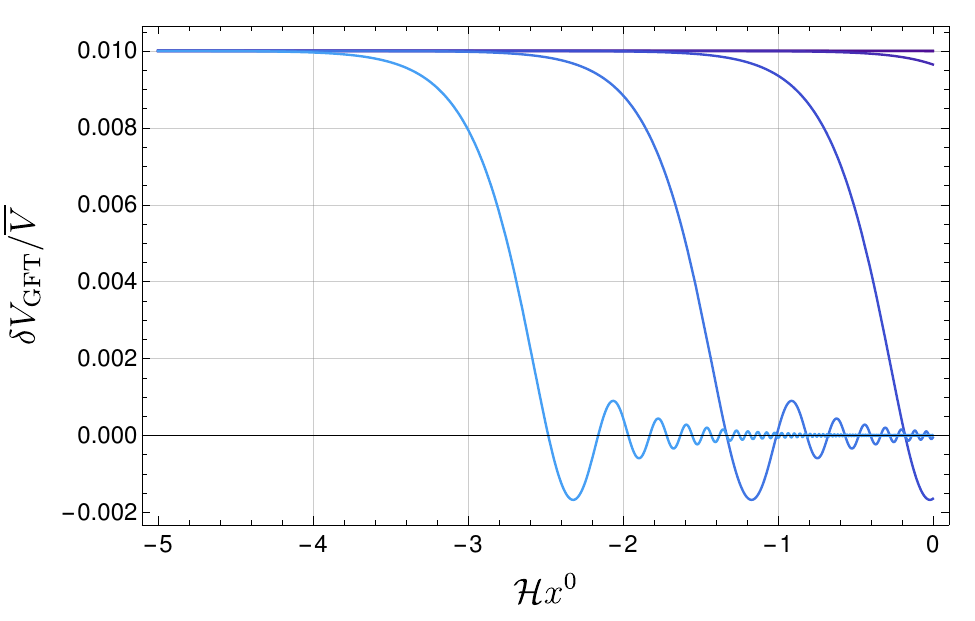}
    \end{subfigure}%
    \begin{subfigure}{0.5\textwidth}
    \includegraphics[width=\linewidth]{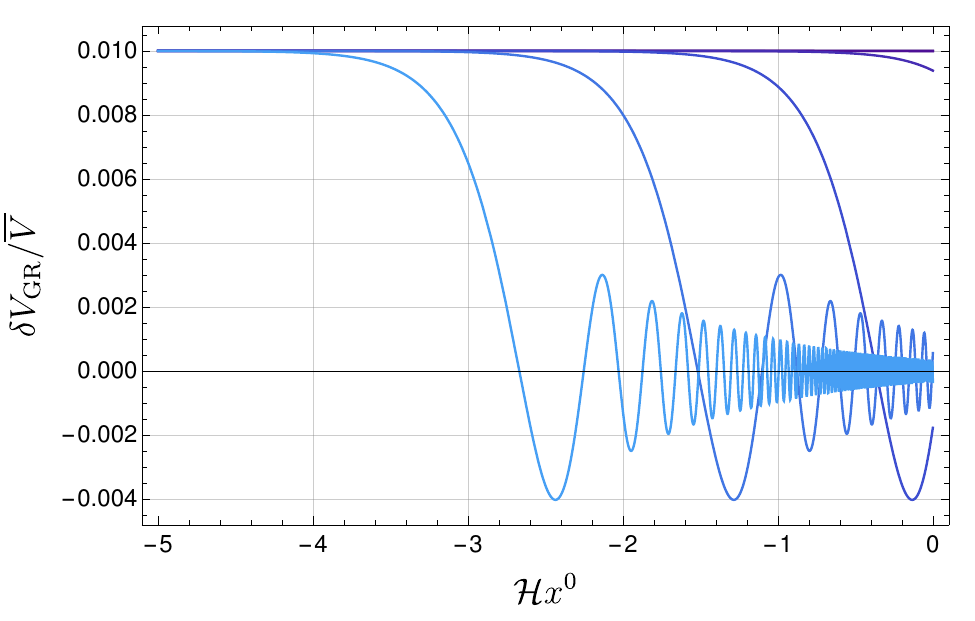}
    \end{subfigure}
    \caption{Evolution of relative volume perturbations $\delta V/\bar{V}$ as dictated by GFT, respectively GR. Here, we set the integration constant $c_1^{(\mathrm{GR})} = 10^{-2}$. Notice, that the quantities $(x^0,\Hubble)$ and $(k,\Hubble)$ only appear as a product, respectively a quotient, setting the two scales for the system $\Hubble x^0$ and $k/\Hubble$. The drawn momenta lie in the range $k/\Hubble\in\{10^{-3},\;...,\;10^2\}$, where lighter blue corresponds to larger modes. Qualitatively, this range captures super-horizon, intermediate and sub-horizon modes, respectively.} 
    \label{fig:volume_pert}
\end{figure}

\paragraph{Volume perturbations.} In momentum space for the rod variable, the dynamics of the relative volume perturbation $\delta V/\bar{V}$ are captured by 
\begin{align}\label{eqn:gftvolumecomp}
\left(\frac{\delta V}{\bar{V}}\right)_\GFT''+a^4 k^2\left(\frac{\delta V}{\bar{V}}\right)_\GFT &= -3\Hubble\left(\frac{\delta V}{\bar{V}}\right)_\GFT',\\[7pt]
\left(\frac{\delta V}{\bar{V}}\right)_\GR''\;\:+a^4 k^2\left(\frac{\delta V}{\bar{V}}\right)_\GR\;\: &= 0,\label{eqn:grvolumeperturbations}
\end{align}
for GFT and GR, respectively. A derivation of the classical equation is given explicitly in Appendix~\ref{sec:Classical perturbation theory - geometry}. These equations can be analytically solved, yielding
\begin{align}  
    \left(\frac{\delta V}{\bar{V}}\right)_\GFT &= e^{-3\Hubble x^0/2}\left[c^\GFT_1 J_{-3/4}\left(\frac{a^2k}{2 \Hubble}\right)+ c^\GFT_2 J_{3/4}\left(\frac{a^2k}{2 \Hubble}\right)\right],\\[7pt]
    \left(\frac{\delta V}{\bar{V}}\right)_\GR\;\: &= c^\GR_1 J_0\left(\frac{a^2k}{2 \Hubble}\right)+ c^\GR_2 Y_0\left(\frac{a^2 k}{2 \Hubble}\right),
\end{align}
where $J_n$ and $Y_n$ are Bessel functions of the first and second kind, respectively. Requiring that classical GR perturbations are constant in the super-horizon regime $a^2 k/\Hubble\ll 1$ imposes the condition $c_2^\GR = 0$. Matching the GFT perturbations in the super-horizon limit, one obtains the two conditions
\begin{equation}
    c_1^\GFT = 0,\qquad    c_2^\GFT = \left(4\frac{\Hubble}{k}\right)^{3/4}\Gamma(7/4)c_1^\GR.
\end{equation}
With these choices of initial conditions, we thus have
\begin{align}
\left(\frac{\delta V}{\bar{V}}\right)_\GFT &= c_1^\GR \Gamma\left(\frac{7}{4}\right)\left(\frac{4\Hubble}{k}\right)^{3/4}\e^{-3\Hubble x^0/2}J_{3/4}\left(\frac{a^2 k}{2\Hubble}\right),\label{eq:GFT solution relative volume perturbation}\\[7pt]
\left(\frac{\delta V}{\bar{V}}\right)_\GR\;\: &= c_1^\GR J_0\left(\frac{a^2k}{2\Hubble}\right)\label{eq:GR solution relative volume perturbation},
\end{align}
depicted in Fig.~\ref{fig:volume_pert} for $c_1^\GR = 0.01$. By assumption, the perturbations are smaller than the background volume and thus, $c_1^{\mathrm{GR}}$ is required to be smaller than $1$. As we are going to show in the following paragraphs, this is consistent with finding good matching of matter and curvature-like perturbations.

The deviations between GFT and GR volume perturbations are two-fold, and become relevant for modes $a^2k/\Hubble\gtrsim 1$. First, as the perturbations cross the Hubble horizon, the ratio $(\delta V/\bar{V})_{\mathrm{GFT}}$ is more strongly suppressed by an additional exponential factor of $\e^{-3\Hubble x^0/2}$. Second, the phase of the two functions is shifted. This becomes apparent by considering an asymptotic expansion of $\delta V/\bar{V}$ in terms of large modes $ a^2 k/\Hubble\gg 1$
\begin{align}
\left(\frac{\delta V}{\bar{V}}\right)_\GFT &\;\underset{a^2 k/\Hubble\gg 1}{\longrightarrow}\;\sqrt{\pi} c_1^\GR\Gamma\left(\frac{7}{4}\right)\left(\frac{4\Hubble}{k}\right)^{5/4}\e^{-5\Hubble x^0/2}\sin\left(\frac{\pi}{8}-\frac{a^2k}{2\Hubble}\right),\\[7pt]
\left(\frac{\delta V}{\bar{V}}\right)_\GR\;\: &\;\underset{a^2 k/\Hubble\gg 1}{\longrightarrow}\; \sqrt{\pi}c_1^\GR\left(\frac{4\Hubble}{k}\right)^{1/2}\e^{-\Hubble x^0}\cos\left(\frac{\pi}{4}-\frac{a^2k}{2\Hubble}\right).
\end{align}
We will give a physical interpretation for this sub-horizon deviations at the end of this section. 

The initial conditions have been chosen so that the matching of $\delta V_\GFT/\bar{V}$ and $\delta V_\GR/\bar{V}$ holds for super-horizon modes. Attempting a matching with different initial conditions naturally leads to matching conditions which are inconsistent under relational time-evolution. More precisely, matching the two perturbations at a certain time $x^0_*$ and at a certain scale $k/\Hubble$ which is not super-horizon, the volume perturbations will only match at this instance of time and show strong deviations for all $x^0\neq x^0_*$. Therefore, we conclude that one obtains the closest matching with initial conditions assigned in the super-horizon regime, which is in fact common practice in standard cosmology \cite{Riotto:2002yw,Baumann:2018muz,Brandenberger:2003vk,Dodelson:2003ft,gorbunov}. In the following paragraphs, we will therefore assume that $\delta V/\bar{V}$ is given by Eqs.~\eqref{eq:GFT solution relative volume perturbation} and~\eqref{eq:GR solution relative volume perturbation} for GFT and GR, respectively. 

\paragraph{Matter perturbations.}

\begin{figure}
    \centering
    \begin{subfigure}{0.5\textwidth}
    \includegraphics[width=\linewidth]{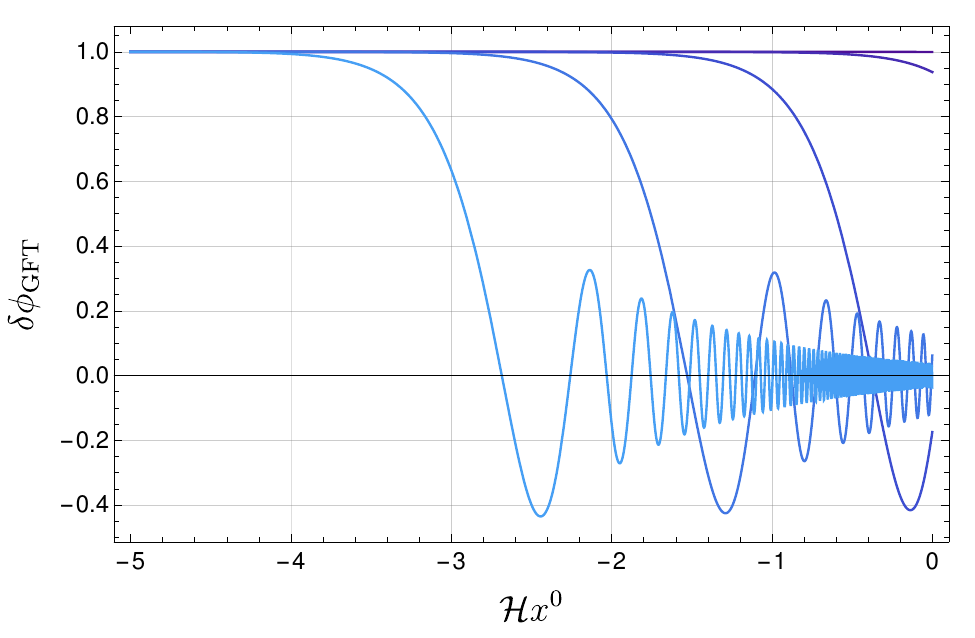}
    \end{subfigure}%
    \begin{subfigure}{0.5\textwidth}
    \includegraphics[width=\linewidth]{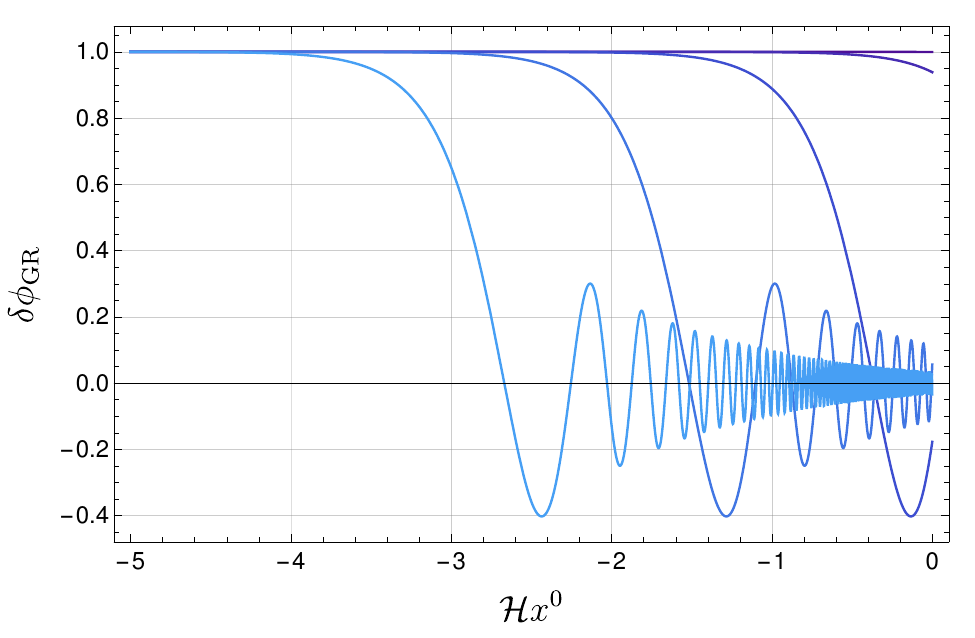}
    \end{subfigure}%
    \caption{Solutions of matter perturbations derived from GFT and GR, respectively, where we set $d_1^\GR = 1$. In this plot, the modes $k/\Hubble$ lie again in the set $\{10^{-3},\;...,\;10^2\}$ with the same color coding as for $\delta V$. Furthermore, the initial condition of the relative volume perturbations is fixed to $c_1^\GR = 10^{-2}$.}
    \label{fig:matter_pert}
\end{figure}

\begin{figure}
    \centering
    \includegraphics[width = 0.65\textwidth]{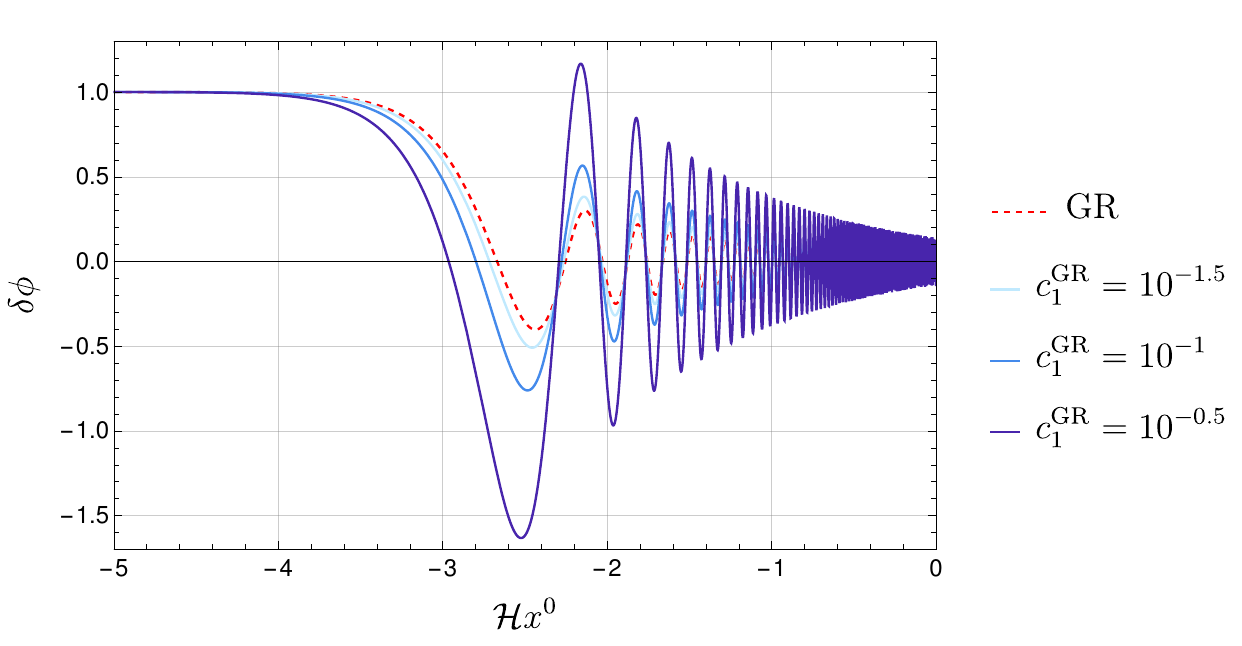}
    \caption{For fixed mode $k/\Hubble = 10^3$, this plot shows a comparison of the GFT (blue) and GR (dashed red) solutions depending on the value of $c_1^\GR$. Clearly, the two solutions closely agree for small $c_1^\GR$.}
    \label{fig:matter_pert_c}
\end{figure}

Matter perturbations in GFT and GR are respectively governed by
\begin{align}\label{eqn:gftmattercomp}
\delta\mf_\GFT''+a^4 k^2\delta\mf_\GFT &= \left(-3\Hubble\bar{\phi}+2\bar{\phi}'\right)\left(\frac{\delta V}{\bar{V}}\right)_\GFT',\\[7pt]
\delta\mf_\GR''\;\:+a^4 k^2\delta\mf_\GR\;\: &= 0,
\end{align}
where we refer to Appendix~\ref{sec:Classical perturbation theory - matter} for a derivation of the classical equation. While the GR matter equation forms a closed system, the matter perturbation equation of GFT does not close, since the relative volume perturbation enters as a source term on the right-hand side. However, as we show in this paragraph, these inhomogeneities are controlled by the parameter $c_1^{\GR}$. To make this statement explicit, notice that the solution of $\delta\mf_\GR$ is given by
\begin{equation}
\delta\mf_\GR(x^0,k) = d_1^\GR J_0\left(\frac{a^2 k}{2\Hubble}\right).
\end{equation}
For simplicity, we set $d_1^\GR = 1$ in the remainder, such that the difference of initial conditions for $\delta\mf$ and $\delta V/\bar{V}$ is captured by $c_1^\GR$. The differential equation for $\delta\mf_\GFT$ is solved numerically. Given initial conditions $\delta\mf_\GFT(-5) = d_1^\GR$ and $\delta\mf_\GFT'(-5) = 0$, the dynamics of GFT and GR matter perturbations show good agreement, as Fig.~\ref{fig:matter_pert} visualizes. A more direct comparison of the two perturbations for a fixed mode, say $k/\Hubble = 10^3$, and for varying constant $c_1^\GR$ is given in Fig.~\ref{fig:matter_pert_c}. 

\paragraph{Curvature-like perturbations.}

Similar to the matter perturbations, the equations for the curvature-like perturbation $\tilde{\mathcal{R}}$
\begin{align}\label{eqn:gftcurvaturecomp}
\tilde{\mathcal{R}}_\GFT''+a^4 k^2\tilde{\mathcal{R}}_\GFT &= \left[3\Hubble-\frac{1}{4 M_{\Pl}^2}\left(\bar{\mf}^2\right)'\right]\left(\frac{\delta V}{\bar{V}}\right)_\GFT',\\[7pt]
\tilde{\mathcal{R}}_\GR''\;\:+a^4k^2\tilde{\mathcal{R}}_\GR\;\: &= 0,
\end{align}
derived from GFT and GR, respectively, differ in a source term on the right-hand side of the GFT equation. Since $\tilde{\mathcal{R}}$ is by definition a linear combination of the volume perturbation ratio $\delta V/\bar{V}$ and the matter perturbation $\delta \mf$, the initial conditions must be chosen accordingly. 

\begin{figure}
    \centering
    \begin{subfigure}{0.5\textwidth}
    \includegraphics[width=\linewidth]{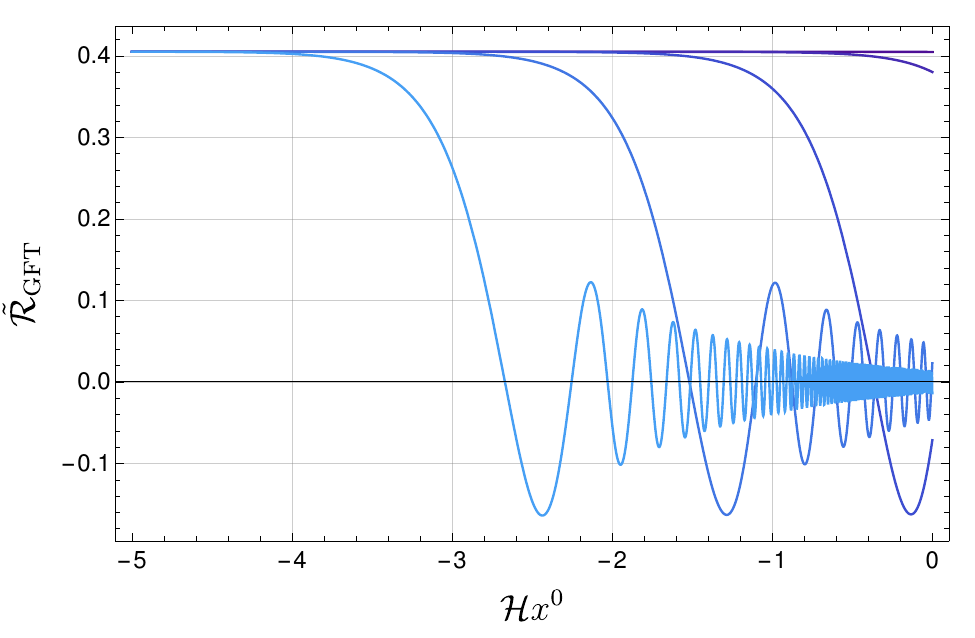}
    \end{subfigure}%
    \begin{subfigure}{0.5\textwidth}
    \includegraphics[width=\linewidth]{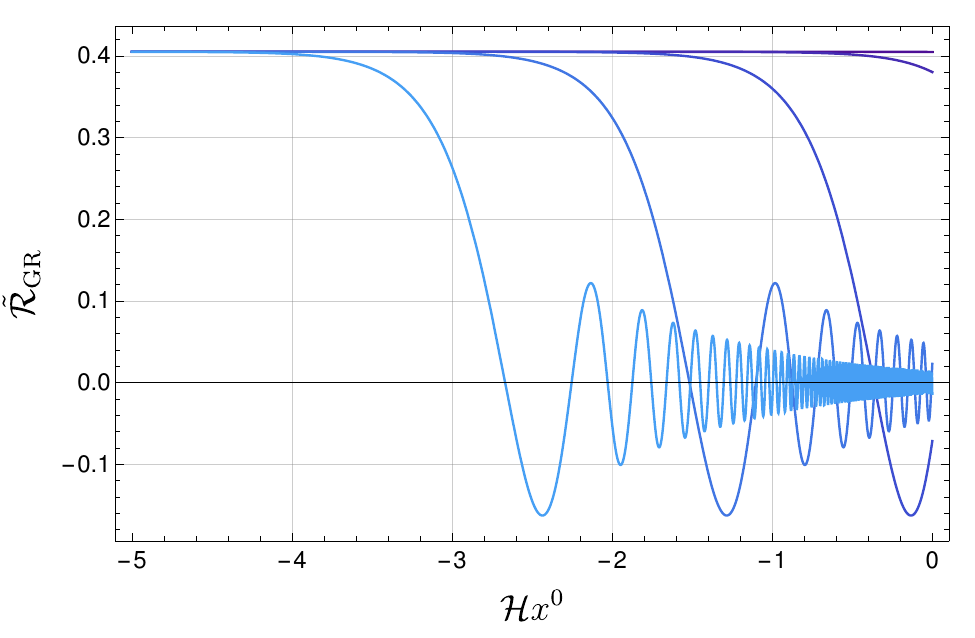}
    \end{subfigure}
    \caption{Solutions of curvature-like perturbations $\tilde{\mathcal{R}}$ in GFT and GR, respectively, where the  integration constant is fixed to $c_1^\GR=1$. The range of modes and the associated coloring in the plots are as for $\delta V$ and $\delta \mf$ above.}
    \label{fig:R_pert}
\end{figure}

With the choices for $\delta V_\GR/\bar{V}$ and $\delta\mf_\GR$ of the previous paragraphs, being $d_1^\GR = 1$ and $c_1^\GR$ not fixed, the curvature-like perturbation equation of GR is solved by
\begin{equation}
\tilde{\mathcal{R}}_\GR = \left(\frac{1}{\sqrt{6}}-\frac{c_1^\GR}{3}\right)J_0\left(\frac{a^2 k}{2\Hubble}\right).
\end{equation}
It is important to notice that $c_1^\GR$ enters the expression of the classical quantity $\tilde{\mathcal{R}}_\GR$ explicitly, in contrast to the classical matter perturbations $\delta\mf_\GR$. To find the behavior of $\tilde{\mathcal{R}}_\GFT$, we solve its governing differential equation numerically, with initial conditions $\tilde{\mathcal{R}}(-5) = \frac{1}{\sqrt{6}}-\frac{c_1^\GR}{3}$ and $\tilde{\mathcal{R}}'(-5) = 0$. The result is depicted on the left-hand side of Fig.~\ref{fig:R_pert}, next to the analytical solution of GR. 

\begin{figure}
    \centering
    \begin{subfigure}{0.5\textwidth}
    \includegraphics[width=\linewidth]{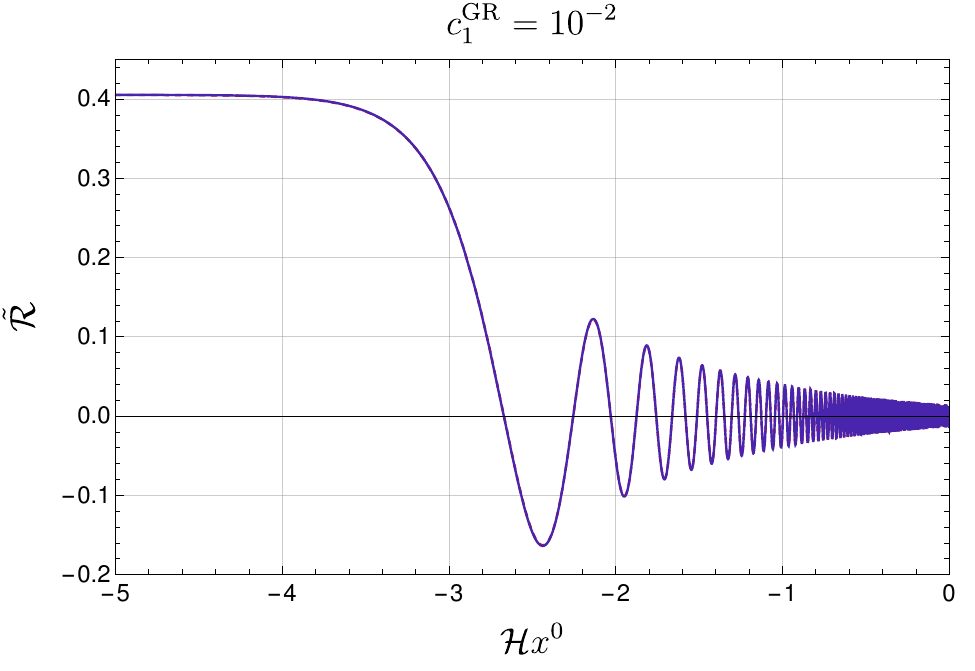}
    \end{subfigure}%
    \begin{subfigure}{0.5\textwidth}
    \includegraphics[width=\linewidth]{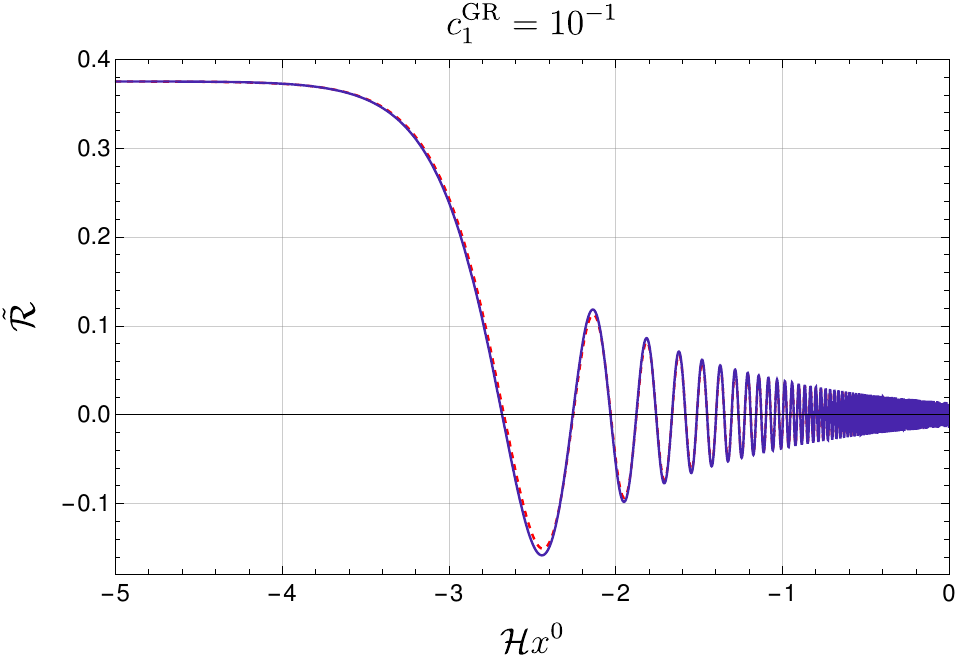}
    \end{subfigure}\\
    \begin{subfigure}{0.5\textwidth}
    \includegraphics[width=\linewidth]{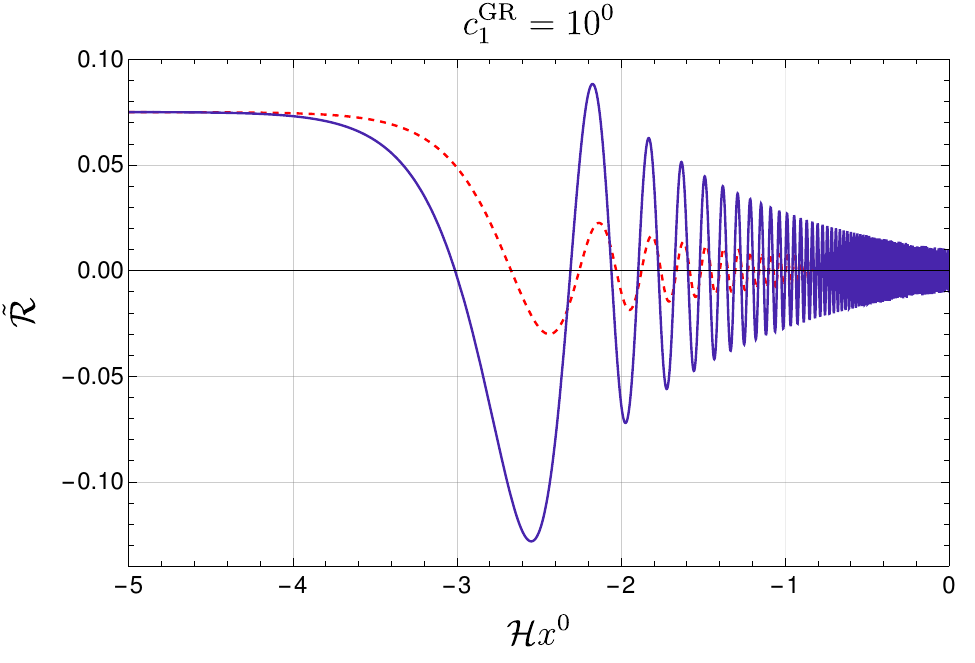}
    \end{subfigure}
    \caption{Comparison of classical (dashed red) and GFT solution (blue) for different values of $c_1^\GR$ at a fixed mode $k/\Hubble = 10^3$. For $c_1^\GR = 10^{-2}$, the curves show almost perfect matching and are therefore visually indistinguishable.}
    \label{fig:R_pert_c}
\end{figure}

Matching of GFT and GR solutions is controlled by the parameter $c_1^\GR$, which now enters both $\tilde{\mathcal{R}}_\GR$ and $\tilde{\mathcal{R}}_\GFT$. A comparison of the classical and the GFT solutions for a fixed mode $k/\Hubble = 10^3$ and different values of $c_1^\GR$ is depicted in Fig.~\ref{fig:R_pert_c}. As explained above, $\tilde{\mathcal{R}}_{\GFT}$ is gauge-invariant as a relational quantity, while $\tilde{\mathcal{R}}_\GR$ changes under gauge transformations as $\tilde{\mathcal{R}}_\GR\mapsto \tilde{\mathcal{R}}_\GR+ k^2\xi$. One can exploit this gauge freedom on the GR side to improve the matching for a fixed mode. However, a consistent improvement of matching between $\tilde{\mathcal{R}}_\GFT$ and $\tilde{\mathcal{R}}_\GR$ by choosing a certain $\xi$ is not possible for all modes. Therefore, there does not exist a gauge where the matching of GFT and GR perturbations is perfect at all modes. 

Summarizing, the GFT perturbation equations for $\delta V/\bar{V}$, $\delta\mf$ and $\tilde{\mathcal{R}}$ do not close in that the relative volume perturbation enters the equations for $\delta\mf$ and $\tilde{\mathcal{R}}$ as a source term. While solutions for $\delta V/\bar{V}$ can be fully matched only in the super-horizon limit, both $\delta\mf$ and $\tilde{\mathcal{R}}$ can be matched for a wide range of modes by requiring the ratio $c_1^\GR/d_1^\GR$ to be a small value (as required for the self-consistency of the perturbative setting). 

It should be emphasized that under the above conditions, the mismatch between GFT and GR solutions is in general only important for trans-Planckian modes, as explicitly shown in \cite{Jercher:2023kfr}. For smaller modes (and in particular for any mode of cosmological interest), the GFT perturbation dynamics are perfectly consistent with those of GR \cite{Jercher:2023kfr}. However, as we emphasized above, the GFT and GR dynamics differ only in the sub-horizon regime. This is because, as we argue below, there is a natural correspondence between the sub-horizon $a^2 k/\Hubble \gg 1$ and trans-Planckian modes when the cosmological system is relationally described in terms of a physical reference frame consisting of four massless scalar fields.

\paragraph{Sub-horizon and trans-Planckian modes.} Particularly important in this paragraph, we work in units of $8\pi G = 1$ and explicitly keep track of factors of Planck masses. Starting this classical analysis, we introduce harmonic coordinates $\rf^\mu =\kappa^\mu x^\mu$ (no summation over $\mu$)~\cite{Marchetti:2021gcv}, with $\kappa^\mu$ dimensionful constants. As explained in Appendix~\ref{sec:Classical perturbation theory}, harmonic coordinates are, at the background, defined by the harmonic gauge condition $a^3/N = c_H = \text{const}$. In this paragraph, we keep the constant $c_H$ arbitrary and set it to unity otherwise. Then, the energy momentum tensor, defined in Eq.~\eqref{eq:classical EM-tensor}, is given by
\begin{equation}
   M_\Pl^2\: T_{\mu\nu}=\sum_{\lambda=0}^3(\kappa^\lambda)^2\left(\delta^\lambda_\mu\delta^\lambda_\nu-\frac{g_{\mu\nu}}{2}g^{\rho\sigma}\delta^\lambda_\rho\delta^\lambda_\sigma\right)+\partial_\mu\phi\partial_\nu\phi-\frac{g_{\mu\nu}}{2}g^{\rho\sigma}\partial_\rho\phi\partial_\sigma\phi\,,
\end{equation}
with background $(00)$-component
\begin{equation}
M_{\Pl}^2 \bar{T}_{00} = \frac{1}{2}\left[(\kappa^0)^2+(\bar{\mf}')^2+3a^4\frac{\kappa^2}{c_H^2}\right].
\end{equation}
which can be expressed equivalently via the canonical conjugate momenta as
\begin{equation}
M_{\Pl}^2\bar{T}_{00} = \frac{1}{2}\left[ \left(\frac{\pi_{\mf}}{c_H}\right)^2+\left(\frac{\pi_{\rf^0}}{c_H}\right)^2+3\frac{(\pi_{\rf^i})^2}{a^4}\right].
\end{equation}
Notice that we set $\kappa \equiv\kappa^i$, motivated by relational isotropy (see Assumption~\ref{ass:ks2}).\footnote{Importantly, we define perturbations at the level of the on-shell energy-momentum tensor rather than perturbing the fields and then splitting $T_{\mu\nu}$. This procedure is more natural from the perspective of perturbing the Einstein equations.}  Because of the particular alignment of coordinates and reference frame, the constants $\partial_\mu \kappa^i x^i$ already enter at the background level. Notice, that this strongly differs to cosmology in coordinates, where the spatial coordinates only enter the perturbed quantities. At zeroth order in perturbations, the Einstein equations are given by\footnote{Notice, that this is the form of the Einstein equations when considering all of the matter momenta. For the argument that we make here, it is important to keep the contributions of the $(\kappa^\mu)^2$. When matching the homogeneous GFT equations to background cosmology, we assume that $\mm$ dominates the matter content, see also~\ref{ass:ds5}.}
\begin{equation}
3\Hubble^2 = \frac{1}{2 M_\Pl^2}\left[(\kappa^0)^2+(\bar{\mf}')^2+3a^4\frac{\kappa^2}{c_H^2}\right].
\end{equation}
As one can check explicitly from e.g. Eq.~\eqref{eq:classical relative perturbed volume equation}, following the definition of $\Hubble^2$ above, the sub-horizon condition is re-expressed as 
\begin{equation}
a^4k^2\gg \Hubble^2\geq \frac{a^4\kappa^2}{2 M_{\Pl}^2c_H^2}.
\end{equation}
Now, in order for the coordinates $x^\mu$ to have mass dimension of $-1$ (given that $[\rf^\mu]=1$), the mass dimension of the $\kappa^\mu$ needs to be $2$. {This is consistent with the fact that $\kappa^\mu$ is proportional to the canonical conjugate momentum which has a mass-dimension of $2$.} Thus, we can re-write $\kappa^\mu$ in terms of the Planck-mass and some dimensionless constant $\tilde{\kappa}^\mu$, i.e. $\kappa^\mu = \tilde{\kappa}^\mu M_{\mathrm{pl}}^2$. The sub-horizon condition can therefore be recast into
\begin{equation}
    k^2\gg \frac{\tilde{\kappa}^2}{c_H^2}{M_{\mathrm{pl}}^2}.
\end{equation}
Choosing a time coordinate that is perfectly adapted to the physical clock suggests $\tilde{\kappa}^0  = 1$. Furthermore, we choose $\tilde{\kappa}/c_H = \tilde{\kappa}^0$, such that the rod contribution to the energy momentum is equal to that of the clock at present time when $a=1$.  In this case, the sub-horizon condition implies that
\begin{equation}
    k^2\gg M_{\mathrm{pl}}^2\,,
\end{equation}
which suggests that the sub-horizon regime is in fact trans-Planckian in the physical reference frame adopted. Importantly, this conclusion is based on the rod contribution to $\bar{T}_{00}$ which, as argued in Appendix~\ref{sec:Change of gauge}, is generic and not only a particular feature of the harmonic coordinate system.  

Notice that this issue becomes only apparent when one considers inhomogeneities, since in the homogeneous setting there is no notion of super- and sub-horizon regimes. This is why choosing $\kappa^\mu=1$ or $\kappa^\mu=M_{\text{pl}}^2$ did not make any practical difference in previous studies (e.g.\ in \cite{Oriti:2016qtz} $\kappa^\mu=1$ was imposed when comparing the effective GFT equations with GR). Clearly, the result that sub-horizon modes are trans-Planckian in the standard effective field
theory treatment of linear cosmological perturbations in the relational frame used here calls for a better understanding and further investigations. Specifically, it should be checked if this also holds for the more realistic case of a dust frame. We comment on the relation to the so-called \textit{Trans-Planckian Censorship Conjecture} (TCC) in Sec.~\ref{sec:Summary}.

\paragraph{Summary.}
Summarizing this section, we compared geometric and matter perturbations in GFT and GR, which we combined to construct a curvature-like variable $\tilde{\mathcal{R}}$. The GFT effective dynamics of these perturbations (when the ratio of initial conditions $c_1^\GR/d_1^\GR$ is small) are in remarkable agreement with GR for non-Planckian modes, as explicitly shown in \cite{Jercher:2023kfr}. Here, we have shown that these  trans-Planckian corrections (with a clear quantum gravitational origin\footnote{Note that because of the single-spin assumption \ref{ass:dc2}, the source term appearing in Eqs.~\eqref{eqn:gftvolumecomp}, \eqref{eqn:gftmattercomp}, and \eqref{eqn:gftcurvaturecomp}, which produces the above corrections, can be equivalently written as proportional to $(\delta V/\bar{V})'$ and $(\delta N_+/\bar{N}_+)'$. Although it is not possible to distinguish between the two quantities within this approximation, we note that there is a crucial conceptual difference between them, since the latter has no classical counterpart. Should further studies identify the source term as proportional to $(\delta N_+/\bar{N}_+)'$, this would provide further evidence that the deviations discussed in this section are quantum gravitational in nature.}) manifest themselves prominently at sub-horizon scales, in virtue of a correspondence between sub-horizon and trans-Planckian
modes due to the relational description that we are adopting. 

Notice, that all perturbation quantities remain small under relational time evolution, implying that the perturbative framework defined here is in fact self-consistent. Finally, classical corrections to the standard cosmological perturbation equations coming from the presence of reference fields only enter proportionally to the rod variable $\kappa^2$, which in turn can be consistently neglected (see Appendix~\ref{sec:Classical perturbation theory - matter} for further details).

\section{Summary and Conclusion}\label{sec:Discussion and Conclusion}

In this section, we provide a summary and discussion of the main results obtained in this article. First, in Sec.~\ref{sec:Summary}, we review the main ideas of this work and the procedures applied to obtain our results. Thereafter, in Sec.~\ref{sec:Discussion}, we discuss in more detail the two main results and how these can serve as an important step towards a better understanding of the microscopic nature of cosmic perturbations and towards connecting GFT-perturbations with cosmological observations. Moreover, we provide an overview of pursuing research directions that we consider fruitful.

\subsection{Summary}\label{sec:Summary}

Throughout this article, our explorations have been guided by two overarching principles, both of which are facilitated by the extended causal structure of the complete BC model~\cite{Jercher:2022mky}. {The first principle is that the causal properties of frame fields should be faithfully transferred to the quantum theory.} {The second principle is that inhomogeneities of cosmological observables emerge from quantum entanglement between GFT quanta}.

Following the first principle, in Sec.~\ref{sec:fockspacestructure} we introduced the complete BC GFT model for Lorentzian quantum gravity, including both spacelike and timelike tetrahedra. This is a minimal framework in which the first principle above can be consistently implemented. The inclusion of timelike tetrahedra necessitates an extension of the Fock space structure, which we realized by means of a tensor product between the spacelike and timelike sector. We extended the algebra of creation and annihilation operators to the tensor product Fock space. Still at the level of kinematics, we introduced in Sec.~\ref{sec:Coupling reference and matter fields} five minimally coupled massless free scalar fields to the GFT model, serving as a relational frame and matter content, respectively. Here, we made the first guiding principle manifest by restricting the kinetic kernels of the GFT in such a way that the clock and the rods only propagate along timelike, respectively spacelike dual edges. 

The second principle was implemented instead in Sec.~\ref{sec:Coherent peaked states and perturbations}, where we introduced the concept of perturbed coherent peaked states. These extend the concept of spacelike CPSs~\cite{Marchetti:2020qsq,Marchetti:2020umh} in two ways. At the level of the background, we complemented the spacelike CPS by introducing a condensate also on the timelike sector. The specific form of the condensate wavefunctions and their peaking properties have been guided by geometric and relational isotropy (see Assumption~\ref{ass:ks2}) as well as to simplify the ensuing computations of perturbed observables. The second important difference between these states and the spacelike CPSs is that in the former, perturbations are associated with the action of three types of two-body operators, which encode correlations between and within spacelike and timelike sectors. This is how the second principle above is implemented in our framework. Thereupon, in Sec.~\ref{sec:Effective relational dynamics}, we derived dynamical equations for the perturbed condensate in a relational manner as the expectation value of the full quantum dynamics with respect to the perturbed CPS. 

In Sec.~\ref{sec:Perturbation equations from quantum gravity}, we derived the background and perturbed dynamics of observables by computing expectation values of GFT operators with respect to the perturbed CPS. To compare such emergent quantities to observables of GR, we introduced matching conditions that relate macroscopic GFT observables with 
cosmological quantities. At the background level, we reproduced the results of~\cite{Marchetti:2020umh,Jercher:2021bie}, showing that for late relational times, the dynamics of the spatial volume $\bar{V}$ and the scalar field $\bar{\mf}$ in GFT and GR agree. We computed the dynamics of the perturbed spatial volume $\delta V$ and the perturbed matter field $\delta\mf$ in Secs.~\ref{sec:Geometric observables} and~\ref{sec:Dynamics of matter observables}, respectively. Their effective equations show a significant improvement compared to~\cite{Marchetti:2020umh} as the harmonic spatial derivative term of GR, entering with a factor of $a^4$, is now faithfully reproduced. {Also, we} combined geometric and matter perturbations into a curvature-like quantity $\tilde{\mathcal{R}}$, the dynamics of which again show an improvement with respect to previous work. Despite the similarities between the effective dynamics of GFT and GR perturbations, deviations in the form of source terms are present in the differential equations for $\delta V,\delta \mf$ and $\tilde{\mathcal{R}}$. To study these differences further we provided a comparison between solutions of perturbations in Sec.~\ref{sec:Solutions of GFT and GR perturbations}. Remarkably, the curvature-like observable $\tilde{\mathcal{R}}_\GFT$ shows very good agreement with $\tilde{\mathcal{R}}_\GR$ 
for sub-Planckian modes if initial conditions are chosen appropriately. Quantum gravity corrections become important instead for trans-Planckian modes~\cite{Jercher:2023kfr}, which we have shown to be naturally associated with sub-horizon scales in the relational framework we employ.

\subsection{Discussion and outlook}\label{sec:Discussion}
In this section we discuss the two main results of this work. First, that from the perspective of the underlying quantum gravity theory, cosmological perturbations can be seen as arising from quantum correlations. Second, that the macroscopic effective dynamics of these perturbations show some deviations from GR at scales that we interpret as trans-Planckian in our physical scalar frame. 

\subsubsection{Cosmological perturbations and quantum entanglement} The first of the two results mentioned above provides crucial insights into the intrinsically quantum nature of cosmological perturbations. Moreover, it concretely substantiates the intuition, shared by many approaches to QG, that non-trivial geometries are associated with entanglement between the fundamental {geometric} degrees of freedom, {see for instance \cite{Giddings:2005id,Ryu:2006bv,VanRaamsdonk:2010pw,Bianchi:2012ev,Maldacena:2013xja,Giddings:2015lla,Cao:2016mst,Donnelly:2016auv,swingle2018spacetime,colafrancheschi2022emergent,Bianchi:2023avf}.}

In this work, entanglement is encoded in relational two-body nearest neighbor correlations between GFT atoms, and it is lifted to the macroscopic level (and thus associated with cosmological quantities) by the properties of two-body coherent states. These states deviate substantially from one-body condensate states and their differences can be seen perturbatively as entangled \qmarks{out-of-condensate} components. We leave it to future research to investigate the associated entanglement entropy.

{We emphasize however that this is the only quantum effect that has been incorporated since the microscopic dynamics are still obtained within a mean-field analysis.} As shown in Appendix~\ref{sec:Going beyond mean-field in the absence of interactions}, this is a very robust approximation as long as GFT interactions are negligible. {Interactions will eventually become important at very late times~\cite{Oriti:2016qtz,Marchetti:2020umh}, and thus their study will be crucial for determining the self-consistency of the framework.}
When interactions are present, the inclusion of out-of-condensate perturbations will require a systematic analysis of the quantum properties of the GFT field. This scenario (in the simpler deparametrized setting of \cite{Wilson-Ewing:2018mrp}) is studied in \cite{Garcia:2024abc} by perturbatively splitting the GFT field into a classical homogeneous part and an inhomogeneous quantum field.

Finally, we leave it as an interesting avenue of future research to extend our analysis to {a causally complete GFT formulation of the EPRL model~\cite{Rovelli:2011eq,Perez:2012wv}}. Since recent results suggest that the BC and EPRL models could lie in the same universality class~\cite{Asante:2020qpa,Dittrich:2021kzs,Asante:2021zzh,Jercher:2021bie,Dittrich:2022yoo,Marchetti:2022igl,Marchetti:2022nrf}, we do not expect substantial differences between the perturbation theory of these two models. {However, once our framework matured sufficiently to model tensor perturbation modes, it will be interesting to investigate if these are anomalously polarized in the EPRL model, as can be expected from the parity-oddness of the Holst term~\cite{Ashtekar:1988sw,Freidel:2005sn,Contaldi:2008yz,Neiman:2011gf,Bethke:2011ru,Bethke:2011kx,Rovelli:2012yy}.}

\subsubsection{Emergent effective dynamics of cosmological perturbations}

{Including timelike tetrahedra turned out crucial to obtain effective dynamics of cosmological perturbations similar to GR.} {However, several parameters associated with the timelike sector of our states were not involved at all in the GR matching and are thus completely unconstrained.} This \qmarks{passive} behavior of timelike quanta, exemplified by the fact that the GR matching forces their perturbed number to be trivial (see Sec.~\ref{sec:Number of tetrahedra}), admits two possible interpretations. {First, it could reflect mathematically that our Universe can be fully described at the background level by evolving spacelike quantities only.} On the other hand, it could be due to the fact that only spacelike quantities have been considered in this work. {In this case,} one would expect that extending the analysis to other types of geometric observables (especially the extrinsic curvature) would constrain the free parameters of the timelike sector of our states. 

However, the presence of such \qmarks{irrelevant} free parameters in the timelike sector is a blessing in disguise. Indeed, it implies that the emergent perturbation equations are determined by only a set of parameters which turn out to be fixed by the requirement of consistency with the background dynamics. In other words, the predictions made by these models under a mean-field approximation are in principle easily falsifiable by comparison with cosmological observations. There are three different levels (in increasing order of conceptual and computational complexity) at which one could try to make contact with observations.

\paragraph{Phenomenology.} First, one could try to phenomenologically incorporate the modified dynamics of trans-Planckian modes arising from GFT into the Standard Cosmological Model. Since, for an inflationary phase slightly longer than the minimum period, all modes observed today were originally trans-Planckian \cite{Brandenberger:2012aj}, this could already produce non-trivial observable effects. Note that this would not technically violate the \emph{Trans-Planckian Censorship Conjecture} (TCC) \cite{Bedroya:2019snp,Bedroya:2019tba,Brandenberger:2021pzy}, since in principle the quantum gravity effects captured by these phenomenological models need not be described in terms of an effective field theory.

\paragraph{Extraction of full cosmological dynamics.} A more systematic approach would be to \emph{derive} the emergent cosmological perturbation dynamics from the full QG theory. However, this requires: (i) the construction of additional operators to extract the full effective (anisotropic) geometry, (ii) the inclusion of the appropriate matter content, and (iii) the generalization of the analysis performed here to early times. Let us discuss the above three points in some more detail.

\begin{enumerate}[label=(\roman*)]
    \item The construction of additional geometric operators is extremely important for several reasons. First, it would allow us to extend our analysis to {other cosmological observables }such as the comoving curvature $\mathcal{R}$, requiring the introduction of anisotropic observables (see Sec.~\ref{sec:mukhanovsasaki}). {Such observables} would also be important to reconstruct the full effective metric, including not only scalar perturbations, but also vector and, most importantly tensor perturbations. {The construction of anisotropic operators would require to represent only relational observables in the GFT Fock space,} {which is, however, a highly non-trivial task~\cite{Marchetti:2020umh}.} {Finally, it is conceivable that modelling tensor perturbations requires the inclusion of lightlike tetrahedra as these are expected to propagate along lightlike dual edges.} {Studies in GFT cosmology mostly focused on minimally coupled massless free scalar fields~\cite{Oriti:2016qtz,Li:2017uao} although recently, the analysis has been extended to include a non-trivial potential~ \cite{Ladstatter:2023abc}.} To move towards a more realistic matter content, one would have to include cosmic fluids. Considerable effort has been devoted to the study of dust in classical and quantum gravity (see e.g.\ \cite{Giesel:2012rb,Domagala:2010bm} and \cite{Giesel:2007wk,Giesel:2020bht,Giesel:2020xnb,Husain:2020uac} for cosmological applications), since it not only constitutes a key component of the Universe, but also serves as a natural physical reference frame. In fact, it is the reference frame in which the Cosmological Principle is formulated, and thus in which background and perturbations are defined. This is particularly important, since the mixing between sub-horizon and trans-Planckian modes emphasized in Sec.~\ref{sec:Solutions of GFT and GR perturbations} {hinges on the }definition of perturbations with respect to our physical scalar frame. Coupling dust to GFT models would therefore allow us to {set up} more realistic cosmological models, in which one could explore delicate issues such as the sub-horizon/trans-Planckian mixing. Importantly, it would also allow us to test the physical covariance (or violation thereof) of the emergent cosmological dynamics. As a final comment on this point, we report that matter components may also emerge due to the underlying GFT dynamics, see \cite{Oriti:2021rvm} for an example.
    \item Generalizing the current analysis to earlier times would be important to understand the imprint of the quantum gravity bounce on the perturbations. Moreover, for the perturbation theory developed here to be self-consistent also in this regime it will be important to check if the energy density of the perturbations remains bounded and small compared to the background quantum geometry so that back reaction effects can be ignored. This concerns in particular perturbations of trans-Planckian wavelength and was dubbed the \enquote{real trans-Planckian issue} in another setting~\cite{Agullo:2012sh,Agullo:2023rqq}. In general, however, the mean-field equations become considerably more complicated as the density of the background condensate decreases \cite{Marchetti:2021gcv}. In particular, recent results \cite{Fischer:2023abc} seem to suggest that, in the super-horizon limit, the early times emergent dynamics cannot be reconciled with that of any modified gravity theory.
\end{enumerate}

\paragraph{Initial conditions from the fundamental theory.} Finally, in both the methods discussed so far, the primordial power spectrum would be obtained by a Fock quantization of the macroscopic degrees of freedom. This is a strategy that is followed, for example, in both condensed matter physics \cite{altland_simons_2010} and loop quantum cosmology~\cite{Fernandez-Mendez:2012poe,Fernandez-Mendez:2013jqa,gomar,CastelloGomar:2016rjj,CastelloGomar:2017kbo,Agullo:2012sh,Agullo:2012fc,Agullo:2013ai,Agullo:2016hap,Ashtekar:2020gec} (see \cite{Wilson-Ewing:2016yan,Agullo:2023rqq} for a review). From the perspective of QG, however, it would be natural to look for a mechanism for generating the cosmological initial conditions that can be derived entirely from the fundamental theory. If the statistical properties of the cosmological inhomogeneities and anisotropies are ultimately due to quantum gravity fluctuations, their description may require going beyond the effective relational framework (which is based on observable averages) used here, and considering relational observables rigorously defined on the GFT Fock space. Similarly, one could explore the possibility that the inflationary mechanism is purely quantum geometric in nature, although this possibility has already been ruled out in some phenomenological GFT models \cite{deCesare:2016rsf,DeSousa:2023tja}.

\

{In closing, we hope that our results on the coupling of a physical Lorentzian reference frame and on the exploration of the link between quantum geometric entanglement and cosmological perturbations will impact on a wider set of quantum gravity approaches. In particular, we believe that our work strengthens the argument for models that incorporate a causally complete set of discrete Lorentzian geometries such as causal dynamical triangulations~\cite{Ambjorn:2012jv,Loll:2019rdj,Loll:1998aj}, causal sets~\cite{Surya:2019ndm}, Lorentzian Regge calculus~\cite{Jercher:2023csk,Asante:2021phx,Asante:2021zzh,Dittrich:2021gww,Dittrich:2023rcr} or Lorentzian spin foams and LQG~\cite{Conrady:2010kc,Conrady:2010vx,Simao:2021qno,Han:2021bln,Han:2021kll}.}

\section*{Acknowledgements}

{The authors are grateful to an anonymous referee for insightful remarks which led to an improvement of the manuscript.} The authors also thank Edward Wilson-Ewing, Steffen Gielen, Daniele Oriti, Kristina Giesel, Renata Ferrero and Viqar Husain for helpful discussions and comments.

AFJ gratefully acknowledges support by the Deutsche Forschungsgemeinschaft (DFG, German Research Foundation) project number 422809950 and by grant number 406116891 within the Research Training Group RTG 2522/1. AGAP acknowledges funding from the DFG research grants OR432/3-1 and OR432/4-1 and the John-Templeton Foundation via research grant 62421. AFJ and AGAP are grateful for the generous financial support by the MCQST via the seed funding Aost 862933-9 granted by the DFG under Germany’s Excellence Strategy – EXC-2111 – 390814868. AGAP gratefully acknowledges support by the Center for Advanced Studies (CAS) of the LMU through its \emph{Junior Researcher in Residence program}. LM acknowledges support from AARMS and the Natural Sciences and Engineering Research Council of Canada, and thanks the Arnold Sommerfeld Center (LMU) for hospitality.

\appendix

\section{List of assumptions}\label{sec:assumptions}

In this appendix, we present an exhaustive list of the assumptions made in order to arrive at the results we obtained in the main body of this work. Naturally, the assumptions can be split into kinematic and dynamic ones. Furthermore, we split the assumptions into two categories, namely structural ones and those stemming from matching with GR perturbations. The former category contains assumptions that are motivated either conceptually or that simplify technically challenging computations. 

\subsection{Kinematic assumptions}

Kinematic approximations are related to the properties of the specific states we are considering.

\paragraph*{Structural.}

\begin{description}

\item[KS1\label{ass:ks1}] $\hspace{-2pt}$($\hspace{-1pt}$\textbf{\itshape{Perturbed condensate states}}): 
In this paper, we model a spatially homogeneous and isotropic spacetime with scalar perturbations by perturbed condensate states, which extend the notion of usual condensate states~\cite{Gielen:2013naa,Gielen:2014ila,Gielen:2014uga,Oriti:2015qva,Gielen:2016dss,Oriti:2016qtz,Pithis:2016cxg,Pithis:2019tvp} in two ways. First, given the causally extended structure including timelike tetrahedra, the background contains an additional timelike condensate, introduced in Sec.~\ref{sec:Timelike CPS}. Second, perturbations are encoded in $2$-body operators that create an entanglement within and between the spacelike and timelike sector. As discussed in Secs.~\ref{sec:Discussion and Conclusion} and Appendix~\ref{sec:Going beyond mean-field in the absence of interactions}, this is a first step in the direction of out-of-condensate perturbations.  

\item[KS2\label{ass:ks2}] $\hspace{-2pt}$($\hspace{-1pt}$\textbf{\itshape{Geometrical and relational isotropy}}): The spacelike and timelike condensate wavefunctions as well as the correlation functions $\delta\Phi,\delta\Psi$ and $\delta\Xi$ are required to satisfy a quantum analogue of isotropy, realized by setting all the area eigenvalues of the faces $f$ of a tetrahedron to the same value, $\rho_f =\rho$. Moreover, the spacelike peaking encoded in the condensate wavefunction $\tau$ is isotropic by choosing the same peaking parameters for all of the three spatial directions. This also ensures that in a derivative expansion for the effective relational equations, only the Laplace operator with respect to the rod variables enters.

\item[KS3\label{ass:ks3}] $\hspace{-2pt}$($\hspace{-1pt}$\textbf{\itshape{Peaking and frame-dependence}}): Following the strategy of~\cite{Marchetti:2020umh,Marchetti:2020qsq,Marchetti:2021gcv}, the condensate wavefunction factorizes into a peaking term and a reduced wavefunction. On the spacelike sector, the peaking function only contains the clock variable, while the timelike condensate is peaked on clock and rods. This ensures that the spatial derivative, acting on the volume perturbations, is only associated to spacelike dual edges. The peaking functions are Gaussians with a non-trivial phase and a small width. As argued in~\cite{Marchetti:2020umh,Marchetti:2020qsq}, a non-vanishing $\epsilon$ as well as a non-vanishing phase are required in order to guarantee that all quantum fluctuations of observables associated to the reference fields are small in the classical regime. Both of the reduced condensate wavefunctions, $\slrcw(\rf^0)$ and $\tlrcw(\rf^0)$, only depend on relational time as they are assumed to be part of the background. For the $2$-body correlations $\delta\Phi,\delta\Psi$ and $\delta\Xi$, no peaking is encoded and an explicit rod-dependence is assumed in order to render these functions part of the perturbations. {Notice that in spite of assuming a rod-dependence of $\delta\Xi$, the dynamical equations, together with Eq.~\eqref{eq:relation of dPsi and dPhi}, render this function only time-dependent as we argue in Sec.~\ref{sec:Perturbed equations of motion}.} 

\item[KS4\label{ass:ks4}] $\hspace{-2pt}$($\hspace{-1pt}$\textbf{\itshape{Signature of tetrahedra and faces}}): We extend the spacelike Barrett-Crane model by timelike tetrahedra and leave the inclusion of lightlike tetrahedra for future research. For the timelike condensate, entering the perturbed CPS in Eq.~\eqref{eqn:perturbedstates}, we assume that the corresponding timelike tetrahedra contain spacelike faces only. This assumption is supported by two reasons: $(i)$ Recent studies in~\cite{Jercher:2024abc} suggest that only spacelike faces contribute to a condensate phase and $(ii)$ timelike and spacelike tetrahedra can interact only via spacelike faces, as shown in~\cite{Jercher:2022mky}. 

\item[KS5\label{ass:ks5}] $\hspace{-2pt}$($\hspace{-1pt}$\textbf{\itshape{Regularization}}): At several points of the analysis, infinities appear that need to be treated appropriately. Extended closure in Eq.~\eqref{eq:closure} yields empty $\SL$-integrations in the definition of the GFT action, $2$-body operators and the equations of motion. On the other hand, due to spatial homogeneity and the fact that $\sigma$ is peaked only on the clock, the spacelike background equations in Sec.~\ref{sec:Background equations of motion} as well as the background expectation value of the $3$-volume operator in Sec.~\ref{sec:Geometric observables} contain empty rod-integrations. Both types of divergencies are considered as unphysical, and we consider only the finite factors of these expressions.

\end{description}

\paragraph{Motivated by classical matching.}

\begin{description}

\item[KC1\label{ass:kc1}] $\hspace{-2pt}$($\hspace{-1pt}$\textbf{\itshape{Peaking on matter momenta}}): We assume the condensate wavefunctions $\sigma$ and $\tau$ to be peaked on the matter momentum $\mm$, realized by a Gaussian function without phase. This is necessary in order to recover the Friedmann equations at the background level as well as to render feasible perturbation equations for the condensate and the observables in Secs.~\ref{sec:Effective relational dynamics} and~\ref{sec:Perturbation equations from quantum gravity}, respectively. As pointed out in~\cite{Marchetti:2021gcv}, a deeper physical intuition  for this peaking may be obtained when considering a scalar field with non-trivial potential.

\item[KC2\label{ass:kc2}] $\hspace{-2pt}$($\hspace{-1pt}$\textbf{\itshape{Local perturbation functions}}): A priori, the $2$-body correlations $\delta\Phi,\delta\Psi$ and $\delta\Xi$ are bi-local functions of the relational reference frame and the matter momentum $\mm$. To make contact to localized perturbations, we assume that the two arguments are identified via a $\delta$-distributions. Following the simplicial gravity picture, this assumption would correspond to correlations within the same $4$-simplex with momentum conservation across tetrahedra.

\item[KC3\label{ass:kc3}] $\hspace{-2pt}$($\hspace{-1pt}$\textbf{\itshape{Peaking parameters}}): To render the computations of Sec.~\ref{sec:Perturbation equations from quantum gravity} feasible and to allow for a more straightforward matching of GFT and GR perturbations, we assume that the peaking parameters $\epsilon^\pm$ and $\pi_0^\pm$ of the spacelike and timelike sector are related as $\epsp = \epsm$ and $\pi_0^+ = -\pi_0^-$. Furthermore, we assume that the phase parameters $\pi_x$ and $\pi_0^+$ satisfy the strong inequality $\pi_x\gg\pi_0^+$, which yields a drastic simplification for the dynamical equation of the perturbed volume in Eq.~\eqref{eq:dV with A}.

\end{description}

\subsection{Dynamic assumptions}

Dynamic approximations are related to the details of the GFT action and on how background and perturbed equations for the condensate and the observables are obtained.

\paragraph{Structural.}

\begin{description}

\item[DS1\label{ass:ds1}] $\hspace{-2pt}$($\hspace{-1pt}$\textbf{\itshape{GFT action and causal building blocks}}): We choose to work with an extension of the Lorentzian Barrett-Crane model~\cite{Jercher:2021bie,Jercher:2022mky} and here study its cosmological implications while including spacelike and timelike tetrahedra and excluding lightlike ones. Thus, the configurations are the two group fields $\varphi_+$ and $\varphi_-$ which are distinguished in their domain and in the form of the simplicity constraint, given in Eq.~\eqref{eq:simplicity}.

\item[DS2\label{ass:ds2}] $\hspace{-2pt}$($\hspace{-1pt}$\textbf{\itshape{Scalar field coupling}}): Clock and rod fields $\rf^\mu$ as well as the matter field $\mf$ are coupled to the GFT such that the Feynman amplitudes correspond to a simplicial gravity path integral with the fields propagating along dual edges. As discussed in~\cite{Li:2017uao}, this coupling is obtained in a semi-classical limit, which is an assumption that should be kept mind. In order to turn the scalar reference frame into a physical Lorentzian reference frame, we align its causal character with that of the geometry by imposing restrictions on the kernels in Eqs.~\eqref{eqn:restrictionkinetic}. As elaborated in Sec.~\ref{sec:Coupling reference and matter fields}, the physical picture of these conditions is that the clock propagates along timelike dual edges and the rods propagate along spacelike dual edges.

\item[DS3\label{ass:ds3}] $\hspace{-2pt}$($\hspace{-1pt}$\textbf{\itshape{Mean-field dynamics}}): We assume the effective dynamics to be well approximated by the mean-field equations. In fact, as we show in Appendix~\ref{sec:Going beyond mean-field in the absence of interactions} and discuss in the conclusion, the mean-field equations solve the higher-order Schwinger-Dyson equations if negligible interactions and linear perturbation theory is assumed.

\item[DS4\label{ass:ds4}] $\hspace{-2pt}$($\hspace{-1pt}$\textbf{\itshape{Negligible interactions}}): When taking expectation values with respect to the perturbed CPS $\ket{\Delta;x^0,\vb*{x}}$,  interaction terms are expected to be negligible in the effective dynamics. It has been shown in~\cite{Oriti:2016qtz}, that the error induced by this assumption grows with the number of GFT particles. This assumption can therefore be consistently implemented at late but not very late times. See also~\cite{Gielen:2013naa} for a complementary discussion on this matter.

\item[DS5\label{ass:ds5}] $\hspace{-2pt}$($\hspace{-1pt}$\textbf{\itshape{Classical matter content}}): We assume that the matter content of the classical theory, given by the five fields $\rf^\mu$ and $\mf$, is dominated by $\mf$. In this way, perturbations of $\rf^\mu$ are considered to be negligible and inhomogeneities of matter and geometry can be defined unambiguously with respect to the clock and rod fields. In Appendix~\ref{sec:Classical perturbation theory - matter}, we compute the corrections to the classical Einstein equations if one would not neglect the frame contributions to the energy-momentum tensor. 

\end{description}

\paragraph{Motivated by classical matching.}

\begin{description}

\item[DC1\label{ass:dc1}] $\hspace{-2pt}$($\hspace{-1pt}$\textbf{\itshape{Mesoscopic regime}}): Classical dynamics are obtained in a mesoscopic regime, where the averaged number of particles of the system is taken to be large enough to allow for both a continuum interpretation  of the expectation values of relevant operators and classical behavior, but not too large that interactions are dominating. Furthermore, the perturbation equations of Sec.~\ref{sec:Effective relational dynamics} and~\ref{sec:Perturbation equations from quantum gravity} simplify significantly in this regime, as the condensate wavefunctions are solved by simple exponential functions, see Eqs.~\eqref{eqn:bkgsigma} and~\eqref{eqn:bkgtau}.

\item[DC2\label{ass:dc2}] $\hspace{-2pt}$($\hspace{-1pt}$\textbf{\itshape{Single-spin dominance}}): We assume that coefficients of $\sigma,\tau$ and the $2$-body correlations $\delta\Phi,\delta\Psi$ and $\delta\Xi$ in spin representation are dominated by a single representation label $\rho$, mostly suppressed in the notation. For the background condensates, this assumption is supported by the studies of Ref.~\cite{Gielen:2016uft,Pithis:2016wzf,Pithis:2016cxg,Jercher:2021bie}, which show that under rather generic assumptions on the kinetic kernels of the GFT, a dominant representation emerges dynamically.

\item[DC3\label{ass:dc3}] $\hspace{-2pt}$($\hspace{-1pt}$\textbf{\itshape{Dynamical freedom}}): Exploiting the dynamical freedom of having two first-order equations for three functions, we relate the spacelike-spacelike correlation $\delta\Phi$ and the spacelike-timelike correlation $\delta\Psi$ via a function $\mathrm{f}$. This function is chosen such that the dynamical equation for $\delta\Psi$ as well as the definition of $\delta V$ simplify drastically. Finally, when matching the $a^4$-term in front of the spatial derivative of $\delta V$ in Eq.~\eqref{eq:dV with A}, the function $\mathrm{f}$ is fixed completely. As we detail in Sec.~\ref{sec:Discussion}, it is left open for future studies to include interactions (see for instance Ref.~\cite{Garcia:2024abc} for further details) such that the number of independent first-order equations matches the number of perturbations functions and no such dynamical freedom is present.  

\item[DC4\label{ass:dc4}] $\hspace{-2pt}$($\hspace{-1pt}$\textbf{\itshape{Constant phase}}): Splitting the perturbation function $\delta\Psi$ into modulus $R$ and phase $\Theta$, we assume the phase to be constant, rendering the dynamical equations of $\delta\Psi$ and $\delta V$ feasible. Moreover, $\Theta$ is set to $n\frac{\pi}{2}$ in Sec.~\ref{sec:Geometric observables}, simplifying the dynamical equation for $\delta V$ in Eq.~\eqref{eq:dV with A}. One can show that if one relaxes the constant-$\Theta$ condition to a pure time dependence, then consistency requires that $\Theta$ is in fact constant. Clearly, the most general case is given for $\Theta$ carrying a space dependence. However, the resulting equations of motion for the perturbation function as well as for observables take a highly intricate form, which does not allow for further analytical studies.

\item[DC5\label{ass:dc5}] $\hspace{-2pt}$($\hspace{-1pt}$\textbf{\itshape{Normalized scale factor}}): As commonly done in standard cosmology, the scale factor $a$ is normalized to $1$ at present relational time, such that it takes smaller values for all earlier relational times. Notice, that this assumption is not in contradiction with a large GFT-particle number or, equivalently, a large volume. That is because the volume is given by $a^3$ times a fiducial volume factor, which can be much larger than one.

\item[DC6\label{ass:dc6}] $\hspace{-2pt}$($\hspace{-1pt}$\textbf{\itshape{Timelike particle number}}): When computing the dynamics of matter observables, we assumed that the effective mass parameters of the spacelike and timelike sector satisfy $\mu_+ > \mu_-$. This assumption is guided by the intuition that the background is predominantly described by the spacelike condensate. Indeed, as a consequence, one finds that the background number of spacelike GFT-particles dominates over the timelike particle number in the classical limit. In Sec.~\ref{sec:Dynamics of matter observables}, this leads to a simplification of the emergent equations for the perturbed matter field $\delta\mf$ and the perturbed matter momentum $\delta\mm$.

\end{description}

\section{Going beyond mean-field in the absence of interactions}\label{sec:Going beyond mean-field in the absence of interactions}

In this appendix, we show that in the absence of interactions, higher order Schwinger-Dyson equations reduce to powers of the lowest mean-field equation. This result is of interest for the perturbed coherent states introduced in Sec.~\ref{sec:Perturbed coherent peaked states}, as these states constitute out-of-condensate perturbations only if interactions are taken into account. Furthermore, this result is exactly the reason for the dynamical freedom of one of the $2$-body correlation functions which we exploited in Sec.~\ref{sec:Perturbation equations from quantum gravity} to match the perturbations of GR.

For a polynomially bounded functional $\mathcal{O}[\varphi_\alpha,\bar{\varphi}_\alpha]$ on field space, where $\alpha$ refers to either the spacelike or the timelike group field, the Schwinger-Dyson equations are given by
\begin{equation}
\expval{\fdv{\mathcal{O}}{\bar{\varphi}_\alpha}-\mathcal{O}\fdv{S}{\bar{\varphi}_\alpha}} = 0.
\end{equation}
The polynomial expansion of the operator $\mathcal{O}$ is given by
\begin{equation}
\mathcal{O}[\varphi_\alpha,\bar{\varphi}_\alpha] = \sum_{k,l,m,n}\mathcal{O}_{klmn}\;\bar{\varphi}_+^k\varphi_+^l\otimes\bar{\varphi}_-^m\varphi_-^n,
\end{equation}
with $\mathcal{O}_{klmn}$ being the kernel. Given the perturbed state $\ket{\Delta}$, defined in Sec.~\ref{sec:Perturbed coherent peaked states}, the expectation value of $\mathcal{O}$ up to first order in perturbations is schematically given by
\begin{equation}
\begin{aligned}
&\expval{\mathcal{O}}{\Delta} = \eval{\mathcal{O}}_{\sigma,\tau}+\left(2\eval{\bar{\varphi}_+\fdv{\mathcal{O}}{\varphi_+}}_{\sigma,\tau}+\eval{\fdv[2]{\mathcal{O}}{\varphi_+}}_{\sigma,\tau}\right)\delta\Phi\;+\\[7pt]
+& \left(\eval{\bar{\varphi}_-\fdv{\mathcal{O}}{\varphi_+}}_{\sigma,\tau}+\eval{\bar{\varphi}_+\fdv{\mathcal{O}}{\varphi_-}}_{\sigma,\tau}\right)\delta\Psi+\left(2\eval{\bar{\varphi}_-\fdv{\mathcal{O}}{\varphi_+}}_{\sigma,\tau}+\eval{\fdv[2]{\mathcal{O}}{\varphi_-}}_{\sigma,\tau}\right)\delta\Xi + \mathrm{c.c.},
\end{aligned}
\end{equation}
where a vertical line indicates that the operator $\mathcal{O}$ is functionally evaluated on the condensate wavefunctions $\sigma$ and $\tau$. 

Building up on this result, we consider now the Schwinger-Dyson equation under the assumption that interactions are absent, i.e. $S = \sum_\alpha \bar{\varphi}_\alpha\mathcal{K}_\alpha\varphi_\alpha$. At zeroth order, the equations then take the form
\begin{equation}
\eval{\fdv{\mathcal{O}}{\bar{\varphi}_\alpha}}_{\sigma,\tau} = \eval{\mathcal{O}\mathcal{K}_\alpha\varphi_\alpha}_{\sigma,\tau},
\end{equation}
which, for $\mathcal{O}=\one$, reproduces the mean-field equation studied in Sec.~\ref{sec:Background equations of motion}. 

At first order, we look at the coefficients in front of the same perturbation function. For instance, the $\delta\Psi$-term is given by
\begin{equation}
\begin{aligned}
\eval{\bar{\varphi}_-\frac{\delta^2\mathcal{O}}{\delta\varphi_+\delta\bar{\varphi}_\alpha}}_{\sigma,\tau}+\eval{\bar{\varphi}_+\frac{\delta^2\mathcal{O}}{\delta\varphi_-\delta\bar{\varphi}_\alpha}}_{\sigma,\tau} = \eval{\bar{\varphi}_-\frac{\delta(\mathcal{O}\mathcal{K}_\alpha\varphi_\alpha)}{\delta\varphi_+}}_{\sigma,\tau}+\eval{\bar{\varphi}_+\frac{\delta(\mathcal{O}\mathcal{K}_\alpha\varphi_\alpha)}{\delta\varphi_-}}_{\sigma,\tau},
\end{aligned}
\end{equation}
with the other coefficients given in a similar form. Clearly, upon solutions of the zeroth-order equation, also the first-order equations are satisfied. Hence, higher-order Schwinger-Dyson equations do not yield additional dynamical equations if one works $(i)$ in a perturbative setting and $(ii)$ in the absence of interactions. {This can be also shown by computing $n$-point Green functions which, in perturbation theory and in the absence of interactions, factorize into one-point functions corresponding to the mean-field. This is demonstrated for a single-sector GFT in~\cite{Gielen:2013naa}.} As a consequence, we find in the analysis of the $2$-body correlation functions in Sec.~\ref{sec:Effective relational dynamics} a dynamical freedom for one of the variables. The results of this appendix suggest that in order to obtain higher-order out-of-condensate equations, one would need to take interactions into account. We comment on this matter further in Sec.~\ref{sec:Discussion and Conclusion}.

\section{Classical perturbation theory}\label{sec:Classical perturbation theory}

In this appendix, we provide an overview of the perturbation equations for geometry and matter in classical general relativity. In order to allow for a simpler comparison with relational GFT results, we mostly use harmonic coordinates $\{x^\mu\}$ which are adapted to the reference field $\{\rf^\mu\}$ via the relation $\rf^\mu = \kappa^\mu x^\mu$ (no summation over $\mu$), where $\kappa^\mu$ are some dimensionful proportionality factors \cite{Gielen:2018fqv}. In harmonic coordinates, the reference fields are assumed to satisfy the Klein-Gordon equation at all orders of perturbations. This can equivalently be re-written as 
\begin{equation}\label{eq:harmonic gauge}
\Gamma^{\lambda}_{\mu\nu}g^{\mu\nu} = 0,
\end{equation}
which poses a condition on the metric.

\subsection{Geometry}\label{sec:Classical perturbation theory - geometry}

At zeroth order, the line element of a spatially flat FLRW spacetime {with signature of $(-,+,+,+)$} is given by
\begin{equation}
\dd{s}^2 = \bar{g}_{\mu\nu}\dd{x}^\mu\dd{x}^\nu = -N^2\dd{t}^2+a^2\dd{\vb*{x}}^2,
\end{equation}
where $N$ is the lapse function, $a$ is the scale factor and $\dd{\vb*{x}}^2$ the line element of $3$-dimensional Euclidean flat space. Imposing harmonic gauge on the background yields $a^3/N = c_H$, where $c_H$ is an integration constant. For the remainder, we set $c_H = 1$, and we assume that the matter content is dominated by the matter field $\mf$ with conjugate momentum $\mm$. 

Within these assumptions, the dynamics of the geometry at background level are captured by
\begin{equation}\label{eq:classical 3H^2 eq}
3\Hubble^2 = \frac{1}{2 M_{\Pl}^2}\bar{\pi}_\mf^2,\qquad \Hubble' = 0,
\end{equation}
where $\Hubble = a'/a$ is the Hubble parameter {in harmonic coordinates} and $\bar{\pi}_\mf$ is the background contribution of the canonical conjugate of the scalar field, defined in Eq.~\eqref{eq:clasical mm def}. Introducing the background volume $\bar{V} = a^3$, the geometric equations can be recast to
\begin{equation}
3\left(\frac{\bar{V}'}{3\bar{V}}\right)^2 = \frac{1}{2 M_{\Pl}^2}\bar{\pi}_\mf^2,\qquad \left(\frac{\bar{V}'}{3\bar{V}}\right)' = 0.
\end{equation}

To derive perturbed volume equations from GR, we consider in the following first-order scalar perturbations of the FLRW metric. Using the background harmonic gauge condition, the line element is given by
\begin{equation}\label{eq:perturbed line element}
\dd{s}^2 = -a^6(1+2A)\dd{t}^2+a^4\partial_iB\dd{t}\dd{x^i}+a^2\left((1-2\psi)\delta_{ij}+2\partial_i\partial_j E\right)\dd{x}^i\dd{x}^j,
\end{equation}
with scalar perturbation functions $A,B,\psi$ and $E$. Einstein's equations at linear order yield~\cite{Marchetti:2021gcv,Battarra:2014tga}
\begin{align}
 \frac{1}{2 M_{\Pl}^2}\bar{\mf}'\delta\mf'+3\Hubble\psi'-a^4\nabla^2\psi-\Hubble\nabla^2\left(E'-a^2 B\right) &= 0,\label{eq:pEFE1}\\[7pt]
\Hubble A + \psi'-\frac{1}{2 M_{\Pl}^2}\bar{\mf}'\delta\mf &= 0,\label{eq:pEFE2}\\[7pt]
 E''-a^4\nabla^2 E &= 0\label{eq:pEFE3},
\end{align}
where $\delta\mf$ is the scalar field perturbation. Combining Eq.~\eqref{eq:pEFE1} and the time-derivative of Eq.~\eqref{eq:pEFE2}, we obtain
\begin{equation}\label{eq:psi''}
\psi'' = -\Hubble A'-3\Hubble\psi'+a^4\nabla^2\psi+\Hubble\nabla^2\left(E'-a^2 B\right).
\end{equation}

To obtain an equation for the perturbed volume, which is an observable accessible also from the GFT side, consider on a slice of constant time the  local volume element
\begin{equation}
\sqrt{-g_{(3)}} = \bar{V} + \delta V = a^3\left(1-3\psi+\nabla^2E\right).
\end{equation}
Thus, we identify the perturbed spatial volume as
\begin{equation}
\frac{\delta V}{\bar{V}} = -3\psi+\nabla^2 E.
\end{equation}
Taking the second derivative of $\delta V/\bar{V}$ and using Eqs.~\eqref{eq:pEFE3} and~\eqref{eq:psi''}, one obtains
\begin{equation}\label{eqn:classicalpertvolumegeneral}
\left(\frac{\delta V}{\bar{V}}\right)'' + 3\Hubble\left(\frac{\delta V}{\bar{V}}\right)'-a^4\nabla^2\left(\frac{\delta V}{\bar{V}}\right) = 3\Hubble\left(A'+a^2\nabla^2 B\right).
\end{equation}

\paragraph{Harmonic gauge.} At first order in perturbations, the harmonic gauge conditions given by~\cite{Battarra:2014tga}
\begin{subequations}\label{eq:pert HGC}
\begin{align}
0 &= A'+3\psi'-\nabla^2(E'-a^2B),\label{eq:pert HGC1}\\[7pt]
0 &= (a^2B)'+a^4(A-\psi-\nabla^2 E)\label{eq:pert HGC2},
\end{align}
\end{subequations}
which, imposed on Einstein's equations~\eqref{eq:pEFE1} - \eqref{eq:pEFE3}, yields~\cite{Battarra:2014tga,Marchetti:2021gcv}
\begin{align}
\psi''-a^4\nabla^2\psi &=0, &&A'' -a^4\nabla^2 A + 4a^4\nabla^2\psi =0\, \\[7pt]
E'' -a^4\nabla^2 E&=0, &&(a^2B)''-a^4\nabla^2(a^2B)-8a^2(a^2\psi)' =0.
\end{align}
Expressed in terms of the volume, the first harmonic gauge condition is expressed as
\begin{equation}
    A'+a^2\nabla^2 B = \left(\frac{\delta V}{\bar{V}}\right)',
\end{equation}
such that the volume equation becomes
\begin{equation}\label{eq:classical relative perturbed volume equation}
\left(\frac{\delta V}{\bar{V}}\right)''-a^4 \nabla^2\left(\frac{\delta V}{\bar{V}}\right) = 0.
\end{equation}
or equivalently
\begin{equation}\label{eq:classical perturbed volume equation}
\delta V'' - 6\Hubble\delta V'+9\Hubble^2\delta V-a^4 \nabla^2\delta V = 0.
\end{equation}
To change to Fourier space in the rod variable, heavily employed in Sec.~\ref{sec:Perturbation equations from quantum gravity}, one can simply perform the substitution $\nabla^2 \rightarrow -k^2$ here and in the following.

Following~\cite{Battarra:2014tga}, there is a residual gauge freedom in performing a coordinate transformation 
\begin{equation}
\xi^\mu \mapsto x^\mu+\xi^\mu,
\end{equation}
with $\xi^\mu = (\xi^0,\partial^i\xi)$  satisfying
\begin{equation}
(\xi^0)''-a^4\nabla^2\xi^0 = \xi''-a^4\nabla^2\xi = 0,
\end{equation}
such that harmonicity is conserved. Under this transformation, the perturbation functions transform as~\cite{Battarra:2014tga,Marchetti:2020umh}
\begin{align}
&\psi \mapsto \psi +\Hubble\xi^0, &&A \mapsto A -(\xi^0)' - 3\Hubble\xi^0,\label{eq:pert gauge trafo1}\\[7pt]
&E \mapsto E -\xi, &&B \mapsto B +a^2\xi^0-a^{-2}\xi'\label{eq:pert gauge trafo2}.
\end{align}
After introducing the matter equations in the following, we combine the geometric and matter quantities in a single fully gauge-invariant quantity, the so-called curvature perturbation $\mathcal{R}$. 

\subsection{Matter}\label{sec:Classical perturbation theory - matter}

The matter content of the classical theory consists of four reference scalar fields $\rf^\mu$ as well as one additional free minimally coupled real scalar field $\mf$, defined by the continuum action
\begin{equation}
S[\rf^\mu,\mf] = -\frac{1}{2 M_{\Pl}^2}\int\dd[4]{x}\sqrt{-g}g^{ab}\left(\partial_a\mf\partial_b\mf+\sum_{\mu=0}^3\partial_a\rf^\mu\partial_b\rf^\mu\right).
\end{equation}
{In this form, the action poses a well-defined variational principle, yielding the Klein-Gordon equations for appropriate boundary conditions. One of such admissible conditions are von Neumann boundary conditions which assume vanishing variation of the gradients at the boundary. For reference fields in harmonic coordinates, as used in the remainder of this subsection, $\rf^\mu = \kappa^\mu x^\mu$, this clearly applies since $\partial_\mu\rf^\nu = \delta_\mu^\nu\kappa^\nu$ is constant and thus has vanishing variation.}

The energy momentum tensor in arbitrary coordinates is given by 
\begin{equation}\label{eq:classical EM-tensor}
  M_{\Pl}^2\; T_{ab}=\sum_{\lambda=0}^3\left(\partial_a\chi^\lambda\partial_b\chi^\lambda-\frac{g_{ab}}{2}g^{mn}\partial_m\chi^\lambda\partial_n\chi^\lambda\right)+\partial_a\phi\partial_b\phi-\frac{g_{ab}}{2}g^{mn}\partial_m\phi\partial_n\phi,
\end{equation}
{which we assume to be dominated by the matter field $\mf$.} The full equations of motion for $\mf$ are given by the massless Klein-Gordon equation
\begin{equation}
\partial_\mu\left(\sqrt{-g}g^{\mu\nu}\partial_\nu\mf\right) = 0.
\end{equation}
Linearizing in both, the scalar field and the metric, we obtain the zeroth order equation
\begin{equation}
\bar{\phi}'' = 0.
\end{equation}
and the first-order perturbation equation
\begin{equation}
\delta\mf''-a^4\nabla^2\delta\mf = \left[A'+3\psi'-\nabla^2E'+a^2\nabla^2B\right]\bar{\mf}',
\end{equation}
respectively. Supplementing the latter with the harmonic gauge condition in Eq.~\eqref{eq:pert HGC1}, $\delta\mf$ satisfies
\begin{equation}\label{eq:classical mf perturbation equation}
\delta\mf''-a^4\nabla^2\delta\mf = 0.
\end{equation}

To define the GR counterpart of the GFT observables $\hat{\Pi}_\mf^\alpha$, defined in Eq.~\eqref{eq:mm operator}, we introduce the momentum conjugate to the scalar field, commonly defined as
\begin{equation}\label{eq:clasical mm def}
\mm^\mu\defeq  \pdv{\tilde{\mathcal{L}}}{(\partial_\mu\mf)} = -\sqrt{-g}g^{\mu\nu}(\partial_\nu\phi),
\end{equation}
where $\tilde{\mathcal{L}}$ is the Lagrangian density, defined by the matter field action above. Expanding up to linear order, the scalar field momentum $\pi_\phi^\mu$ is given by
\begin{equation}
\pi^\mu_\phi = -\overline{\sqrt{-g}}\bar{g}^{\mu 0}\partial_0\bar{\phi}-\delta\sqrt{-g}\bar{g}^{\mu 0}\partial_0\bar{\phi}-\overline{\sqrt{-g}}\left(\delta g^{\mu 0}\partial_0\bar{\phi}+\bar{g}^{\mu\nu}\partial_\nu\delta\phi\right),
\end{equation}
which can be split into background and perturbed part
\begin{equation}
\pi_\phi^\mu = \bar{\pi}_\phi^\mu+\delta\pi_\phi^\mu.
\end{equation}

At the background level and in harmonic gauge, $\bar{\pi}_\mf^\mu$ is given by
\begin{equation}
\bar{\pi}_\phi^0 = \partial_0\bar{\phi},\qquad \bar{\pi}_\phi^i = 0.
\end{equation}
The perturbed part of $\mm^\mu$ is given by
\begin{align}
\delta\mm^0 &= \left(-A-3\psi+\nabla^2 E\right)\bar{\mf}'+\delta\mf',\label{eq:classical perturbed mm 0}\\[7pt]
\delta\mm^i &= a^2\bar{\mf}'\partial^iB-a^4\partial^i\delta\mf.\label{eq:classical perturbed mm i}
\end{align}
Applying the zeroth- and first-order equations for $\mf$, the perturbed momentum satisfies the relativistic energy-momentum conservation equation
\begin{equation}
\partial_\mu\delta\mm^\mu = \left(\delta\mm^0\right)'+\partial_i\delta\mm^i = 0.
\end{equation}
Re-expressing this equation in terms of observables that are available in GFT, being $\bar{V},\delta V,\bar{\mf}$ and $\delta\mf$, we find
\begin{equation}\label{eq:classical perturbed mm equation}
\left(\delta\mm^0\right)'-\delta\mf''-\left(\frac{\delta V}{\bar{V}}\right)'\bar{\mf}' = -A'\bar{\mf}'.
\end{equation}
While the left-hand side is given in terms of variables available in GFT, the right-hand side contains the variable $A$, which is not accessible by the GFT observable that are available at the present state. In Sec.~\ref{sec:Discussion}, We comment on the importance of defining additional geometrical observables in GFT.

\paragraph{Classsical Mukhanov-Sasaki-like equation.}

As the transformations of Eqs.~\eqref{eq:pert gauge trafo1} and~\eqref{eq:pert gauge trafo2} show, the harmonic gauge conditions leaves a residual gauge freedom. Under these transformations, the perturbed scalar field $\delta\mf$ changes as
\begin{equation}
\delta\mf\mapsto \delta\mf -\bar{\mf}'\xi^0.
\end{equation}
Given this transformation behavior, one can combine $\psi$ and $\delta\mf$ to a fully gauge-invariant quantity, the so called gauge-invariant curvature perturbation
\begin{equation}\label{eq:classical R}
\mathcal{R} \defeq \psi + \Hubble\frac{\delta\mf}{\bar{\mf}'}.
\end{equation}
Since in harmonic gauge, $\psi$ and $\delta\mf$ satisfy the same equation, $\mathcal{R}$ satisfies~\cite{Battarra:2014tga}
\begin{equation}
\mathcal{R}''-a^4\nabla^2 \mathcal{R} = 0.
\end{equation}

In the context of GFT, one does not have direct access to the quantity $\psi$ but rather to the perturbed volume $\delta V$. For comparison of classical and GFT mechanics, we define the \textit{curvature-like perturbation} $\tilde{\mathcal{R}}$ as
\begin{equation}\label{eq:classical Rtilde}
\tilde{\mathcal{R}}\defeq -\frac{\delta V}{3\bar{V}}+\Hubble\frac{\delta\mf}{\bar{\mf}'}.
\end{equation}
Again, since $\delta V/\bar{V}$ and $\delta\mf$ satisfy the same equation, $\tilde{\mathcal{R}}$ obeys
\begin{equation}\label{eq:classical R perturbation equation}
\tilde{\mathcal{R}}''-a^4\nabla^2\tilde{\mathcal{R}} = 0.
\end{equation}
Notice however, that $\tilde{\mathcal{R}}$ is not gauge invariant but changes as
\begin{equation}
\tilde{\mathcal{R}}\mapsto \tilde{\mathcal{R}}-\nabla^2\xi.
\end{equation}
Still, since $\xi$ is assumed to satisfy the equation above, the equation for $\tilde{\mathcal{R}}$ does not change under gauge transformations. 

\paragraph{Reference field corrections to the perturbation equations.} In this Appendix, we have so far assumed that the matter content is dominated by the scalar field $\mf$, neglecting contributions of the clock and rod fields $\rf^\mu$. However, in order to rule out the possibility that the source term in the GFT perturbation equations~\eqref{eq:GFT perturbed relative volume equation},~\eqref{eq:GFT perturbed mf equation} and~\eqref{eq:GFT perturbed R} is associated to the a potential contribution of they physical reference frame, it is instructive to consider the background and perturbed field equations with all matter terms, similarly to what has been done e.g.\ in~\cite{Husain:2020uac} (albeit with a different physical frame). 

At the background level, the Einstein equations take the form
\begin{equation}
3\Hubble^2 = \frac{1}{2 M_{\Pl}^2}\left(\bar{\mf}'^2+(\kappa^0)^2+3\kappa^2 a^4\right),\qquad  \Hubble' = \frac{1}{M_{\Pl}^2}\kappa^2 a^4,
\end{equation}
which are readily dependent. Clearly, if one assumes $\mm^2$ to dominate, consistency requires $\Hubble' = 0$. 

To compute the perturbed Einstein equations, we add the frame contributions to the perturbed energy-momentum tensor, the components of which are explicitly given by
\begin{align}
M_{\Pl}^2\;\tensor{(\delta T^\rf)}{^0_0} &= \frac{(\kappa^0)^2}{a^6}A-\frac{\kappa^2}{a^2}(3\psi-\nabla^2 E),\\[7pt]
M_{\Pl}^2\;\tensor{(\delta T^\rf)}{^0_i} &= \frac{\kappa^2}{a^4}\partial_i B,\\[7pt]
M_{\Pl}^2\;\tensor{(\delta T^\rf)}{^i_j} &= \delta^i_j\left(-\frac{(\kappa^0)^2}{a^6}A+\frac{\kappa^2}{a^2}(-\psi+\nabla^2 E)\right)-2\frac{\kappa^2}{a^2}\partial^i\partial_j E.
\end{align}
Using the Einstein tensor components of~\cite{Battarra:2014tga}, the perturbed equations of motion are given by
\begin{align}
\frac{1}{2 M_{\Pl}^2}\bar{\mf}'\delta\mf'+3\Hubble\psi'-a^4\nabla^2\psi-\Hubble\nabla^2(E'-a^2B)+\frac{1}{2 M_{\Pl}^2}a^4\kappa^2(3(A+\psi)-\nabla^2 E) &= 0,\\[7pt]
\Hubble A + \psi'+\frac{1}{2 M_{\Pl}^2}\left(\kappa^2a^2B -\bar{\mf}'\delta\mf\right)&= 0,\\[7pt]
E''-a^4\nabla^2 E + \frac{2}{M_{\Pl}^2}\kappa^2 a^4 E &= 0\label{eq:EFE3 rods},
\end{align}
complemented by the harmonic gauge conditions in Eqs.~\eqref{eq:pert HGC}. Already at this point, one observes that the clock contributions, entering with $(\kappa^0)^2$, will cancel out upon the background equations as these only enter with the perturbation function $A$. We leave a clarification of this intriguing cancellation open for future investigations. 

Following similar steps as above, we obtain a differential equation for $\psi$
\begin{equation}
\psi''-a^4\nabla^2\psi+\frac{2}{M_{\Pl}^2}\kappa^2 a^4(A+\psi) = 0\label{eq:psi eq rods},
\end{equation}
that is modified by the presence of the rod contribution, entering with $\kappa^2$. Using the definition of the relative perturbed volume as well as Eqs.~\eqref{eq:EFE3 rods} and~\eqref{eq:psi eq rods}, we obtain 
\begin{equation}
\left(\frac{\delta V}{\bar{V}}\right)''-a^4\nabla^2\left(\frac{\delta V}{\bar{V}}\right) = -\frac{2}{M_{\Pl}^2}\kappa^2a^4\left(\frac{\delta V}{\bar{V}}-3A\right),
\end{equation}
which is modified compared to Eq.~\eqref{eq:classical relative perturbed volume equation} by a term that is controlled via the rod variable $\kappa$. 

The equation for the perturbed matter field $\delta\mf$ is not altered by the additional clock and rod contributions. Thus, using the volume and matter equation, we obtain an equation for $\tilde{\mathcal{R}}$ in the presence of frame contributions:
\begin{equation}
\tilde{\mathcal{R}}''-a^4\nabla^2\tilde{\mathcal{R}} = \frac{2}{M_{\Pl}^2}\kappa^2 a^4\left[\tilde{\mathcal{R}}-A+3\frac{\Hubble}{\bar{\mf}'}\delta\mf+2\frac{\delta\mf'}{\bar{\mf}'}\right].
\end{equation}
Clearly, all of the corrections compared to Eq.~\eqref{eq:classical Rtilde} enter with the rod-variable $\kappa$ and contain a mixture of geometric and matter perturbations. Structurally, these corrections are therefore quite different from those obtained in~\cite{Husain:2020uac}, possibly as a result of the different gauge-fixing. Importantly, this result substantiates the interpretation of the source terms in the perturbation equations obtained from GFT that we give in Sec.~\ref{sec:Solutions of GFT and GR perturbations}. That is, the source terms are not artifacts of reference frame contributions but are rather interpreted as genuine quantum gravity corrections. 

\if
Assuming that the geometrical side of the Einstein equations is correctly given in~\cite{Battarra:2014tga}, we compute here the perturbed energy momentum tensor of the four reference fields $\rf^\mu$ and the matter field $\mm$ to see how rods and clocks enter the perturbed Einstein equations. In accordance with~\cite{Battarra:2014tga}, the components of the $\mf$-part of the energy-momentum tensor, $\tensor{(T^\mf)}{^\mu_\nu}$, are given by
\begin{align}
\tensor{(\delta T^\mf)}{^0_0} &= \frac{1}{a^6}\left[\bar{\mf}'^2 A -\bar{\mf}'\delta\mf'\right],\\[7pt]
\tensor{(\delta T^\mf)}{^0_i} &= -\frac{1}{a^6}\bar{\mf}'\partial_i\delta\mf,\\[7pt]
\tensor{(\delta T^\mf)}{^i_j} &= \frac{\delta^i_j}{a^6}\left[\bar{\mf}'\delta\mf'-\bar{\mf}'^2A\right].
\end{align}
For the reference fields $\rf^\mu$, we plug the solution $\rf^\mu = \kappa^\mu x^\mu$\footnote{Notice that we pick $\kappa^i = \kappa$ as an isotropy condition on the solutions of the reference frame.} into the tensor and then perform the splitting into background and perturbation. In this way we find
\begin{equation}
\begin{aligned}
\tensor{(\delta T^\rf)}{^\mu_\nu} &= \sum_\lambda\left[\delta g^{\mu\sigma}\partial_\sigma\rf^\lambda\partial_\nu\rf^\lambda-\frac{1}{2}\delta^\mu_\nu\delta g^{\alpha\beta}\partial_\alpha\rf^\lambda\partial_\beta\rf^\lambda\right]\\[7pt]
&=
\sum_\lambda(\kappa^\lambda)^2\left[\delta g^{\mu\lambda}\delta^\lambda_\nu-\frac{1}{2}\delta^\mu_\nu\delta g^{\lambda\lambda}\right],
\end{aligned}
\end{equation}
Importantly, indices of perturbed quantities are \emph{not} raised and lowered by the background metric. To see that, consider a generic $2$-tensor $A$. The relation of $\tensor{\delta A}{^\mu_\nu}$ and $A_{\mu\nu}$ is given by
\begin{equation}
\tensor{A}{^\mu_\nu} = g^{\mu\alpha}A_{\alpha\nu} = \bar{g}^{\mu\alpha}\bar{A}_{\alpha\nu}+\bar{g}^{\mu\alpha}\delta A_{\alpha\nu}+\delta^{\mu\alpha}\bar{A}_{\alpha\nu},
\end{equation}
where we identify
\begin{equation}
\tensor{\delta A}{^\mu_\nu} = \bar{g}^{\mu\alpha}\delta A_{\alpha\nu}+\delta g^{\mu\alpha}\bar{A}_{\alpha\nu}.
\end{equation}
For the e.m.-tensor, I checked explicitly that this is correct. 
\fi

\subsection{Change of gauge}\label{sec:Change of gauge} In ordinary cosmology, formulated without the use of reference fields as relational coordinates, the most commonly used coordinates are so-called conformal-longitudinal (CL). In this section, we discuss the change from harmonic to CL-coordinates and the consequences for the interpretation of clock and rod fields. To start with, the line element is given by~\cite{maggiorebook,Dodelson:2003ft}
\begin{equation}\label{eq:CL line element}
\dd{s}^2 = a^2(\tau)\left[-(1+2\Psi(\tau,\vb*{x}))\dd{\tau}^2+(1-2\Phi(\tau,\vb*{x}))\dd{\vb*{x}}^2\right],
\end{equation}
in CL-coordinates, where $\Psi$ and $\Phi$ are the gauge-invariant Bardeen variables. In order to transform from harmonic coordinates $\{x_\H^\mu\}$ to CL-coordinates $\{x_\CL^\mu\}$, we have to perform two transformations, first at the background to conformal time and then a transformation at the level of perturbations. For the background transformation we consider
\begin{equation}
x_\H^0\rightarrow \tau(x_\H^0) = \int\limits_{0}^{x_\H^0}\dd{\tilde{x}^0}a^2(\tilde{x}^0),    
\end{equation}
such that $a^2 \dd{x_\H^0} = \dd{\tau}$. 

Following Eqs.~\eqref{eq:pert gauge trafo1} and~\eqref{eq:pert gauge trafo2}, the perturbation functions change upon the conformal transformation as
\begin{align}
&\psi \mapsto \psi +\mathcal{H}\zeta^0, &&A \mapsto A -\dv{}{\tau}\zeta^0 - 3\mathcal{H}\zeta^0,\\[7pt]
&E \mapsto E -\xi, &&B \mapsto B +\zeta^0-\dv{}{\tau}\xi,
\end{align}
where we introduced the re-scaled variable $\zeta^0 = a^2\xi^0$ {and the Hubble parameter in conformal time, $\mathcal{H} = \frac{1}{a}\dv{a}{\tau}$}. The line element of Eq.~\eqref{eq:CL line element} is then obtained via the transformation
\begin{align}
    \zeta^0 &= E'-B,\\[7pt]
    \xi &= E.
\end{align}
where A and $\psi$ transform to
\begin{align}
A &\longmapsto A -3\mathcal{H}(E'-B)-(E'-B)'\equiv \Psi,\\[7pt]
\psi &\longmapsto \psi+\mathcal{H}(E'-B) \equiv -\Phi.
\end{align}
Indeed as one can easily verify, the perturbation functions $\Psi$ and $\Phi$ are fully gauge-invariant.  

In the new coordinates, the gauge-invariant curvature satisfies
\begin{equation}
\dv[2]{\mathcal{R}}{\tau}+2 \mathcal{H} \dv{\mathcal{R}}{\tau}-\nabla^2\mathcal{R} = 0.
\end{equation}
Since $\tilde{\mathcal{R}}$ is, in contrast to $\mathcal{R}$, not gauge-invariant, it changes under the transformation above to
\begin{equation}
\tilde{\mathcal{R}}\longmapsto \tilde{\mathcal{R}}-\nabla^2 E.
\end{equation}
However, since the equation for $E$ in Eq.~\eqref{eq:pEFE3} is valid manifestly (not only in harmonic gauge) and is the same as that of $\mathcal{R}$,  the equation for $\tilde{\mathcal{R}}$ in conformal-longitudinal gauge is given by
\begin{equation}
   \dv[2]{\tilde{\mathcal{R}}}{\tau}+2\mathcal{H}\dv{\tilde{\mathcal{R}}}{\tau}-\nabla^2\tilde{\mathcal{R}} = 0. 
\end{equation}
As a result, $\tilde{\mathcal{R}}$ satisfies the same equation in harmonic and CL-coordinates, but solutions of $\tilde{\mathcal{R}}$ change accordingly.

We consider next the four reference fields $\{\rf^\mu\}$, which satisfy the Klein-Gordon (KG) equation
\begin{equation}
\partial_a\left(\sqrt{-g}g^{ab}\partial_b\right)\rf^\mu = 0.
\end{equation}
In harmonic coordinates, we chose adapted solutions of the form $\rf^\mu = \kappa^\mu x^\mu$. Plugging this ansatz into the KG equation, one re-obtains the harmonic gauge conditions 
\begin{equation}
c^\mu = \partial_\alpha\left(\sqrt{-g}g^{\alpha\beta}\partial_\beta\right)\kappa^\mu x^\mu\overset{!}{=}0.
\end{equation}
Crucially, since coordinates do in general not transform as scalars nor do the four constraints $c^\mu$. That can also be seen by the relation of $c^\mu$ and the Christoffel-symbols, $c^\mu = \Gamma^\mu_{\alpha\beta}g^{\alpha\beta}$, which are well-known to not transform as tensors. As a consequence, $c^\mu = 0$ is only satisfied in harmonic gauge and is violated in general in other coordinates $\{x^a\}$, i.e. $c^\mu(x^a)\neq 0$. 

The behavior of clocks and rods differs significantly in different coordinates. To see that explicitly, note that the ansatz $\rf^0 = \kappa^0 x^0$ is only a solution of the KG-equation in harmonic coordinates and is invalid in particular in conformal-longitudinal coordinates. In contrast, plugging the ansatz $\rf^i = \kappa^i x^i_\CL$ into the KG-equation in conformal-longitudinal coordinates yields
\begin{equation}
    \partial_i\left(\sqrt{-g}g^{ij}\partial_j\chi^l\right)=\kappa^l\partial_i\left[a^2(1+\Psi-\Phi)\delta^{ij}\partial_jx^l_{\text{CL}}\right]=a^4\kappa^l\partial_l(\Psi-\Phi)=0\,,
\end{equation}
where we expanded the geometric quantities up to first order. The last equation above holds true on-shell and by assuming vanishing shear which is the case in our system. 

Consequently, when plugging the on-shell solution $\rf^i = \kappa^i x^i_\CL$ into the energy-momentum tensor, the background components $\bar{T}_{\mu\nu}(x_\CL)$ contain a contribution of the rods. Explicitly, we find in CL-coordinates the inequalities
\begin{equation}
k^2\gg \mathcal{H}^2\geq \frac{\kappa^2}{2 M_{\Pl}^2},
\end{equation}
where $\mathcal{H}$ is again the Hubble parameter in conformal time, such that for $\kappa = M_{\Pl}^2$, one obtains
\begin{equation}
k^2\gg M_{\Pl}^2.
\end{equation}
Following the arguments of the last paragraph of Sec.~\ref{sec:Solutions of GFT and GR perturbations}, these inequalities show the mixing of sub-horizon and trans-Planckian modes. Concluding, the computations of this section demonstrate that the observations of Sec.~\ref{sec:Solutions of GFT and GR perturbations} are not a particular feature of harmonic coordinates but are also present using the commonly used conformal-longitudinal coordinates.   

\section{Derivation of condensate dynamics}\label{sec:Derivation condensate dynamics}

In this appendix we provide the detailed derivations of the dynamical equations for the perturbed condensate introduced in Sec.~\ref{sec:Perturbed coherent peaked states}. To that end, we consider an expansion of the kinetic kernels
\begin{align}
\mathcal{K}_+((\rf^{0})^2,\mm^2) &= \sum_{n = 0}^\infty\frac{\mathcal{K}_+^{(2n)}(\mm^2)}{(2n)!}(\rf^0)^{2n},\label{eq:K+ expansion}\\[7pt]
\mathcal{K}_-(\abs{\vb*{\rf}}^2,\mm^2) &= \sum_{n = 0}^\infty\frac{\mathcal{K}_-^{(2n)}(\mm^2)}{(2n)!}\abs{\vb*{\rf}}^{2n}\label{eq:K- expansion},
\end{align}
the existence of which is supported by the studies of~\cite{Li:2017uao}. Notice that the reference fields are coupled to the GFT model via the Eqs.~\eqref{eq:spacelike kernel restriction} and~\eqref{eq:timelike kernel restriction}, such that their expansion differs slightly from that discussed in~\cite{Marchetti:2021gcv}. The reduced condensate wavefunctions $\slrcw$ and $\tlrcw$ are expanded in derivatives
\begin{align}
\slrcw(\rf^0+x^0,\mm) &= \sum_{n = 0}^\infty \frac{\slrcw^{(n)}(x^0,\mm)}{n!}(\rf^0)^n,\label{eq:slrcw expansion}\\[7pt]
\tlrcw(\rf^0+x^0,\mm) &= \sum_{n=0}^\infty\frac{\tlrcw^{(n)}(x^0,\mm)}{n!}(\rf^0)^n,\label{eq:tlrcw expansion}
\end{align}
where $\slrcw^{(n)}$ denotes the $n$-th derivative with respect to the clock argument, applying similarly to $\tlrcw$. These expansion will be employed both, for the background as well as the perturbed part of the equations of motion.

\subsection{Background equations}\label{sec:Derivation of background equations}

\paragraph{Spacelike part.} Using the expansions of Eqs.~\eqref{eq:K+ expansion} and~\eqref{eq:slrcw expansion}, the (regularized) spacelike background equation~\eqref{eq:spacelike bkg eom} evaluates to
\begin{equation}
\begin{aligned}
0 &= \int\dd{\rf^0}\mathcal{K}_+((\rf^0)^2,\mm^2)\slrcw(\rf^0+x^0,\mm)\eta_{\epsp}(\rf^0,\pip)\\[7pt]
&= 
\sum_{m,n}\frac{\mathcal{K}_+^{(2m)}(\mm^2)\slrcw^{(n)}(x^0,\mm)}{(2m)!n!}\int\dd[4]{\rf}\eta_{\epsp}(\rf^0,\pip)(\rf^0)^{2m+n}\\[7pt]
&\approx
\mathcal{K}_+^{(0)}\left[\left(I_0+I_2\frac{\mathcal{K}_+^{(2)}}{2\mathcal{K}_+^{(0)}}\right)\slrcw(x^0,\mm)+I_1\partial_0\slrcw(x^0,\mm)+\frac{1}{2}I_2\partial_0^2\slrcw(x^0,\mm)\right]\int\dd[3]{\rf},
\end{aligned}
\end{equation}
Following~\cite{Marchetti:2021gcv}, we introduced the function $I_{2m+n}(\epsp,\pip)$, defined as the $\rf^0$ integration, which can be explicitly evaluated to
\begin{equation}\label{eq:In}
I_{n}(\epsp,\pip) =\mathcal{N}_{\epsp}\sqrt{2\pi\epsp}\left(i\sqrt{\frac{\epsp}{2}}\right)^{n}\e^{-z_+^2}H_n\left(\sqrt{\frac{\epsp}{2}}\pip\right),
\end{equation}
where $H_n$ are the Hermite polynomials and $z_+^2 = \epsp(\pip)^2/2$. We truncated the expansion at order $\epsp$, leading to the condition that only terms with $2m+n\leq 2$ contribute. Introducing the quantities
\begin{align}
E_+^2(\mm) &\defeq \frac{2}{\epsp(2z_+^2-1)}-\frac{\mathcal{K}_+^{(2)}}{\mathcal{K}_+^{(0)}},\\[7pt]
\tpip &\defeq\frac{\pip}{2z_+^2-1},
\end{align}
we finally obtain
\begin{equation}
\partial_0^2\slrcw(x^0,\mm)-2i\tpip\partial_0\slrcw(x^0,\mm)-E_+^2(\mm)\slrcw(x^0,\mm) = 0.
\end{equation}

\paragraph{Timelike part.} On the timelike sector, the procedure to obtain the equations of motion differs slightly because of the different peaking properties of $\tau$ and the mere rod-dependence of the timelike kernel $\mathcal{K}_-$. Starting with Eq.~\eqref{eq:timelike bkg eom} and inserting the expansions of Eqs.~\eqref{eq:K- expansion} and~\eqref{eq:tlrcw expansion}, we obtain
\begin{equation}
\begin{aligned}
0 &= \int\dd[4]{\rf}\mathcal{K}_-(\abs{\vb*{\rf}}^2,\mm^2)\tlrcw(\rf^0+x^0,\mm)\eta_{\epsm}(\rf^0,\pim)\eta_{\delta}(\abs{\vb*{\rf}},\pi_x)\\[7pt]
&= \sum_n\frac{\tlrcw^{(n)}(x^0,\mm)}{n!}\int\dd{\rf^0}\eta_{\epsm}(\rf^0,\pim)(\rf^0)^{n}\int\dd[3]{\rf}\mathcal{K}_-(\abs{\vb*{\rf}}^2,\mm)\eta_{\delta}(\abs{\vb*{\rf}},\pi_x).
\end{aligned}
\end{equation}
Assuming that the spatial integral is non-zero, the equations factorize. Truncating at linear order in $\epsm$ finally yields
\begin{equation}
I_0^-\tlrcw(x^0,\mm)+I_1^-\partial_0\tlrcw(x^0,\mm)+\frac{1}{2}I_2^-\partial_0^2\tlrcw(x^0,\mm)\approx 0,
\end{equation}
where $I_n^-$ is defined equivalently to Eq.~\eqref{eq:In} but evaluated on the timelike peaking parameters $\epsm$ and $\pim$. Introducing
\begin{align}
E_-^2 &\defeq \frac{2}{\epsm(2z_-^2-1)},\\[7pt]
\tpim &\defeq \frac{\pim}{2z_-^2-1},
\end{align}
the background equation for the timelike reduced condensate wavefunction reads
\begin{equation}
\partial_0^2\tlrcw(x^0,\mm)-2i\tpim\partial_0\tlrcw(x^0,\mm)-E_-^2\tlrcw(x^0,\mm) = 0.
\end{equation}
Notice that due to the interplay of peaking and kernel dependencies, the quantity $E_-$ does not carry a matter momentum dependence, in cotrast to $E_+(\mm)$.

\subsection{Perturbation equations}\label{sec:Derivation of perturbation equations}

Continuing the analysis of the equations of motion, we derive in this section the perturbed equations of motion for the spacelike and then the timelike sector. 

\paragraph{Spacelike part.} Starting point is Eq.~\eqref{eq:pert eom sl}, which we complement by the peaking properties of the condensate wavefunctions $\sigma$ and $\tau$. As for the background, we expand $\mathcal{K}_+,\slrcw$ and $\tlrcw$ according to Eqs.~\eqref{eq:K+ expansion},~\eqref{eq:slrcw expansion} and~\eqref{eq:tlrcw expansion}. Also we use the relation of $\delta\Psi$ and $\delta\Phi$ in Eq.~\eqref{eq:relation of dPsi and dPhi} and the relation of peaking parameters in Eq.~\eqref{eq:peaking parameter relations}. Truncating then at linear order in $\epsp$ and $\delta$, one obtains
\begin{equation}
\begin{aligned}
0 &= \mathcal{K}_+^{(0)}(\pmm^2)\Bigg{[}\left(I_0+I_2\frac{\mathcal{K}_+^{(2)}}{2\mathcal{K}_+^{(0)}}\right)\delta\Psi\left(J_{0,\vb*{0}}\bar{\tlrcw}+f\e^{i\theta_f}\bar{\slrcw}\right)+I_1\partial_0\left(\delta\Psi(J_{0,\vb*{0}}\bar{\tlrcw}+f\e^{i\theta_f}\bar{\slrcw})\right)\\[7pt]
&+\frac{I_2}{2}\partial_0^2\left(\delta\Psi(J_{0,\vb*{0}}\bar{\tlrcw}+f\e^{i\theta_f}\bar{\slrcw})\right) + \bar{\tlrcw}I_0\frac{J_{0,(0,0,2)}}{2}\nabla_{\vb*{x}}^2\delta\Psi\Bigg{]}.
\end{aligned}
\end{equation}
All fields, $\slrcw$, $\tlrcw$ and $\delta\Psi$ are evaluated at $x^0$, respectively $x^i$ and the peaked matter momentum $\pmm$. Notice that the first-order time derivative enters with a coefficient $I_1$ and not its complex conjugated because of the the relation between $\pip$ and $\pim$ as well as the phase factor of the function f in Eq.~\eqref{eq:def of f}. The functions $I_0$ and $I_2$ are the functions of temporal peaking parameters defined in the section above, evaluated on the $+$-parameters. Due to the spatial peaking of the timelike condensate $\tau$, coefficients $J_{m,(n_1,n_2,n_3)}$ appear in the expression above, defined as
\begin{equation}
J_{m,(n_1,n_2,n_3)} = \int\dd[3]{\rf}\eta_\delta(\abs{\vb*{\rf}},\pi_x)\abs{\vb*{\rf}}^{2m}\prod_{i=1}^3(\rf^i)^{n_i}.
\end{equation}
The relevant coefficients for the derivation of the equations of motion are $J_{0,\vb*{0}}$, $J_{2,\vb*{0}}$ and $J_{0,(0,0,2)}$, explicitly defined as~\cite{Marchetti:2021gcv}
\begin{align}
J_{0,\vb*{0}} &= -2\mathcal{N}_\delta\sqrt{2\pi}\pi^2\delta^{3/2}z^2\e^{-z^2},\\[7pt]
J_{2,\vb*{0}} &= 4\mathcal{N}_\delta\sqrt{2\pi}\pi^2\delta^{5/2}z^4\e^{-z^2},\\[7pt]
J_{0,(0,0,2)} &= \frac{16}{3}\mathcal{N}_\delta\sqrt{2\pi}\pi\delta^{5/2}z^4\e^{-z^2},
\end{align}
keeping only first-order contributions in the peaking parameter $\delta$, where $z^2 = \delta\pi_x^2/2$. 

Factorizing $I_2/2$ from the spacelike perturbed equations of motion above, we finally obtain
\begin{equation}
\begin{aligned}
0 &= \partial_0^2\left(\delta\Psi(J_{0,\vb*{0}}\bar{\tlrcw}+f\e^{i\theta_f}\bar{\slrcw})\right)-2i\tpip\partial_0\left(\delta\Psi(J_{0,\vb*{0}}\bar{\tlrcw} +f\e^{i\theta_f}\bar{\slrcw})\right)+\\[7pt]
&- E_+^2\delta\Psi\left(J_{0,\vb*{0}}\bar{\tlrcw}+f\e^{i\theta_f}\bar{\slrcw}\right) + \alpha\tlrcw\nabla_{\vb*{x}}^2\delta\Psi,
\end{aligned}
\end{equation}
where the parameter $\alpha$ is defined as
\begin{equation}\label{eq:alpha definition}
\alpha\defeq \frac{I_0 J_{0,(0,0,2)}}{I_2}.
\end{equation}

\paragraph{Timelike part.} To derive the perturbed condensate equation on the timelike sector, given in Eq.~\eqref{eq:dXi equation}, our starting point is Eq.~\eqref{eq:pert eom tl}. We use the expansions of Eqs.~\eqref{eq:K- expansion},~\eqref{eq:slrcw expansion} and~\eqref{eq:tlrcw expansion} as well as the relations of Eqs.~\eqref{eq:relation of dPsi and dPhi} and~\eqref{eq:peaking parameter relations} to arrive at
\begin{equation}
\begin{aligned}
0 &= \int\dd[3]{\rf}\mathcal{K}_-(\abs{\rf}^2,\pmm^2)\left[I_0\delta\Psi\bar{\slrcw}+\bar{I}_1\partial(\delta\Psi\bar{\slrcw})+\frac{I_2}{2}\partial_0^2\left(\delta\Psi\bar{\slrcw}\right)\right]\\[7pt]
  &+ \mathcal{K}_-^{(0)}(\pmm^2)\Bigg{[}\left(I_0J_{0,\vb*{0}}+I_0J_{2,\vb*{0}}\frac{\mathcal{K}_-^{(2)}}{\mathcal{K}_-^{(0)}}\right)\delta\Xi\bar{\tlrcw}+J_{0 ,\vb*{0}}I_1\partial\left(\delta\Xi\bar{\tlrcw}\right)\\[7pt]
  &+ J_{0,\vb*{0}}\frac{I_2}{2}\partial_0^2\left(\delta\Xi\bar{\tlrcw}\right)+I_0\frac{J_{0,(0,0,2)}}{2}\bar{\tlrcw}\nabla_{\vb*{x}}^2\delta\Xi\Bigg{]},
\end{aligned}
\end{equation}
where the coefficients $I_0,I_2,J_{0,\vb{0}},J_{2,\vb*{0}}$ and $J_{0,(0,0,2)}$ are defined as above. Using the background equations of motion in the classical limit, with solutions given by Eqs.~\eqref{eqn:bkgsigma} and~\eqref{eqn:bkgtau}, and factorizing $J_{0,\vb*{0}}I_2/2$, one obtains
\begin{equation}
\begin{aligned}
0 &= \bar{\slrcw}\int\dd[3]{\rf}\mathcal{K}_-(\abs{\vb*{\rf}},\pmm^2)\left[\left(\frac{2I_0}{I_2}+(\pip)^2+\mu_+^2\right)\delta\Psi+2\mu_+\partial_0\delta\Psi+\partial_0^2\delta\Psi\right]\\[7pt]
&+\mathcal{K}_-^{(0)}J_{0,\vb*{0}}\bar{\tlrcw}\Bigg{[}\left(\frac{2I_0}{I_2}+\frac{I_0J_{2,\vb*{0}}}{I_2J_{0,\vb*{0}}}\frac{\mathcal{K}_-^{(2)}}{\mathcal{K}_-^{(0)}}+(\pip)^2+\mu_-^2\right)+2\mu_-\partial_0\delta\Xi+\partial_0^2\delta\Xi+\frac{\alpha}{J_{0,\vb*{0}}}\nabla_{\vb*{x}}^2\delta\Xi\Bigg{]}.
\end{aligned}
\end{equation}
Using the definition of $\mu_-^2$ and introducing
\begin{align}
\beta\defeq  -\frac{I_0J_{2,\vb*{0}}}{I_2J_{0,\vb*{0}}}\frac{\mathcal{K}_-^{(2)}}{\mathcal{K}_-^{(0)}},\qquad \gamma\defeq \frac{\alpha}{J_{0,\vb*{0}}},
\end{align}
the perturbed equation of motion on the timelike sector is finally given by
\begin{equation}
\begin{aligned}
0 &=  \bar{\slrcw}\int\dd[3]{\rf}\mathcal{K}_-(\abs{\vb*{\rf}},\pmm^2)\left[\partial_0^2\delta\Psi+2\mu_+\partial_0\delta\Psi-\frac{\mathcal{K}_+^{(2)}}{\mathcal{K}_+^{(0)}}\delta\Psi\right]\\[7pt]
&+ 
\bar{\tlrcw}\mathcal{K}_-^{(0)}J_{0,\vb*{0}}\Bigg{[}\partial_0^2\delta\Xi+2\mu_-\partial_0\delta\Xi-\beta\delta\Xi+\gamma\nabla_{\vb*{x}}^2\delta\Xi\Bigg{]}.
\end{aligned}
\end{equation}

\bibliographystyle{JHEP}
\bibliography{references.bib}

\end{document}